\documentclass[aps,prb,onecolumn,showpacs,preprintnumbers,amsmath,amssymb, groupedaddress]{revtex4}
\usepackage{dcolumn}
\usepackage{bm}

\begin{document}

\title{A Mathematically Controlled Alternative to the Supersymmetric Sigma Model of Disorder}

\author{Vincent E. Sacksteder IV}
 \email{vincent@authors-last-name.com}
 \homepage[]{www.sacksteder.com}
 \affiliation{Asia Pacific Center for Theoretical Physics, Hogil Kim Bldg. \#501, POSTECH, San 31, Hyoja-dong, Namgu, Pohang, Gyeongbuk 790-784, Korea  }

\date{\today}

\begin{abstract}
This paper shows how to obtain non-rigorous mathematical control over models of loosely coupled disordered grains; it lays groundwork for rigorous proofs and provides new information about saddle point structure and perturbative corrections.  Both the Wegner model and a variant due to Disertori are transformed to matrix models which are similar to the supersymmetric model of disorder, having two matrices $Q^f$ and $Q^b$ which correspond to the two bosonic sectors of the SUSY matrix.  However the Grassman (fermionic) sector of the SUSY matrix is omitted, and compensated by a spectral determinant.  The transformation is exact for Disertori's model, while for the Wegner model it involves an integral which can be approximated while maintaining mathematical control.  After this transformation a saddle point approximation is used to integrate the matrix eigenvalues, resulting in a sigma model.  Previous derivations of  sigma models of disorder assumed a spatially uniform saddle point independent of the Goldstone bosons and found that corrections are well controlled in the large $N$ limit.  This paper takes into account spatial fluctuations of the Goldstone bosons and finds that corrections to the sigma model approximation in extended systems  are controlled by powers of the inverse conductance $1/g$.   The sigma model approximation is well controlled only well within the localized regime and far from the band edge.  In other treatments energy band information is repackaged within phenomenological parameters; here this information is preserved explicitly, with clear prescriptions for incorporating it into calculations of observables.  After performing the sigma model approximation this paper specializes to the weak localization regime, where the kinetics are diffusively small but at the same time dominate the disorder.  In this regime Disertori's model exhibits remarkable simplifications and is completely controlled by perturbative expansions in various small parameters.  The Wegner model likely can also be controlled but this would require further analysis of the saddle point equations and the determinant.  The simplified weak localization model seems to be equivalent to the SUSY sigma model.  The standard weak localization results of the supersymmetric sigma model, including anomalously localized states, are reproduced and extended. 
\end{abstract}


\maketitle

\section{\label{Introduction}Introduction}
Fifty-one years ago Anderson argued that materials with sufficiently strong disorder - i.e. complex irregularities at a fine scale - do not conduct \cite{Anderson58}.    After many theoretical developments, rigorous mathematical control has been obtained only in the fully localized phase \cite{Frohlich85}, and in the tails of the conduction band.  Wegner's model of weakly coupled disordered grains with $N$ orbitals per grain \cite{Schafer80}, along with the supersymmetric sigma model \cite{Efetov97} which can be derived from it, are believed to capture the essential physics of both disordered conduction and localization, but have remained mathematically tractable only in the conducting phase with small disorder, or in special geometries.  Even here full mathematical rigor is not preserved except in $D = \{0,1\}$ dimensions.

Wegner's model defines an ensemble of random Hermitian\footnote{Here we develop the unitary ensemble characterized by Hermitian matrices.  This work can easily be generalized to the orthogonal ensemble\cite{Fyodorov02a} and probably also the symplectic ensemble.} Hamiltonians and prescribes that observables be averaged over that ensemble.    It is conventional to adopt a field theory approach oriented toward computing this theory's observables using certain generating functions which can be written as path integrals. For instance, the energy level correlator $R_2(E_1, E_2)$ may be written as the second derivative with respect to $\hat{E}$ of a generating function: 
\begin{equation}
{\frac{{{det}{({{\hat{E}^f}} - H)}}}{{det}{({{\hat{E}^b}} - H)}}} \propto  \int {{d\overline{\psi}}{d{\psi}}} {dS}  e^{\frac{\imath}{2}  S L {({{\hat{E}^b}} - H)} {S^{*}} +  \frac{\imath}{2}\psi {({{\hat{E}}^f - H} )}  \overline{\psi} }
\nonumber 
\label{SUSYGeneratingFunctionExact}
\end{equation}
 The $\psi_{vni}$  vector contains $2N $ complex Grassman variables at each site, while $S_{vni}$ contains $2N$ complex scalars at each site. $n$ specifies the orbital, $v$ specifies the lattice site, and the two possible values of $i = \{1,2\}$ correspond to the two energy levels, whose energies are given by $\hat{E}_i$. $L$ is diagonal sign matrix introduced for convergence reasons.  
 
When $\hat{E}$ is proportional to the identity, the Wegner model has an exact symmetry under unitary transformations $\psi_i \rightarrow \sum_{i_2} U_{i i_2} \psi_{i_2}$, and another exact matrix symmetry under hyperbolic transformations of $S$.  Therefore one expects that if the energy level splitting $\omega = E_1 - E_2$ is small enough, then the physical degrees of freedom are not $2N$-component vectors  but instead two $2\times 2$ matrices: one Hermitian and the other a member of the pseudo-Hermitian group corresponding to hyperbolic transformations \cite{Fyodorov02}.

Efetov pursued a program of first averaging over all Hamiltonians and then making an exact change of variables from the original vectors to matrices.  His result has at each site a single $4 \times 4$ graded matrix $Q$ which contains the two required Hermitian and pseudo-Hermitian matrices, along with additional Grassman variables in the off-diagonal blocks \cite{Efetov97}.  A non-local and nonlinear action $\frac{N}{2}{Tr}(Q^2) + N  {Tr}(Q \hat{E}) + N \,{Tr}(\ln (Q + \epsilon k ) ) $ controls the graded matrices \cite{Mirlin99}.   The  $N$ multiplying the action invites a saddle point approximation which would constrain $Q$'s eigenvalues and leave only angular variables as dynamical degrees of freedom.  Although a correct application of the saddle point technique would require taking into account fluctuations in the saddle point induced by the fluctuations in the angular variables, until now all sigma model derivations have neglected such fluctuations and have assumed that the saddle point is spatially uniform \cite{Schafer80, Verbaarschot85, Efetov97,   Mirlin99}.  It has also been conventional during this procedure to avoid explicit consideration of the details of the band structure, and to instead introduce phenomenological constants.  After fixing the eigenvalues and making a Taylor series expansion of the logarithm in powers of $Q$'s fluctuations, Efetov arrived at the famous supersymmetric sigma model.

In this article I follow the slightly different path of Fyodorov, as developed by Disertori.  In the case of a single grain, Fyodorov showed how to make an exact transformation \cite{Fyodorov02, Fyodorov02a} from the original Wegner model  to $2 \times 2$ matrices $Q^f$ and $Q^b$.  Disertori followed Fyodorov very closely but considered Hamiltonians which are entirely random, and showed that Fyodorov's exact transformation succeeds in any dimension on a very wide class of lattices \cite{DisertorisModel}.  The principal difference between the two models is that Wegner's density of states is determined by the kinetic operator $k$, while Disertori's density of states is just the semicircular distribution of random matrix theory.  We analyze Disertori's model and consider also the Wegner model.  The latter model contains a third matrix degree of freedom  in addition to the $Q^f$ and $Q^b$ matrices.   These $W$ matrices exhibit spontaneous symmetry breaking and therefore can be integrated while maintaining mathematical control.

The matrix models produced by Fyodorov's transformations are similar to Efetov's model: the Lagrangian's terms have direct correspondences to Efetov's action $\frac{N}{2}{Tr}(Q^2) + N  {Tr}(Q \hat{E}) + N \,{Tr}(\ln (Q + \epsilon k ) ) $.     However there are no Grassman variables in Fyodorov's matrices, and in compensation the path integral contains a determinant coupling $Q^f$ with $Q^b$.  The overall effect is the same as if one had started with Efetov's model prior to any approximations and then somehow managed to integrate exactly the Grassman variables.  This route has been until now impassable because Efetov's logarithm contains all powers of the Grassman variables.  Fyodorov's alternative procedure bypasses this difficulty and is still exact in Disertori's case and well controlled in Wegner's case.

Having converted to matrix variables, we analyze Disertori's model in detail, taking care to maintain mathematical control at each step.  There are two principal challenges. The first is to integrate the matrix eigenvalues in a controlled manner, in contrast to the conventional SUSY procedure which presumes a spatially uniform saddle point. In contrast to previous treatments we find that the sigma model approximation to both the Wegner model and the Disertori model is corrected by powers of $1/g$ not $1/N$, and fails near the band edge and in the localized regime. The second challenge is to control the $Q^f - Q^b$ coupling.  This is a very intricate object, but simplifies drastically in the weak localization regime, where fluctuations in $Q^f$ and $Q^b$ are small and the kinetics dominate the disorder.  In this regime the $Q^f - Q^b$ coupling simplifies to a zero-momentum coupling, a normalization constant, and corrections proportional to $1/g$.  In the end we reproduce and extend the standard SUSY results about the two level correlator and anomalously localized states, including the correct normalization constant.  Our results go beyond the SUSY results by incorporating explicit band information.  Additionally we find in the two level correlator an extra term controlled by on site elements of the Green's function, but this term may have been simply overlooked by previous authors, or could possibly be cancelled by higher orders of $1/g$ perturbation theory.   A partial analysis of the Wegner model suggests that it also agrees with SUSY results for observables in the weak localization regime, but completion of this analysis waits on further perturbation theory results about the $dW$ integral and more precise solutions of the saddle point equations.
 
Our results suggest that the simplified weak localization version of the  models treated here is equivalent to the SUSY sigma model.  Corrections to the sigma model approximation and explicit band structure information could likely be obtained within the SUSY formalism.    What seems more doubtful is whether outside of the weak localization regime SUSY could correctly reproduce the present article's derivation of the correct $Q^f - Q^b$ coupling.

Section \ref{ModelDerivation} of this article performs exact derivations of matrix models equivalent to Wegner's and Disertori's models.  The rest of this paper concentrates on Disertori's model, but includes asides about the Wegner model.  Section \ref{SaddlePoint} performs the saddle point approximation, section \ref{WeakLocalization} obtains control of the $Q^f - Q^b$ coupling in the weak localization regime, section \ref{Dimensionality} integrates fluctuations in $Q^f$ and $Q^b$, section \ref{Observables} shows that the standard SUSY results for weak localization are reproduced and extended, and section \ref{MathematicalControl} reviews how we maintained mathematical control throughout.  Lastly we briefly discuss new vistas for understanding disorder outside of the weak localization regime.
 
 \section{Derivation of the Matrix Models\label{ModelDerivation}}
 In the next pages we will develop Fyodorov's approach \cite{Fyodorov02} to disordered systems.  Beginning with a general model of loosely coupled grains, we will move by exact steps to a path integral with vector degrees of freedom and then to matrix degrees of freedom.  The Wegner model and Disertori's model will be considered together through most of the derivation.  Both models prescribe random ensembles of Hermitian Hamiltonian matrices modelling conduction through a system.  The system is modelled as a graph with $V$ sites, and we will avoid taking the continuum limit in order to preserve flexibility and mathematical rigor.  There are $N$ basis elements at each site, so the total basis size is $N \times V$. The lower case letters $n, v$ will denote respectively the orbital index and the position.

\subsection{The Observables}

The field theory approach to disordered systems computes generating functions which are tailor made for specific observables; therefore one must start by choosing the observable to be computed.  In this paper we will calculate the averaged two point correlator, averaging over the ensemble of Hamiltonians:
 \begin{equation}
  {R_{2}{(E_{1}, E_{2})}} \equiv {\frac{\overline{{{Tr}({\delta{(E_{1} - H)}})} \, {{Tr}({\delta{(E_{2} - H)}})} }}{\overline{\rho{(E_{1})}} \quad  \overline{\rho{(E_{2})}}}}, \;\; \overline{\rho{(E)}} \equiv {{Tr}{(\overline{\delta{(E - H)}})}}
 \label{TwoPointDefinitionExact}
 \end{equation}
 $\delta$ in this context is the matrix delta function.  There are also spatially resolved versions of these quantities; the first line of the next equation writes them in terms of matrix delta functions, while the second line gives them  in terms of $H$'s eigenvalues $\epsilon$ and eigenfunctions $\psi_\epsilon$.
  \begin{eqnarray}
 \overline{\rho{(E, v)}} & \equiv & {\sum_n {(\overline{\delta{(E - H)}})_{nn vv}}}, \;\; {R_{2}{(E_{1}, E_{2}, v_1, v_2)}} \equiv {\frac{\sum_{n_1 n_2} \overline{{ ({\delta{(E_{1} - H)}})_{n_1 n_1 v_1 v_1}} \, {({\delta{(E_{2} - H)}})_{n_2 n_2 v_2 v_2} }}}{\overline{\rho{(E_{1}, v_1)}} \quad  \overline{\rho{(E_{2}, v_2)}}}}
 \nonumber \\
 \overline{\rho{(E, v)}} &=& \sum_\epsilon \delta(E - \epsilon) | \langle \psi_\epsilon | v \rangle |^2, \;\; {R_{2}{(E_{1}, E_{2}, v_1, v_2)}} \equiv {\frac{\sum_{\epsilon_1 \epsilon_2} \overline{{ {\delta{(E_{1} - \epsilon_1)}}} \, {{\delta{(E_{2} - \epsilon_2)}} } | \langle \psi_{{\epsilon}_1} | v_1 \rangle |^2 | \langle \psi_{{\epsilon}_2} | v_2 \rangle |^2}}{\overline{\rho{(E_{1}, v_1)}} \quad  \overline{\rho{(E_{2}, v_2)}}}}
 \label{SpatialTwoPointDefinitionExact}
 \end{eqnarray}
 
  In the zero-dimensional gaussian unitary ensemble $R_2$ has been calculated using many different techniques.  Taking $\tilde{\epsilon}$ as the characteristic scale of the disorder, it is
  \begin{equation}
  {R_{2}{(E + \omega/2, E - \omega/2)}} = {{\delta{(\omega \overline{\rho(E)})}} + {1} - {\frac{\sin^2{(\pi \omega \overline{\rho(E)})}}{\pi^2 \omega^2 (\overline{\rho(E)})^2}}}, \;\; \overline{\rho{(E)}} = {\frac{N}{\pi \tilde{\epsilon}} \sqrt{1 - \frac{E^{2}}{4{\tilde{\epsilon}}^{2}}} }
 \label{DOSZeroCorrectResultExact}
 \end{equation}

 \subsection{Green's Functions and their Generating Functions}
In the field theory approach one rewrites the delta functions  in terms of Green's functions. 
 It is important to distinguish between the advanced and retarded Green's functions: the former is ${G_{A}{(E)}} \equiv {(E - {\imath \nu }- H)}^{-1}$, while the latter is ${G_{R}{(E)}} \equiv {(E + {\imath \nu }- H)}^{-1}$.  $\nu$ is infinitesimally small, positive, and destined to be set to zero sometime during the calculation.  The scalar $\delta$ function can be represented as ${\frac{1}{ \pi} \lim_{\nu \rightarrow 0} {{Im}({(E - {\imath \nu} - \acute{E})}^{-1})}}$; therefore ${\rho{(E)}} = {\frac{1}{ \pi} \lim_{\nu \rightarrow 0} {{Im}{({Tr}{(G_{A}{(E)})})}}}$.  Similarly, the two point correlator can  be extracted from product of two Green's functions:
\begin{eqnarray}  
{{\rho{(E)}}{\rho{(\acute{E})}}} & = & {\frac{1}{ {\pi}^{2}} \lim_{\nu \rightarrow 0} {{Im}{({Tr}{(G_{A}{(E)})})}} {{Im}{({Tr}{(G_{A}{(\acute{E})})})}}}
\nonumber \\
& = & {-\frac{1}{4 {\pi}^{2}} \lim_{\nu \rightarrow 0} [ {({{Tr}{(G_{A}{(E)}})} - {{Tr}{(G_{R}{(E)}})})} \times {{({{Tr}{(G_{A}{(\acute{E})})}}  - {{Tr}{(G_{R}{(\acute{E})})}})}} ]}
\nonumber \\
& = & -\frac{1}{4 {\pi}^{2}} \lim_{\nu \rightarrow 0}  [{{{Tr}{(G_{A}{(E)}})}{{Tr}{(G_{A}{(\acute{E})})}}} 
+ {{{Tr}{(G_{R}{(E)}})}{{Tr}{(G_{R}{(\acute{E})})}}}
\nonumber \\ 
&  & \qquad\qquad - {{{Tr}{(G_{A}{(E)}})}{{Tr}{(G_{R}{(\acute{E})})}}} 
- {{{Tr}{(G_{R}{(E)}})}{{Tr}{(G_{A}{(\acute{E})})}}}]
\\ \nonumber 
& = & {\frac{{{Re}{(R_{RA})}} - {{Re}{(R_{AA})}}}{2 {\pi}^{2} \quad \overline{ \rho{(E_{1})}} \quad  \overline{\rho{(E_{2})}}}}, \;\;
R_{RA}  =  \lim_{\nu \rightarrow 0} \overline{{Tr}{(G_{R}{(E)})}{{Tr}{(G_{A}{(\acute{E})})}}}, \;\; R_{AA} = \lim_{\nu \rightarrow 0}  \overline{{Tr}{(G_{A}{(E)})}{{Tr}{(G_{A}{(\acute{E})})}}}
\label{TwoLevelZeroGreensAExact}
\end{eqnarray}
In general, calculating an $n$-point correlator requires averaging products of $n$ Green's functions.  

It is mathematically convenient to calculate not Green's functions but instead their generating functions.  First consider a function which is sufficient for generating a single Green's function:
\begin{equation}
{Z{(E^f, E^b, J)}} = {\frac{{{det}{(E^f - H - J^f)}}}{{det}{(E^b - H - J^b)}}}
\end{equation}
$J$ is a source matrix ${J} = {{J}_{v_{1} v_{2} } \delta_{n_{1} n_{2}}  }$. $E^b$ has a small imaginary part $\nu$ which one chooses to be either negative if one wants an advanced Green's function or instead positive if one wants a retarded Green's function.  All other quantities are  Hermitian.  When the source $J$ is set to zero and the two energies are set to be equal, the generating function ${Z{({E^f = E^b}, {J^f = J^b = 0})}}$ is equal to one.  Using the identity  ${{det}{(A)}} = {{e}^{{Tr}{(\ln{(A)})}}}$ one can easily prove that ${\frac{dZ}{dJ^b_{ v_{1} v_{2}}}} = {Z \times \sum_n { {{\langle n v_{2} |} {(E^b - H  - J^b)}^{-1} {| n v_{1} \rangle} }}}$.  

One can obtain a Green's function from $Z$ by first taking its derivative with respect to $J^b$ and then setting $E^f = E^b, J^f = J^b = 0$.  Similarly, one can take a derivative with respect to $J^f$, set $E^f = E^b, J^f = J^b = 0$, multiply by $-1$, and obtain the exact same result.  The identity of the $J^f$ derivative  with the $J^b$ derivative is a  Ward identity for this theory, a manifestation of the symmetry between the determinant in the numerator and the determinant in the denominator.  The numerator-denominator symmetry is part of the graded matrix supersymmetry of the SUSY sigma model, and is very important for understanding disordered systems.
  
If one wants globally averaged quantities one may take derivatives with respect to $E$;  $J$ is useful only for spatially resolved quantities.  We will put a tilde over $\tilde{J}$ as a reminder of this.

The generating function for a product of $I$ Green's functions is a product of $I$ ratios:
\begin{equation}
Z  \equiv \frac{\prod_{i=1}^{I} \det(E^f_i - H - \tilde{J}^f_i)}{\prod_{j=1}^{I} \det(E^b_j - H - \tilde{J}^b_j)}
\end{equation}

Instead of keeping track of all $2I$ determinants separately, it is convenient to merge them into one determinant in the numerator and one in the denominator.  For instance, consider two matrices $A$ and $B$ which each inhabit a basis with $M$ basis elements.  One can invent a new basis with $2M$ basis elements - $M$ elements for $A$ and $M$ for $B$. In the new basis $AB = BA = 0$ and therefore ${{\det{(A)}} \times {\det{(B)}}} = {\det{(A+B)}}$.  This is a sleight of hand: one avoids the complication of having two determinants at the expense of using a basis which is twice as large.  We apply this trick to $Z$, where each determinant inhabits a basis with $N V$ basis elements.  In order to group the numerator's determinants together, we create a new basis with $NVI$ basis elements, denoted by the indices $nvi$.  The denominator's basis is similar, denoted by the indices $nvj$.   The new generating function is
\begin{eqnarray}
Z & = &  \frac{\det({{\hat{E}}^{f}} - H - \tilde{J}^f)}{\det({{\hat{E}}^{b}} - H - \tilde{J}^b)}, \;\;
{\hat{E}}^f  \equiv  E^f_i \, \delta_{v_1 v_2} \delta_{i_1 i_2}  \delta_{n_1 n_2}, \; {\hat{E}}^b \equiv \, E^b_j \delta_{v_1 v_2} \delta_{j_1 j_2}  \delta_{n_1 n_2},
\nonumber \\
{\tilde{J}}^f & \equiv & \tilde{J}^f_{i v_1 v_2}  \delta_{i_1 i_2}  \delta_{n_1 n_2}, \; \tilde{J}^b \equiv J^b_{j v_1 v_2} \delta_{j_1 j_2}  \delta_{n_1 n_2},\;\;
H^f  \equiv  H_{v_1 v_2} \delta_{i_1 i_2} \delta_{n_1 n_2}, \; H^b  \equiv  H_{v_1 v_2} \delta_{j_1 j_2} \delta_{n_1 n_2}
\end{eqnarray}

We will consider a more general generating function with $I^f$ determinants in the numerator and $I^b$ determinants in the denominator:
\begin{eqnarray}
Z & \equiv &\frac{\prod_{i=1}^{I^f} \det(E^f_i - H - \tilde{J}^f_i)}{\prod_{j=1}^{I^b} \det(E^b_j - H  - \tilde{J}^b_j)}
= \frac{\det({{\hat{E}}^{f}} - H - \tilde{J}^f)}{\det({{\hat{E}}^{b}} - H- \tilde{J}^b)}
\end{eqnarray}
The extra generality can be useful for obtaining a toy model of unquenched QCD, for checking the theory's constants and scaling, and for investigating various special cases.

The next step is to rewrite the generating function as a path integral, converting the determinant in the numerator into an integral over Grassman variables and the determinant in the denominator into an integral over bosonic variables.  If $A$ is an Hermitian $M \times M$ matrix, $\psi, \overline{\psi}$ are vectors of $M$ Grassman variables,  and $S$ is a vector of $M$ complex bosonic variables, then
\begin{equation}
{ {{det}{(A)}}} = {{(-2\imath)}^{M}{\int{{{d\overline{\psi}}{d{\psi}}} e^{\frac{\imath}{2} \psi A \overline{\psi}}}}}, \;\; {{ {{det}^{-1}{(A)}}}} = {{(2 \pi \imath)}^{-M}{\int{{{dS^{R}}{dS^{I}}}} e^{\frac{\imath}{2} S A {S^{*}}}}}, \;\; { d \overline{\psi}}{ d {\psi}} \equiv {\prod_{m}^{M} {d\overline{\psi}}_{m} d{\psi}_{m} }
\label{FermionDeterminantExact}
\end{equation}
However the bosonic integral diverges if the imaginary part of $A$ is not positive definite.  This seemingly easy point causes some of the interesting mathematical intricacies in the supersymmetric theory.  Using these identities, we rewrite the generating function:
\begin{eqnarray}
{Z} & = & {(2 \pi)}^{-V N I^{b}}  {(\imath / 2)}^{-V N I^{f}} {(\imath)}^{- N V I^b} {(\det L)}^{NV}
 {\int {d\overline{\psi}}{d{\psi}} {{dS^{R}}{dS^{I}}}  e^{\frac{\imath}{2} S L {({{\hat{E}}^{b}} - H - \tilde{J}^b)} {S^{*}}} e^{\frac{\imath}{2} \psi {({{\hat{E}}^{f}} - H - \tilde{J}^f)} \overline{\psi}} }
\nonumber \\
 {L_{n_{1} n_{2} v_{1} v_{2} j_{1} j_{2}}} &\equiv &{{{sign}{(Im{({E}^{b}_{j})})}} {\delta_{n_{1} n_{2}}} {\delta_{v_{1} v_{2}}} {\delta_{j_{1} j_{2}}}}
\label{SUSYGeneratingFunction}
\end{eqnarray}
 The $\psi_{vni}$  vector contains $I^f N $ complex Grassman variables at each site, while $S_{vnj}$ contains $I^b N$ complex scalars at each site.   Both vectors have dimensions of $[ Energy ]^{-1/2}$.  $L$ is a diagonal sign matrix introduced to ensure the convergence of the bosonic integrals.  Its signs correspond to the choice of retarded vs. advanced Green's functions, and are $+1$ for the retarded case and $-1$ for the  advanced case.
 
 One could also introduce a sign matrix similar to $L$ in the fermionic integrals, and their guaranteed convergence allows one to choose any combination of signs one likes.  Verbaarschot et al\cite{Verbaarschot85} explored this freedom in the context of the supersymmetric sigma model, and discovered that in that context one must choose the signs to be all the same, and thus obtain a compact representation for the fermionic variables.  In the approach being developed here, one can choose any signs one likes, but soon one integrates out the fermions and then the fermionic sign matrix factors out entirely; one is again forced to use a compact representation.

\subsection{The Models}

We will analyze models of disordered grains with $V$ grains and $N$ orbitals in each grain.  The Hermitian Hamiltonian $H = \epsilon_0 H_0 + \tilde{\epsilon} \tilde{H}$ has a deterministic part $H_0$ and a random part $\tilde{H}$ which is fully described by the second moment ${\overline{{\tilde{H}_{n_{1} n_{2} v_{1} v_2}} {\tilde{H}_{n_{3} n_{4} v_{3} v_4}} }} = {{N}^{-1} {(1-k)_{v_1 v_2}} {\delta_{n_{1} n_{4}}} {\delta_{n_{2} n_{3}}} {\delta_{v_{1} v_{4} }} {\delta_{v_{2} v_{3} }} }$.   The system geometry, including the number of dimensions and all other structural details, is encoded in the positive indefinite kinetic operator $k$ and in $H_0$.  We  work in the diffusive limit $\langle \vec{s} | k|\vec{s} \rangle \ll 1 \,\, \forall \, \vec{s}$, where $|\vec{s}\rangle$ are the momentum basis functions.  We also require that $k$ and $H_0$ be Laplacians, meaning that $k|\vec{0}\rangle = 0$ and $ H_0 | \vec{0} \rangle = 0$.  $\epsilon_0$ and $\tilde{\epsilon}$ are the energy scales of the two operators.

The traditional Wegner model \cite{Schafer80} is obtained by setting $H_0 = -\nabla^2$  and $k = 0$, so that the kinetic term is deterministic and the random potential is local.  Disertori \cite{DisertorisModel} proposed a more tractable model where $H_0 = 0$ and both the kinetics and the potential are random.  We will analyze both models. 
 
\subsection{Averaging over the Disorder}

We want to calculate the average of products of Green's functions, so the next step is to average the generating function $Z$ over realizations of the random potential.    One expands formula \ref{SUSYGeneratingFunction} as a power series in the random potential $\tilde{H}$, counts pairings of $\tilde{H}$, and then substitutes the second moment for each pairing.  The term to be averaged is
${\overline{\exp{(\frac{-i}{2} \tilde{\epsilon} {({SL \tilde{H} S^{*}} + {\psi \tilde{H} \overline{\psi}})})}}} = {\sum_{l=0}^{\infty}{\frac{1}{l!} {(\frac{-\imath \tilde{\epsilon}}{2})}^{l}  \overline{{({SL \tilde{H} S^{*}} + {\psi \tilde{H} \overline{\psi}})}^{l}}}}
$. Odd moments of $\tilde{H}$ average to $0$, while even moments average to constants times powers of the second moment;
\begin{equation}{\sum_{l=0}^{\infty}{\frac{N_p(2l)}{(2l)!} {{(\frac{-\imath \tilde{\epsilon}}{2})}^{2l}
  { \left[ \overline{{({SL\tilde{H}S^{*}} + {\psi \tilde{H} \overline{\psi}})}^{2}} \right]}^{l}}}} 
  = e^{-2^{-3} {\tilde{\epsilon}^2} \overline{{({SLHS^{*}} + {\psi H \overline{\psi}})}^{2}}}
, \; {N_{p}{(2a)}} 
={{(2 \pi)}^{-1/2} {\int {d\alpha} {\alpha}^{2a} e^{-\frac{1}{2}{\alpha}^{2}}}}
= {(2a - 1)!}
\end{equation}
Inserting the formula for $\tilde{H}$'s second moment completes the averaging process;
\begin{eqnarray}
{\bar{Z}} &=& {\gamma {\int {d\overline{\psi}}{d{\psi}} {{dS^{R}}{dS^{I}}}{\exp}{(\mathcal{L})}}}, \;\; \gamma  \equiv  {{(2 \pi)}^{-V N I^{b}}  {(\imath  / 2)}^{-V N I^{f}} {(\imath)}^{- N V I^b} {(\det L)}^{NV}}
\nonumber \\
{\mathcal{L}} & \equiv & {\imath N \tilde{\epsilon}^{-1} {Tr}_{v j}( \hat{S} L {({{\hat{E}}^{b}} - \epsilon_0 H_0 - {\tilde{J}}^b)}) } + {\frac{\imath}{2} \psi {({{\hat{E}}^{f}} - \epsilon_0 H_0 - {\tilde{J}}^f)} \overline{\psi}}  
\nonumber \\
& - & {\frac{N}{2}{(-{\sum_{v_1 v_2} {(1-k)_{v_1 v_2}}{Tr}{({\hat{\psi}}_{v_1} {\hat{\psi}}_{v_2})} } + {\sum_{v_1 v_2} {(1-k)_{v_1 v_2}} {Tr}{({\hat{S}}_{v_1 v_1} L {\hat{S}}_{v_2 v_2} L)}} + 2X )}}
\nonumber \\
 { X} & \equiv & \frac{\tilde{\epsilon}^2}{2N} {\sum_{v_1 v_2 i_1 j_1 n_1 n_2} {(1-k)_{v_1 v_2}} {\psi_{n_{1} v_{1} i_{1}}} {\overline{\psi}_{n_{2} v_{2} i_{1}}} {S_{n_{2} v_{2} j_{1}}} {S^{*}_{n_{1} v_{1} j_{1}}} {L_{j_1}}}
 \nonumber \\
{\hat{\psi}_{v_{1} v_{2} i_{1} i_{2}}} & \equiv & \frac{\tilde{\epsilon}}{2N} {\sum_{n} {\overline{\psi}_{n v_{1} i_{1}}} {\psi_{n v_{1} i_{2}}} {\delta_{v_{1} v_{2}}}}, \;
 {\hat{S}_{v_{1} v_{2} j_{1} j_{2}}} \equiv \frac{\tilde{\epsilon}}{2N} { \sum_{n} {S^{*}_{n v_{1} j_{1}}} {S_{n v_2 j_{2}}} } 
\end{eqnarray}
The minus sign in the quartic $\psi$ term was caused by the anticommutation of fermions and is important for the theory's convergence. The new dimensionless matrices $\hat{S}$ and $\hat{\psi}$ give an early hint at this theory's matrix structure.

\subsection{Hubbard-Stratonovich Conversion of the Fermionic Variables}
Up to this point we have kept company with Efetov  in his development of the supersymmetric sigma model \cite{Efetov97}.  Now we part from him, taking a parallel path.  Both paths convert from vector variables $S$ and $\psi$ to matrices and then perform various approximations.  The distance between the paths is caused by a different choice of matrix variables: Efetov chose graded matrices, while we follow Fyodorov in choosing ordinary matrices. 

First we do an exact Hubbard-Stratonovich transformation of the fermionic vector $\psi$ into a bosonic $I^{f} \times I^{f}$ Hermitian matrix $Q^{f}$.   If $1-k$ is positive definite, 
\begin{eqnarray}
{{\exp}{(\frac{N}{2} {\sum_{v_1, v_2} (1-k)_{v_1, v_2} {Tr}{({\hat{\psi}}_{v_1}{\hat{\psi}}_{v_2})}})}} & = & {(N / 2 \pi)}^{V {I^{f}}^{2} / 2} 2^{V I^{f} {(I^{f} -1)} / 2} {(\det (1-k))}^{{I^{f}}^{2} / 2}
\nonumber \\
& \times & {\int {{dQ} {{\exp}{(-{\frac{ N}{2} {\sum_{v_1 v_2} (1-k)^{-1}_{v_1 v_2} {Tr}(Q_{v_1} Q_{v_2})}} \pm {N{ \sum_{v} {Tr}{(Q^{f}_v {\hat{\psi}}_v )}}})}} }},
\nonumber \\
dQ & = & (\prod_l {dQ_{ll}}) (\prod_{l < m} {dQ^R_{lm}} {dQ^I_{lm}}) \; =  \Delta_{VdM}^2(x) \, {dU} \, {dx} , \;\; Q = U x U^\dagger
\end{eqnarray}
$U x U^\dagger$ gives $Q$'s decomposition into eigenvalues $x$ and a unitary matrix $U$. $\Delta_{VdM}(x) = \prod_{i_1 < i_2}(x_{i_1} - x_{i_2})$ is the Van der Monde determinant.  We apply this transformation at each site in the system's volume, with a different $Q^{f} = U x^f U^\dagger$ at each point.  
\begin{eqnarray}
{\bar{Z}} & = & {\gamma {\int { {dU} \, {dx}} \, {d\overline{\psi}}{d{\psi}} {{dS^{R}}{dS^{I}}} \, \prod_v \Delta_{VdM}^2(x^f_v) \; {\exp}{(\mathcal{L})}}}
\nonumber \\
{\mathcal{L}} & \equiv & {\imath N  {Tr}_{vj}( \hat{S} L {({\tilde{\epsilon}^{-1} {\hat{E}}^{b}} - \epsilon H_0 - \tilde{\epsilon}^{-1} {\tilde{J}}^b)} )} + {\frac{\imath}{2} \psi {({{\hat{E}}^{f}} - \epsilon_0 H_0 - {\tilde{J}}^f)} \overline{\psi}}  
\nonumber \\
& - &{ \frac{N}{2} \sum_{v_1 v_2} {(1-k)^{-1}_{v_1 v_2}}{Tr}{({Q^f}_{v_1} {Q^f}_{v_2})} } \pm {N{ \sum_{v} {Tr}{(Q^{f}_v {\hat{\psi}}_v )}}} - {\frac{N}{2}{(  {\sum_{v_1 v_2} {(1-k)_{v_1 v_2}} {Tr}{({\hat{S}}_{v_1 v_1} L {\hat{S}}_{v_2 v_2} L)}} + 2X )}}
\nonumber \\
\gamma & = & {{(2 \pi)}^{-V N I^{b}}  {(\imath  / 2)}^{-V N I^{f}} {(\imath)}^{- N V I^b}  {(\det L)}^{NV} {(N / 2 \pi)}^{{I^{f}}^{2}V / 2} 2^{I^{f} {(I^{f} -1)} V / 2}  {(\det (1-k))}^{{I^{f}}^{2} / 2}}
\label{HSTransformationPlusMinus}
\end{eqnarray}

The new Lagrangian is linear in $\psi$ and $\overline{\psi}$, so we use equation \ref{FermionDeterminantExact} to integrate these variables.  At the same time we rescale $S \rightarrow (2 N /\tilde{\epsilon})^{-1/2} S$ to remove energy units from the integral.  $\epsilon = \epsilon_0 / \tilde{\epsilon}$ is the strength of $H_0$ relative to the disorder. 
\begin{eqnarray}
{\bar{Z}} & = &{\gamma {\int {dU \, dx^f} {{dS^{R}}{dS^{I}}} \;\prod_v \Delta_{VdM}^2(x^f_v)\; {\exp}{(\mathcal{L})}}} \;\; {\det({A_0 {\delta_{n_{1} n_{2}}}} {- \tilde{\epsilon} (2N)^{-1} \sum_{j_1} {(1-k)_{v_1 v_2}} {S_{n_{1} v_{1} j_{1}}} {L_{j_{1}}} {S^{*}_{n_{2} v_{2} j_{1}}}  {\delta_{i_{1} i_{2}}}} )} , 
\nonumber \\
{\mathcal{L}} & = & {\imath N  {Tr}_{v j}(\hat{S} L {({\tilde{\epsilon}^{-1} {\hat{E}}^{b}} - \epsilon H_0 - \tilde{\epsilon}^{-1} {\tilde{J}}^b)} )} - {\frac{N}{2}{{\sum_{v_1 v_2} {(1-k)_{v_1 v_2}} {Tr}{({\hat{S}}_{v_1} L {\hat{S}}_{v_2} L)}}}} 
\nonumber \\
& - & { \frac{N}{2} \sum_{v_1 v_2} {(1-k)^{-1}_{v_1 v_2}}{Tr}{({Q^f}_{v_1} {Q^f}_{v_2})} }
\nonumber \\
{A^0_{ v_{1} v_{2} i_{1} i_{2}}} & \equiv & 
 { {Q^{f}_{v_{1} i_{1} i_{2}}}  {\delta_{v_{1} v_{2}}}}
+ { \imath  {\delta_{i_{1} i_{2}}} {({\tilde{\epsilon}^{-1} {\hat{E}}^{f}_{i_{1}}{\delta_{v_{1}v_{2}}}} -{\epsilon H_{0 v_1 v_2} } - \tilde{\epsilon}^{-1} {{\tilde{J}}^f_{i_1 v_{1} v_{2}}})  }}
\nonumber \\
 {\hat{S}_{v_{1} v_{2} j_{1} j_{2}}} &\equiv& (2N)^{-1} { \sum_{n} {S^{*}_{n v_{1} j_{1}}} {S_{n v_{2} j_{2}}} }, \; Q^f = U x^f U^\dagger
\nonumber \\
\gamma & = & {{(2 \pi)}^{-V N I^{b}}  {(\imath)}^{-  V N I^b - V N I^f} {(\det L)}^{NV} {(N / 2 \pi )}^{{I^{f}}^{2}V / 2} 2^{I^{f} {(I^{f} -1)} V / 2} {(\det (1-k))}^{{I^{f}}^{2} / 2} } (2N)^{V N I^b} 
\label{FermionlessGammaExact}
\end{eqnarray}

Fyodorov showed  how to rewrite the determinant to depend on the $\hat{S}$ matrix rather than the $S$ vector  \cite{Fyodorov02, Fyodorov02a}.  The following lines generalize his proofs to the Wegner model, where $H_0$ and $\tilde{J}^f$ are nonlocal.  The $S$ vectors may be understood as an $N V \times I^b V$ matrix (diagonal in $v$), and  $\sum_j S_{n_1  v_1 j} L_j (1 -k)_{v_1 v_2} S^*_{n_2 v_2 j}$ can be rewritten as the multiple of matrices: $S \acute{k} S^\dagger$, where $\acute{k}_{v_1 v_2 j_1 j_2} =  (1-k)_{v_1 v_2} \delta_{j_1 j_2} L_{j_1}$.  Furthermore $S$ may be written in its Singular Value Decomposition $S_{v v n j} = W s C $, where $W$ is $N V \times N V$ and unitary, $C$ is $I^b V \times I^b V$ and unitary, and the diagonal $I^b V \times I^b V$ matrix $s$ is composed of $S$'s singular values \cite{Golub83}.  With this notation the determinant is $ \det({A^0 {\delta_{n_{1} n_{2}}}} - { \tilde{\epsilon} (2 N)^{-1} {\delta_{i_{1} i_{2}}}   S (1-k)L S^\dagger } ) = {(\det A^0)}^N \det(1 - {  \tilde{\epsilon} (2N)^{-1}  {(A^0)^{-1} W s C \acute{k} C^\dagger s^\dagger W^\dagger } })$.  We would like to use the cyclic properties of the determinant to move $C^\dagger s^\dagger W^\dagger$ around to the left side, but this is not allowed because $C$ and $s$ do not have the same rank as the argument of the determinant.  Therefore we define augmented $N V \times NV $ matrices $\check{C},\,\check{S}$, and $\check{L}$, which are padded with ones on the diagonal and zeros everywhere else.  We also define a diagonal matrix $\theta$ which has ones in the first $I^b V$ diagonal entries and zeros everywhere else.   One may verify that $s C (1 - k) L C^\dagger s^\dagger = \check{s} \check{C} (1 - k) \check{L} \theta  \check{C}^\dagger \check{s}^\dagger$.  This leaves us free to continue rearranging the determinant: 
 \begin{eqnarray}
 & \, & {(\det A^0)}^N  \det(1 - {  \tilde{\epsilon} (2N)^{-1}  {(A^0)^{-1} W \check{s} \check{C} (1 - k) \check{L} \theta  \check{C}^\dagger \check{s}^\dagger W^\dagger } })
 \nonumber \\
 & = & {(\det A^0)}^N \det(1 - {  \tilde{\epsilon} (2N)^{-1}  {\check{C}^\dagger \check{s}^\dagger W^\dagger (A^0)^{-1} W \check{s} \check{C} (1 - k) \check{L} \theta   } })
 \nonumber \\
 & = & {(\det A^0)}^N \det(1 - {  \tilde{\epsilon} (2N)^{-1}  S^\dagger (A^0)^{-1} S (1 - k) L    } )
 \nonumber \\
 & = & {(\det A^0)}^N \det(1 - {  \sum_{v_2} {(A^0)^{-1}_{i_1 i_2 v_1 v_2}}  {\hat{S}_{v_1 v_2 j_1 j_2}} {L_{j_{1}}} {(1-k)_{v_2 v_3}}  } )
\nonumber \\
 & = & {(\det A^0)}^{N - I^b} \det(A^0 {\delta_{j_{1}j_{2}}} - {  A^1  } ), \;
 A^1 \equiv \sum_{i_1 v_1 v_2} {A^0_{i_0 i_1 v_0 v_1}} {(A^0)^{-1}_{i_1 i_2 v_1 v_2}}  {\hat{S}_{v_1 v_2 j_1 j_2}} {L_{j_{1}}} {(1-k)_{v_2 v_3}}
\end{eqnarray}
The $\theta$ in the second line prevents the $I^f \times I^b \times V $ sector of the determinant's argument from coupling with the rest of the basis.  Between the second and third lines the operator inside the determinant moves from living in a basis of size $I^f \times N \times V $ basis to living in a basis of size $I^f \times I^b \times V$.

\subsection{Conversion of the Bosonic Variables}
In the previous section we integrated out the fermionic vectors $\psi$ and $\overline{\psi}$ associated with the determinant in the  numerator of the generating function $Z$.  The result was a new matrix $Q^{f}$, which I will call the fermionic matrix even though it contains no Grassman variables.  Now we will extract an $I^b \times I^b$ Hermitian matrix $Q^b$ from the $S$ vectors.  This extraction process was done first by David, Duplantier, and Guitter \cite{David93}.  Fyodorov  re-derived their result and applied it in the context of random matrix theory \cite{Fyodorov02, Fyodorov02a}, and Spencer and Zirnbauer \cite{Spencer04} proved the special case $I^b = N$. 

 We start with the singular  value decomposition $S = W s C$ and assume that $N \geq I^b$.  In contrast with the previous SVD, now each of the matrices $S_v, W_v, s_v,$ and $C_v$  are local.  The integration measure becomes \cite{Spencer04} ${dS^R} {dS^I} = {ds^2} {d\acute{W}} {dC} {\Delta_{VdM}^2(s^2)} (\det{s^2})^{N - I^b }$. We define the positive indefinite matrix $Q^b \equiv (2N)^{-1} C^\dagger s^2 C$; ${dS^R} {dS^I} = {dQ^b} {dW} \theta(Q^b)  (\det Q^b)^{N - I^b }$.  Fyodorov\cite{Fyodorov02a} computed $\int {dW}$.  Adjusting  for factors of two caused by Fyodorov's use of the integration measure ${dS}{dS^*}$ versus our use of ${dS^R}{dS^I}$, this constant is $\int {dW} = { {{(\prod_{l=N-t}^{N-1}{l!})}^{-1}}  {\pi}^{{-I^b{(I^b-1)}/2} + I^bN} }$. 

As a consequence of the original theory's use of $L$ to make the $dS$ integrals converge,  $Q^b$ always occurs in combination with $L$.  Fyodorov \cite{Fyodorov02} showed  that $Q^bL$ factors into $Q^b L = T x^b T^{-1}$, where $x^b$ is diagonal and constrained by $x^b L \geq 0$, $T$ is a member of the pseudo-unitary hyperbolic group $U(n_+, n_-)$, and $n_+$ and $n_-$ are the numbers  of plus and minus signs in $L$.  In the special case where all of $L$'s entries have the same sign, $T$ is an ordinary unitary matrix.  With the exception of this special case, $U(n_+, n_-)$ is not compact, and as a consequence has parameterizations in which one or more of its parameters is unbounded.  Fyodorov's integration measure is $dQ^b = dx^b \, dT \, \Delta^2_{VdM}(x^b)$.   In these coordinates the path integral is
\begin{eqnarray}
{\bar{Z}} & = & \gamma \int_{x^b L \geq 0} { dx^f \, dU}  { dx^b \, dT} \;\prod_v \Delta_{VdM}^2(x^f_v)\; \Delta_{VdM}^2(x^b_v) \; e^{\mathcal{L} + {{\mathcal{L}}_{eff}}}  
\nonumber \\
{{e}^{{\mathcal{L}}_{eff}}}
& \equiv & {\int {dW} 
\exp{(-\imath N  {Tr}_{vj}(\hat{S} L {( \epsilon H_0 + \tilde{\epsilon}^{-1} {\tilde{J}}^b)} ))} \; {\det(A^0 {\delta_{j_{1}j_{2}}} - { {A^1(\hat{S}) }  } )} }
\nonumber \\
{\mathcal{L}}  & = &  - {\frac{N}{2}{{\sum_{vk}   x_{vk}^2  }}} + {(N - I^b) \sum_{vj} \ln  x^b_{vj}} + {{\imath N}{\tilde{\epsilon}}^{-1} \sum_v {Tr}(T_{v} x_v^b T_{v}^{-1} {{\hat{E}}^{b}}   )} 
+ {{\imath N}{\tilde{\epsilon}}^{-1} \sum_v {Tr}(  U_{v} x_v^f U^{\dagger}_{v}  {{\hat{E}}^{f}} )} 
 \nonumber \\ 
 & - & { \frac{N}{2} \sum_{v_1 v_2} {(k/(1-k))_{v_1 v_2}}{Tr}{(U_{v_1} x_{v_1}^f U^{\dagger}_{v_1}  U_{v_2} x_{v_2}^f U^{\dagger}_{v_2}  )} } 
 + {\frac{N}{2}{{\sum_{v_1 v_2} {{k_{v_1 v_2}} {Tr}{(T_{v_1} x_{v_1}^b T_{v_1}^{-1} T_{v_2} x_{v_2}^b T_{v_2}^{-1}  )}}}}}
 \nonumber \\
 & + & (N - I^b) {Tr}(\ln A^0)
\nonumber \\
\gamma & = & {{(2 \pi)}^{-V N I^{b}}  {(\imath)}^{-V N  I^b - V N I^f} {(\det L)}^{V I^b} {(N / 2 \pi )}^{{I^{f}}^{2}V / 2} 2^{I^{f} {(I^{f} -1)} V / 2} {(\det (1-k))}^{{I^{f}}^{2} / 2} } (2 N)^{V N I^b} 
\nonumber \\
A^0_{ v_{1} v_{2} i_{1} i_{2}} & \equiv & 
 { {Q^{f}_{v_{1} i_{1} i_{2}}}  {\delta_{v_{1} v_{2}}}}
  +  {\imath  {\delta_{i_{1} i_{2}}} {({\tilde{\epsilon}^{-1} {\hat{E}}^{f}_{i_{1}}{\delta_{v_{1}v_{2}}}} -{\epsilon H_{0 v_1 v_2}}- {\tilde{\epsilon}^{-1} {\tilde{J}}^f_{i_1 v_{1} v_{2}}})   }}
 \nonumber \\
 A^1 &\equiv& \sum_{i_1, v_1, v_2} {A^0_{i_0 i_1 v_0 v_1}} {(A^0)^{-1}_{i_1 i_2 v_1 v_2}}  {\hat{S}_{v_1 v_2 j_1 j_2}} {L_{j_{1}}} {(1-k)_{v_2 v_3}}, \;\; 
 \nonumber \\
\hat{S}_{v_1 v_2} L & = & (2N)^{-1} C^\dagger_{v_1} s_{v_1} W^\dagger_{v_1} W_{v_2} s_{v_2} C_{v_2} L, \;\; (2N)^{-1} C^\dagger_v s^2_v C_v L = T_v x^b_v T_v^{-1} 
\end{eqnarray}

\subsection{Disertori's Model}
Disertori \cite{DisertorisModel}  applied Fyodorov's ideas to a model in which the Hamiltonian is entirely random;  $H_0 = 0$.  She also required implicitly that the source $\tilde{J}$ be  diagonal in the position index $v$, which restricted the theory to calculating  on-site elements of the Green's function $G_{vv}$.  With these restrictions in place, $A^0$ is local and $\det A^0$ factorizes site by site.  $A^1$ simplifies to $  {Q^b_{v_1}} {L_{j_{1}}} {(1-k)_{v_1 v_2} {{\delta}_{i_1 i_2}}}$, so that there is no dependence on $W$; the remaining integral in $e^{\mathcal{L}_{eff}}$ is just $(\int dW)^V$.

Shifting $Q^f \rightarrow Q^f -  {\imath {\tilde{\epsilon}}^{-1} {\hat{E}}^{f}} $,  we obtain Disertori's model:
\begin{eqnarray}
{\bar{Z}} & = & {\gamma {\int_{x^b L \geq 0} { dU \, dx^f} \; { dT \, dx^b} \; \;   e^{\mathcal{L}}  }} \; { \prod_v  {\Delta}_{VdM}^2(x_v^f)}  \; { {\Delta}_{VdM}^2(x_v^b)} 
\; {\det({ {U_{v_1} x_{v_1}^f U_{v_1}^{\dagger}} {\delta_{v_1 v_2}} {\delta_{j_1 j_2}} } - {  {T_{v_1} x_{v_1}^b T_{v_1}^{-1}} (1 - k)_{v_1 v_2} {\delta_{i_1 i_2}}  }   )} 
\nonumber \\
{\mathcal{L}}  & = &  - {\frac{N}{2}{{\sum_{vk}   x_{vk}^2  }}} + {(N - I^b) \sum_{vk} \ln  x_{vk}} + {{\imath N}{\tilde{\epsilon}}^{-1} \sum_v {Tr}(T_{v} x_v^b T_{v}^{-1} ({\hat{E}}^{b} - \tilde{J}^b)  )} 
+ {{\imath N}{\tilde{\epsilon}}^{-1} \sum_v {Tr}(  U_{v} x_v^f U^{\dagger}_{v}  {{\hat{E}}^{f}} )} 
 \nonumber \\ 
 & - & { \frac{N}{2} \sum_{v_1 v_2} {(k/(1-k))_{v_1 v_2}}{Tr}{(U_{v_1} x_{v_1}^f U^{\dagger}_{v_1}  U_{v_2} x_{v_2}^f U^{\dagger}_{v_2}  )} } 
 + {\frac{N}{2}{{\sum_{v_1 v_2} {{k_{v_1 v_2}} {Tr}{(T_{v_1} x_{v_1}^b T_{v_1}^{-1} T_{v_2} x_{v_2}^b T_{v_2}^{-1}  )}}}}}
\nonumber \\
\gamma & = & {  N^{{{N I^{b} V } } + {I^{f} I^{f} V / 2}}  {(\det L)}^{V I^b}}  { {{{(\prod_{l=N-I^{b}}^{N-1}{l!})}}^{-V}}  2^{ - {I^{f} V / 2}}   {\pi}^{{-I^{b}{(I^{b}-1)}V/2} - {I^{f} I^{f} V / 2}}   } {(\det (1-k))}^{{I^{f}}^{2} / 2} 
\nonumber \\
& \times & e^{{\frac{NV}{2 {\tilde{\epsilon}}^2} {Tr}({\hat{E}}^f {\hat{E}}^f)}   } \imath^{-VNI^b - VNI^f}
\label{DisertoriModel}
\end{eqnarray}

This completes the sequence of exact steps which converts Disertori's model to matrix coordinates.  The action is remarkably like the exact SUSY action $\frac{N}{2}{Tr}(Q^2) + N  {Tr}(Q \hat{E}) + N \,{Tr}(\ln (Q + \epsilon k ) ) $; the primary difference is that the kinetics are displaced from the logarithm to the quadratic terms, in keeping with the model's semicircular band structure.

 After the shift $Q^f \rightarrow Q^f -  {\imath {\tilde{\epsilon}}^{-1} {\hat{E}}^{f}} $ one must remember that $Q^{f}$ depends on $ {\hat{E}}^{f}$  when making use of the fermionic sources.  Neglect of this dependence will result in incorrect prefactors and will break the Ward identities associated with the numerator-denominator symmetry, even when calculating the density of states.  To avoid these intricacies, from now on we will use  prefactors instead of sources. Derivatives with respect to the local sources $ {\tilde{J}}$ produce the following prefactors:
\begin{eqnarray}
{\frac{d}{d{\tilde{J}}^{f}_i}} &\rightarrow & {-\imath (N - I^b) {\tilde{\epsilon}}^{-1} F^1_{v ii}} - {\imath {\tilde{\epsilon}}^{-1} \sum_j F^2_{vv ii jj}}
\nonumber \\
\frac{d^2}{{d{\tilde{J}}^{f}_{v_1 i_1}}{d{\tilde{J}}^{f}_{v_2 i_2}}} & \rightarrow & {{\frac{d}{d{\tilde{J}}^{f}_{v_1 i_1}}} \otimes {\frac{d}{d{\tilde{J}}^{f}_{v_2 i_2}}}} + { (N - I^b) {\tilde{\epsilon}}^{-2} \delta_{v_1 v_2} F^1_{v_1 i_1 i_2} F^1_{v_1 i_2 i_1}} 
+ { {\tilde{\epsilon}}^{-2} \sum_{j_1 j_2} F^2_{v_1 v_2 i_1 i_2 j_1 j_2} F^2_{v_2 v_1 i_2 i_1 j_2 j_1}}
\label{FermionicPrefactor2}
\nonumber \\
F^1 & \equiv  & {{(U x^f U^\dagger )}^{-1}}, \; \; F^2 \equiv  {{(U x^f U^\dagger   - {   T x^b T^{-1}  (1-k) }  )}^{-1}}
\nonumber \\
{\frac{d}{d{\tilde{J}}^{b}_{vj}}} &\rightarrow & -\imath N  {\tilde{\epsilon}}^{-1} (T x^b T^{-1})_{v jj}, \;\; \frac{d^2}{{d{\tilde{J}}^{b}_{v_1 j_1}}{d{\tilde{J}}^{b}_{v_2 j_2}}}  \rightarrow  {{\frac{d}{d{\tilde{J}}^{b}_{v_1 j_1}}} \otimes {\frac{d}{d{\tilde{J}}^{b}_{v_2 j_2}}}} 
\label{DisertoriObservables}
\end{eqnarray}

When one is interested in calculating off-site elements of the Green's function $G_{v_1 v_2}$ these prefactors require modification to adjust for non-local sources.  If one uses bosonic sources, $\tilde{J}^b$ must have an off-diagonal element connecting $v_1$ and $v_2$. Therefore the $W_{v_1}$ integral in $\mathcal{L}_{eff}$ does not decouple from the $W_{v_2}$ integral.  The correction to the path integral looks like
\begin{eqnarray}
(\int {dW})^{-2} &\;& \int {dW_1} {dW_2} \exp(- \imath \tilde{\epsilon}^{-1} N {Tr}_{v = \{v_1, v_2\}}(\hat{S} L \tilde{J}^b)),
\nonumber \\
 \hat{S}_{v_1 v_2} L & =  & (2N)^{-1} C^\dagger_{v_1} s_{v_1} W^\dagger_{v_1} W_{v_2} s_{v_2} C_{v_2} L, \;\; (2N)^{-1} C^\dagger_v s^2_v C_v L = T_v x^b_v T_v^{-1} 
\end{eqnarray}
Evaluation of this integral lies outside the scope of this paper.

\subsection{The Wegner Model}
In the Wegner model the random potential is purely local and the Laplacian is deterministic.  Within this paper's formalism it can be obtained by setting $k = 0$ and $H_0$ equal to the Laplacian.  I take the liberty of renaming $H_0 = k$.  After absorbing $\hat{E}^f$ into $Q^f$,
\begin{eqnarray}
{\bar{Z}} & = & \gamma \int_{x^b L \geq 0} {dx^f \, dU}  {dx^b \, dT} \; {dW} \;\prod_v \Delta_{VdM}^2(x^f_v)\; \Delta_{VdM}^2(x^b_v) \; e^{\mathcal{L} }  \; {\det(A^0 {\delta_{j_{1}j_{2}}} - { {A^1(\hat{S}) }  } )}
\nonumber \\
{\mathcal{L}}  & = &  - {\frac{N}{2}{{\sum_{vk}   x_{vk}^2  }}} 
+ {{\imath N}{\tilde{\epsilon}}^{-1} \sum_v {Tr}(  U_{v} x_v^f U^{\dagger}_{v}  {{\hat{E}}^{f}} )} 
  + (N - I^b) {Tr}_{vi}(\ln (U x^f U^\dagger - \imath \epsilon k -\imath \tilde{\epsilon}^{-1} \tilde{J}^f))
 \nonumber \\
 & + &    {{\imath N}{\tilde{\epsilon}}^{-1} \sum_v {Tr}(T_{v} x_v^b T_{v}^{-1} {{\hat{E}}^{b}}   )} -\imath N  {Tr}_{vj}(\hat{S} L {( \epsilon k + \tilde{\epsilon}^{-1} {\tilde{J}}^b)} )  + {(N - I^b) \sum_{vj} \ln  x^b_{vj}}
\nonumber \\
\gamma & = & {{(2 \pi)}^{-V N I^{b}}  {(\imath)}^{-V N  I^b - V N I^f} {(\det L)}^{V I^b} {(N / 2 \pi )}^{{I^{f}}^{2}V / 2} 2^{I^{f} {(I^{f} -1)} V / 2} {(\det (1-k))}^{{I^{f}}^{2} / 2} } (2 N)^{V N I^b} 
 e^{{\frac{NV}{2 {\tilde{\epsilon}}^2} {Tr}({\hat{E}}^f {\hat{E}}^f)}   }
\nonumber \\
A^0_{ v_{1} v_{2} i_{1} i_{2}} & \equiv & 
 { {Q^{f}_{v_{1} i_{1} i_{2}}}  {\delta_{v_{1} v_{2}}}}
  - {\imath \epsilon  {\delta_{i_{1} i_{2}}}  k_{ v_1 v_2}   } - \imath {\delta_{i_{1} i_{2}}} \tilde{\epsilon}^{-1} \tilde{J}^f, \;
 A^1 \equiv \sum_{i_1, v_1} {A^0_{i_0 i_1 v_0 v_1}} {(A^0)^{-1}_{i_1 i_2 v_1 v_2}}  {\hat{S}_{v_1 v_2 j_1 j_2}} {L_{j_{1}}} , \;\; 
 \nonumber \\
\hat{S}_{v_1 v_2} L & = &  C^\dagger_{v_1} s_{v_1} W^\dagger_{v_1} W_{v_2} s_{v_2} C_{v_2} L, \;\;  C^\dagger_v s^2_v C_v L = T_v x^b_v T_v^{-1}
\label{TransformedWegnerModel}
\end{eqnarray}

 This action is similar to the exact SUSY action $\frac{N}{2}{Tr}(Q^2) + N  {Tr}(Q \hat{E}) + N \,{Tr}(\ln (Q + \epsilon k ) ) $, where $Q$ contains an implicit $\imath$. However $Q^b$'s kinetics are hidden in the $dW$ integral.  This integral is similar to a logarithm: its derivatives contain Green's functions, and its contribution to the Lagrangian is a logarithm at leading order.  Moreover the Ward identity $\frac{d\bar{Z}(\hat{E}^f = \hat{E}^b, \tilde{J} = 0)}{d \epsilon} = 0$ ensures that the $dW$ integral has the same physics as the $Q^f$ logarithm in equation \ref{TransformedWegnerModel}.  The $Q^f$ logarithm corresponds to multiplying the path integral by $(Q^f - \imath \epsilon k)^{N- I^b}$; when $\epsilon k \gg Q^f$ this multiplier scales with $\epsilon^{N - I^b}$, and in the opposite case $\epsilon k \ll Q^f$ it is constant with respect to $\epsilon$.  The Ward identity ensures that this multiplier must be cancelled by some other contribution, and the only term which can fill this role is the $dW$ integral.  Therefore the $dW$ integral must scale as $\epsilon^{I^b - N}$ when $\epsilon k \gg Q^f$ and be roughly constant with respect to $\epsilon$ when $\epsilon k \ll Q^f$.

The Wegner model and Disertori's model are very much alike.  Section \ref{WegnerSaddle} examines the similarities of the two Lagrangians and their saddle point approximations, which work out roughly the same.  The principal difference is that Disertori's energy band is the semicircular one of random matrix theory, while Wegner's energy band can have any shape at all, according to the kinetic operator which one chooses.  Another superficial difference is that Wegner's model allows easy tuning of the relative strength of the kinetic operator, while in Disertori's model a single energy scale controls both the kinetics and the disorder.  This is merely superficial: $N^{1/2} k$ can be used to tune Disertori's model.   Section \ref{WegnerDeterminant} discusses the $Q^f - Q^b$ determinant, which is wholely responsible for interactions between $Q^f$ and $Q^b$.   In both models the argument of the determinant is the sum of the kinetic operator and $Q^f$, minus a term proportional to $Q^b L$; it is $ Q^b L k + Q^f - Q^bL$ in Disertori's case and $-\imath \epsilon k + Q^f - A^0 ((A^0)^{-1} \cdot  \hat{S} L )$ in Wegner's case.   This form describes competition of the kinetics versus $Q^f$ and $Q^b$.  We will concentrate on the case where $k$ dominates the determinant, otherwise known as the weak localization regime.  In this regime at leading order Wegner's $Q^f - Q^b$ coupling is the same as Disertori's coupling, and both are independent of $W$.  

We turn to the $dW$ integration.  The Wegner model depends only on the first $I^b$ columns of $W$.  Taken in isolation from the rest of $W$, these columns form the Stiefel manifold $V_{I^b}(C^N)$, which is defined as the set of all possible combinations of $I^b$ orthonormal complex $N$-vectors.  This manifold forms an homogeneous (continuous and transitive) space corresponding to the unitary group; $V_{I^b}(C^N) \cong U(N) / U(N - I^b)$.  Integration over the other $N - I^b$ columns in $W$ produces a constant.

The $dW$ integral exhibits an exact continuous symmetry under global rotations of $W$, and is governed by a kinetic term $-\imath N \epsilon {Tr}_{vj}(\hat{S} L  k  )$ which regulates fluctuations in $W$. Therefore $W$'s symmetry is spontaneously broken in $D > 2$ dimensions, and in small volumes displays an effective SSB even in $D = \{1,2\}$ dimensions\cite{Posazhennikova06}.   In the SSB phase $W$ takes on a spatially uniform value, with small fluctuations.   Therefore we will do perturbation theory in $W$'s fluctuations.

We can assess the validity of the $W$ perturbation theory, and of the SSB assumption itself, by calculating corrections to the leading order results. As an example, consider the $I^b = 1$ case where the Stiefel manifold is a sphere  and $W$ is a unit vector $\hat{e}$.  The free energy $F$ is proportional to 
$\ln \, \int {d\hat{e}} \exp(-\imath \epsilon N \sum_{v_1 v_2} k_{v_2 v_1} \sqrt{x_{v_1}^b} \,\, \hat{e}^{\dagger}_{v_1} \cdot \hat{e}_{v_2} \sqrt{x_{v_2}^b})$.  At leading order $F$ scales with the number of degrees of freedom, which is $2N - 1$.  The first correction $F_1$ is proportional to $ N^2 \epsilon \, x^b   \sum_{v_1 v_2} k_{v_1 v_2}  G_{v_1 v_2}^2$, where $G$ is the bare propagator $\langle (\hat{e}^{\dagger}_{v_1} \cdot \hat{j}_1) (\hat{j}_2 \cdot \hat{e}_{v_2})  \rangle$.  The numerical value of this correction may be calculated either numerically or by analytic approximations. When SSB occurs  $G = (N \epsilon\, x^b k)^{-1}_{v_1 v_2} \delta_{j_1 j_2}$, and the relative strength of the first correction to the free energy is controlled by $(N  x^b \epsilon k)^{-1}$.  We will find that the saddle point value of $x^b$ is independent of $N$.  In the $N \rightarrow \infty$ limit SSB can be frustrated only when $D = \{1,2\}$ and only when $V \rightarrow \infty$, in which case the propagator's diagonal elements  diverge\footnote{In addition to the condition $N x^b \epsilon k \gg 1$, $W$'s spontaneous symmetry breaking may also require that $Q^b$ exhibits SSB.}.  

 In the SSB phase prefactors are not affected at leading order by the $W$ fluctuations, so we move the bosonic source $\tilde{J}^b$ outside of the $dW$ integral.  This step is exactly correct if one calculates only local observables like $G_{vv}$. If one wants non-local results then perturbative corrections may be calculated in powers of the bare propagator.    In the weak localization regime the $Q^f - Q^b$ determinant is almost entirely independent of $\hat{S}$ and is best treated as a prefactor,  so we move it also outside of the $dW$ integral.   Here too perturbative corrections may be calculated if desired.

We assume SSB and calculate the $dW$ integral at leading order.  We parameterize $W$ as a unitary matrix; i.e. $W = e^Y$, where $Y = -Y^\dagger$ is anti-Hermitian.   If one calculated corrections one would find an additonal $O(\Sigma)$ term added to $\mathcal{L}_1$'s $h-\bar{h}$, where $ \Sigma \propto N$ is $Y$'s self-energy.
\begin{eqnarray}
\prod_{vj} & \, & (x^b_{vj})^{N-I^b} \; \int {dW}  \exp{(-\imath N  {Tr}_{vj}(\hat{S} L {( \epsilon k + \tilde{\epsilon}^{-1} {\tilde{J}}^b)} ))} \;{\det(A^0 {\delta_{j_{1}j_{2}}} - { {A^1(\hat{S}) }  } )}
\nonumber \\ &=& 
 ( 2 \pi /\imath  N)^{VI^b(N - I^b/2)} {\det(A^0 {\delta_{j_{1}j_{2}}} - { {A^1(\hat{S}) }  } )} \; e^{\mathcal{L}_1 + \mathcal{L}_2}, \; \mathcal{L}_1 \equiv - {(N - I^b/2){Tr}(\ln(h - \bar{h}))}  + (N-I^b) \sum_{vj} \ln x^b_{vj}
\nonumber \\
\mathcal{L}_2 & = &  -\imath N  {Tr}_{vj}(\hat{S} L {( \epsilon k + \tilde{\epsilon}^{-1} {\tilde{J}}^b)} + {\frac{\imath}{2}\sum_{n \leq I^b} \vec{h}_n^\dagger \cdot (h - \bar{h})^{-1} \cdot \vec{h}_n})
\nonumber \\
 h_{v_1 v_2 j_1 j_2}& =& \epsilon k_{v_1 v_2} (s_{v_1} c_{v_1} L c_{v_2}^\dagger s_{v_2})_{j_1 j_2}, \; \bar{h}_{v_1 v_2 } = \frac{1}{2} \delta_{v_1 v_2} \sum_{v_3} (h_{v_1 v_3 } + h_{v_3 v_1 }), \; \vec{h}_{v j n} = \frac{1}{2} \sum_{v_1} (h_{v v_1 j n} - h_{v_1 v j n}),
 \nonumber \\
\hat{S}_{v_1 v_2} L & = &  C^\dagger_{v_1} s_{v_1}  s_{v_2} C_{v_2} L, \;\;  C^\dagger_v s^2_v C_v L = T_v x^b_v T_v^{-1}, \;\; \langle Y^\dagger Y \rangle  =  (1- I^b/(2N)) (\imath  \epsilon)^{-1} (h - \bar{h} + \imath \Sigma)^{-1}
\label{dWIntegral}
\end{eqnarray}
This simplifies tremendously when $s$ and $C$ are spatially uniform and $\epsilon k \gg 1$: 
\begin{eqnarray}
\mathcal{L}_1 + \mathcal{L}_2 & = &  -\imath N  \tilde{\epsilon}^{-1}  {Tr}_{vj}(\hat{S} L {\tilde{J}}^b) - {(N - I^b/2){Tr} \, \ln( L \epsilon k )} - I^b/2 \sum_{vj} \ln x^b_{vj}
\label{dWIntegralSimplified}
\end{eqnarray}
When $\epsilon k \ll 1$ the correct result for $\mathcal{L}_1 + \mathcal{L}_2$ retains the original factor of $(N-I^b) \sum_{vj} \ln x^b_{vj}$.  The $W = e^Y$ parameterization can not be used for other computations because it mixes the Stiefel manifold with the rest of $U(N)$.  $\Sigma$, $\langle W^\dagger W \rangle$, and corrections to the $dW$ integral require a parameterization adapted specifically to small fluctuations on the Stiefel manifold.  In the case of $I^b = 1$ the $\hat{e}$  unit vector parameterization produces easy results for the bare propagator and the leading order self-energy; $G \propto (\epsilon N x^b k)^{-1},\;\Sigma(\vec{s}) \propto \sum_{\vec{s}_1 \neq 0} k^{-1}(\vec{s} - \vec{s}_1) \, k(\vec{s}_1)$.  Yet even here $\langle W^\dagger W \rangle$ contains all even powers of $\hat{e}$'s components.   Further consideration of these issues lies outside the scope of this paper.   We will focus mainly on Disertori's model, and will at intervals discuss how how similar results can be obtained from the Wegner model. 

\subsection{Is the $Q^f - Q^b$ Model Supersymmetric?}
It is important to distinguish between different meanings of the term supersymmetry:
\begin{itemize}
\item Both the Wegner model and Disertori's model possess continuous symmetries associated with global rotations in $Q^f \rightarrow U_0 Q^f U_0^\dagger$ and $Q^b L \rightarrow T_0 Q^b L T_0^{-1}, \, \hat{S} L \rightarrow T_0 \hat{S} L T_0^{-1}$; these symmetries are exact when $\hat{E}$ is proportional to the identity.  
\item We have already discussed another symmetry which is always exact: the symmetry between the determinants in the numerator and the determinants in the denominator.  The numerator-denominator symmetry is not a continuous symmetry.  Instead it manifests itself in any number of  Ward identities, including ones relating $Q^b$ observables to $Q^f$ observables.  
\item In the supersymmetric sigma model there is a single degree of freedom, the graded matrix $Q$ which contains both Grassman and scalar variables.  The scalar sector is roughly equivalent to this paper's $Q^f$ and $Q^b$ matrices.  In this model  both the rotational and the numerator-denominator symmetries are subsumed into a graded continuous symmetry connected to $Q$'s rotations in a graded group; this is the traditional meaning of supersymmetry.  The supersymmetric model's numerous successes, including its noteworthy non-perturbative results, are attributed to the graded symmetry.
\end{itemize}

Fyodorov's conversion to $Q^f - Q^b$ coordinates has been discounted because it does not maintain a unified approach; in particular there is no symmetry transforming $Q^f$ variables into $Q^b$ variables or vice versa.  Despite the exactitude of its derivation, one fears that without the graded symmetry's protection later approximations will produce uncontrolled results.  The obvious differences between $Q^f$ and $Q^b$ make this fear more tangible.  In Disertori's case the Lagrangian has stray $1 - k$ factors in the $Q^f - Q^b$ determinant and in $Q^f$'s kinetics.  The $Q^f$ observable (equation \ref{DisertoriObservables}) is written in terms of both $(Q^f)^{-1}$ and $(Q^f - Q^b L (1-k))^{-1}$, while the $Q^b$ observable is just $-\imath N \tilde{\epsilon}^{-1} Q^b L$.  In Wegner's case the differences between $Q^f$ and $Q^b$ are even more daunting.   And yet we will see later in this paper that Fyodorov's $Q^f - Q^b$ formalism is able to reproduce and extend the SUSY results, including even the correct normalization factors.

I submit that the desire for an obvious symmetry in the Lagrangian is misguided.  Should supersymmetry mean symmetry between $Q^f$ and $Q^b$, or instead symmetry between these variables and the $Q^f - Q^b$ determinant?  The Grassman variables were responsible for the determinants in the numerator; recall that when we integrated them the result was the  product of the $Q^f - Q^b$ determinant and $(\det(Q^f))^{N - I^b}$.  When calculating the two point correlator we will see that integration of $Q^f$ and $Q^b$ produces a matching determinant which almost cancels the $Q^f - Q^b$ determinant. The correct normalization factors come from cancellation between the two.  

Fyodorov's formalism possesses a hidden supersymmetry which is guaranteed by its exact derivation, protected by the Ward identities, and manifested in every detail of the Lagrangian.  When calculating the density of states one finds that the saddle point is $Q^b \propto e^{\imath \phi}, \; Q^f \propto - e^{-\imath \phi}$; we see that the inverse relationship between the $Q^b$ and $Q^f$ observables preserves the Ward identities. The $k$ in the determinant turns out to be crucial for extended systems, and its property $k | \vec{0} \rangle = 0$ is directly linked to the Wigner-Dyson statistics seen in the weak localization regime.  Even the $I^b$ in $(N - I^b) \ln x$ plays a role in obtaining correct overall signs.

\section{The Saddle Point Approximation\label{SaddlePoint}}

Analysis of this theory must begin with identification of the dominant parts of the Lagrangian and of the dominant values of $Q^f$ and $Q^b$.  We will procede in two stages, first treating the eigenvalues $x$ and only later treating the angles $U$ and $T$.  Concerning the eigenvalues, it is important to discern carefully which terms determine the saddle point values.  A good first approximation neglects the determinants in equation \ref{DisertoriModel}; in later sections we will take those determinants into account properly. Unlike previous authors, we do not assume that the saddle point is spatially uniform.  The saddle point equations are
\begin{eqnarray}
0 &=& -N x_{vk} + N (1 - I^b N^{-1}) x_{vk}^{-1} + 2 \imath N (1 - I^b N^{-1})^{1/2} \sin \phi_{vk},
\nonumber \\
 \sin \phi^f_{vi} & \equiv & 2^{-1} (1 - I^b N^{-1})^{-1/2} (U^{\dagger}_v {({{ {\tilde{\epsilon}}^{-1} {\hat{E}}^f }} + \imath \sum_{v_2} {(( k/(1-k))}_{v_1 v_2} U_{v_2} x_{v_2}^f U^{\dagger}_{v_2})} U_v)_{ii}
 \nonumber \\
 \sin \phi^b_{vj} & \equiv & 2^{-1} (1 - I^b N^{-1})^{-1/2} (T^{-1}_v {({{ {\tilde{\epsilon}}^{-1}{\hat{E}}^b}}  - \imath \sum_{v_2} k_{v_1 v_2} T_{v_2} x_{v_2}^b T^{-1}_{v_2})} T_v)_{jj}
 \label{SaddlePoints}
 \end{eqnarray}

Each component of $x$ has two solutions controlled by a sign.  We will encapsulate the sign signature of the saddle point in the diagonal matrix $\acute{L}_k = \pm 1$.  The solutions of the saddle point equation are  $x_{0 vk} = \acute{L}_k \sqrt{1 - I^b N^{-1}} e^{ \imath \acute{L}_k {\phi}_{vk}} = {\acute{L}_k \sqrt{1 - I^b N^{-1}} \cos{(  {\phi}_{vk})}}  + {\imath \sqrt{1 - I^b N^{-1}} \sin{(  {\phi}_{vk})}}$.  The constraint $x^b L > 0$ implies that $\acute{L}^b_j = L_j$.  

 These equations describe the static equilibrium of $I^b + I^f$ identical particles moving in a quadratic potential divided by an infinite logarithmic wall.  Each particle chooses to live close to one of the two minima of the potential, according to $\acute{L}$.  The position of each particle within the potential is determined by the external force $2 \sin \phi_{vk}$.  From a different perspective, these equations describe an energy band of width $4 \tilde{\epsilon} \; \sqrt{ {1 - I^b N^{-1}}}$, and the position of $2 \sin \phi_{vk}$ within that band determines the phase $\phi_{vk}$.  
 
These saddle points dominate the model.  All statistically significant configurations of $Q^f$ and $Q^b$ have $x$ close to $x_0$.  Typical deviations can be estimated from the Lagrangian - at the saddle point its local quadratic part is $-\frac{N \eta}{2} {(x - x_0)}^2$, where $\eta_{vk} =  {2 \cos\phi_{vk}}{{exp}{(-\imath \acute{L}_{vk} {\phi}_{vk} )}}$.  Imagining an added Langevin style of dynamics in the model, one sees that any deviations away from the saddle point values are subject to a restoring force equal to $-N \eta (x - x_0)$.  Therefore fluctuations away from the saddle points will be of order ${(2 N \cos \phi)}^{-1/2}$ or smaller.  The $\cos \phi$ signals that the saddle point approximation breaks down when $2 \sin \phi$ is close to the band edge.

 At first blush these solutions permit $2^{I^f V}$ saddle points; i.e. a different $\acute{L}$ at each site.  In reality (1) $\acute{L}$ does not not fluctuate from site to site, and (2) only $I^f + 1$ out of the $2^{I^f}$ saddle points of $\acute{L}$ can be distinguished one from another.  The latter point is the most accessible.  Define $P$ as the group of permutations of $\acute{L}$ which conserves ${Tr}(\acute{L})$. $P$'s only significance  in the saddle point equations is to decide which eigenvalue $x_{iv}^f$ is coupled to which force $2 \sin \phi_{kv}$.  This choice is physically significant only at the global level, and then only if $\hat{E}^f$ is not proportional to the identity.  Therefore only a single global choice of $P$ is required.  This is a consequence of the theory's symmetry under rotations of $U$. 
 
Going further, all members of $P$ can be transformed into each other by finite rotations $x^f_0 \rightarrow U_p x^f_0 U_p^{\dagger}$.  Therefore if one eventually performs a full (non-perturbative) integration over $U$ then only a single member of $P$ need be considered even at the global level.  
 
What of the possibility that ${Tr}(\acute{L})$ might fluctuate from site to site?  We will return to this question in section \ref{SaddlePointFluctuations}, where we will establish that there is a single optimal value of $\acute{L}$ and that deviations from that value are penalized on a per-site basis by a free energy cost proportional to $\ln  k$.  

The saddle point approximation breaks down when $2 \sin \phi$ is close to the band edge.  The $T$ variables in $\sin \phi^b_{vj}$ are unbounded, and therefore a complete integration of $T$ always includes the band edge and beyond.  Therefore we must either do the $T$ integrations prior to the saddle point approximation, or else regulate $T$. This paper will rely heavily on the mechanism of spontaneous symmetry breaking, which regulates fluctuations in $T$, requiring them to be small. 

The global value of $T$ is not regulated by SSB. Obviously this is not a problem if $\hat{E}^b = \bar{E}^b$, in which case $Z$ contains information about self-correlations of single eigenfunctions.   However one often wants to know about correlations of energy levels or eigenfunctions, and therefore chooses $\hat{E}^b \neq \bar{E}^b$; in this case one needs some way to control the global $T$ integral.  There are two alternatives; the first is to perform the global integral prior to the saddle point approximation, and will be discussed in a little more depth in section \ref{SmallOmega}.  The second alternative is to pin $T$.  One pinning mechanism is to eliminate $T$ ($T = U = 1$) by calculating just the  density of states ($I^f = I^b = 1$).  Another pinning mechanism is to break $T$'s continuous symmetry by choosing energy levels that are far apart compared to the level spacing $\Delta$; i.e. $|\hat{E}^b - \bar{E}^b| \gg \Delta$.  In the main we will depend on the latter pinning mechanism.

\subsection{Spatially Uniform Saddle Point and $1/g$ Corrections}

The saddle point solutions given in equation \ref{SaddlePoints} are not quite satisfactory because they are not spatially invariant: $\sin \phi_k$ depends on $U$ and $T$ which fluctuate from site to site.  Therefore we choose a spatially invariant saddle point and rely on perturbation theory to manage the difference between it and the correct saddle point.   We extract the zero momentum component of $U$ and $T$: $U_v \rightarrow U_v U_0, \;\; T_v \rightarrow T_v T_0$, and then choose a saddle point which depends only on $U_0$ and $T_0$; $ \sin \phi^f_i =  2^{-1} \tilde{\epsilon}^{-1} (1 - I^b N^{-1})^{-1/2} ({U_0^{\dagger} {\hat{E}}^f U_0})_{ii},\;\;  \sin \phi^b_j = 2^{-1} \tilde{\epsilon}^{-1} (1 - I^b N^{-1})^{-1/2}  (T^{-1}_0  {\hat{E}}^b T_0)_{jj} $.  The difference between this choice of $\sin \phi$ and the correct choice is handled during the integration around the saddle point:
\begin{eqnarray}
\int & {dx} & \exp({-\frac{N \eta}{2} (x - x_0)^2} + {2 \imath N (1 - I^b N)^{1/2}} \delta(\sin \phi))
\nonumber \\
& = & \int {dx} \exp({-\frac{N \eta}{2} (x - x_0 - x_g)^2} + \frac{N \eta}{2} x_g^2), \;\;  x_g = {2 \imath {\eta}^{-1} (1 - I^b N)^{1/2} \delta(\sin \phi)}
\end{eqnarray}
Therefore the perturbative prescription is simply to add $x_g$ wherever $x$ occurs ($x \rightarrow x_0 + x_g + \tilde{x}$), and to add a term $\frac{N}{2} \sum_{vk} \eta_{k} x_{gvk}^2$ to the Lagrangian.  These $x_g$ corrections correspond to  $1/g$ corrections, where $g$ is the conductance.  Although $1/g$ corrections have been calculated in the supersymmetric literature \cite{Blanter97}, it seems likely that these particular corrections have been neglected because the standard SUSY procedure is to assume that the saddle point is spatially uniform despite fluctuations in $U$ and $T$.  This amounts to simply dropping $x_g$.

The logarithm in the Lagrangian plays a crucial role in this perturbative approach.  It generates cubic and higher vertices which are proportional to $N (\frac{x}{x_0}-1)^k$; the resulting perturbative corrections are well controlled as long as $\langle (\frac{x}{x_0}-1)^2 \rangle \leq  1/N$.  Since $x \propto x_0 + x_g + (N \eta)^{-1/2}$, the logarithmic diagrams are in control only when $x_g/x_0 \ll N^{-1/2}$.  This is essentially a restriction that fluctuations in $U$ and $T$, when multiplied by $k$, must be smaller than $N^{-1/2}$; i.e. a requirement of spontaneous symmetry breaking.  Additionally, the size of $\hat{E} - \bar{E}$ is constrained.   The perturbation theory breaks down and one must use the spatially-varying saddle points given in equation \ref{SaddlePoints} if SSB is not observed or $\hat{E}- \bar{E}$ is too big.

With a little algebra the Lagrangian at the uniform saddle point can be simplified :
\begin{eqnarray}
&-& { \frac{N V}{2}  \sum_{k} {x_{ 0k}^2  } } + {(N - I^b) V \sum_{k} \ln {  x_{0k}}    } + {{2 \imath NV}{\tilde{\epsilon}}^{-1} (1 - I^b N^{-1})^{1/2} \sum_{k } x_{0k} \sin \phi_{0k} }
\nonumber \\
& = & -2^{-1}NV(I^b + I^f) - NV \sum_k \sin^2 \phi_k + (N - I^b) V \sum_k \ln(\acute{L}_k)    + \imath \acute{L}_k  (  \phi_k + \cos\phi_k \sin \phi_k)
\end{eqnarray}

\subsection{Fluctuations in the Eigenvalues}
Having identified the saddle points, we begin a process of integrating out the eigenvalues $x$.   Fluctuations away from the saddle point will be called $ {\tilde{x}}_v$, giving the decomposition  $ x_v = {x_0 + x_{gv} + {\tilde{x}}_v}$.   We decompose the Lagrangian into the part ${\mathcal{L}}_0 +  \mathcal{L}_1$ which does not depend on ${\tilde{x}}_v$ vs. the part $\tilde{\mathcal{L}}$ which does depend on these fluctuations.  We also approximate $\det(1-k) \approx 1$.
\begin{eqnarray}
{\bar{Z}} & = &  \gamma \int {dU_0} \; {dT_0} \; {dU} \; {dT}  \;\;e^{\mathcal{L}_0 + \mathcal{L}_1 + \tilde{\mathcal{L}}}
\nonumber \\
{\mathcal{L}}_0  & = & 
  - NV \sum_k \sin^2 \phi_k     + \imath  (N - I^b) V \sum_k \acute{L}_k  (  \phi_k + \cos\phi_k \sin \phi_k) 
\nonumber \\
\mathcal{L}_1 & = & {{\imath N}{\tilde{\epsilon}}^{-1} \sum_v {Tr}(T_0^{-1} (T_{v}^{-1} \hat{E}^b T_{v} - \hat{E}^b) T_0 x^b_0   )} 
+ {{\imath N}{\tilde{\epsilon}}^{-1} \sum_v {Tr}(U_0^{\dagger} (  U_{v}^{\dagger} \hat{E}^f U_{v} - \hat{E}^f) U_0  x^f_0 )} + \frac{N}{2} \sum_{vk} \eta_{k} x_{g vk}^2
\nonumber \\
& - &    { \frac{N}{2} \sum_{v_1 v_2} {(k/(1-k))_{v_1 v_2}}{Tr}{(U_{v_1} x^f_0 U^{\dagger}_{v_1}  U_{v_2} x^f_0 U^{\dagger}_{v_2}  )} } 
 +  {\frac{N}{2}{{\sum_{v_1 v_2} {{k_{v_1 v_2}} {Tr}{(T_{v_1} x^b_0 T_{v_1}^{-1} T_{v_2} x^b_0 T_{v_2}^{-1}  )}}}}}
\nonumber \\
e^{\tilde{\mathcal{L}}} & \equiv &
 \int {d\tilde{x}} \; e^{\mathcal{L}_I} \; \exp{(\sum_{vk} -{\frac{N }{2} \eta_{k} {\tilde{x}}_{vk}^2}  )}  \; \det( \alpha + \kappa)\;\; { \prod_v  {\Delta}_{VdM}^2(x^f_{0 } + x^f_{g v} +  \tilde{x}^f_v)}  \; { {\Delta}_{VdM}^2(x^b_{0 } + x^b_{g v} + \tilde{x}^b_v)},
 \nonumber \\
  \alpha & \equiv & {    {((x^f_{0} + x^f_{g v_1} +   \tilde{x}^f_{v_1} )_{i_1 }-(x^b_{0} + x^b_{g v_1} + \tilde{x}^b_{v_1} )_{j_1} ) {\delta_{i_1 i_2}}{\delta_{j_1 j_2}}{\delta_{v_1 v_2}}} }, \;\; \kappa \equiv  {  { ( x^b_{0}  + x^b_{g v_1} + \tilde{x}^b_{v_1})  k_{v_1 v_2} T_{v_1} T_{v_2}^{-1} U_{v_1} U^{\dagger}_{v_2}}    }
\nonumber \\
\mathcal{L}_I & \equiv &   - { \frac{N}{2} \sum_{v_1 v_2} {(k/(1-k))_{v_1 v_2}}{Tr}{(U_{v_1} \tilde{x}^f_{v_1} U^{\dagger}_{v_1}  U_{v_2}   \tilde{x}^f_{v_2}  U^{\dagger}_{v_2}  )} } 
+ {\frac{N}{2}{{\sum_{v_1 v_2} {{k_{v_1 v_2}} {Tr}{(T_{v_1}  \tilde{x}^b_{v_1} T_{v_1}^{-1} T_{v_2}  \tilde{x}^b_{v_2}  T_{v_2}^{-1}  )}}}}}
\nonumber \\
& - & {\sum_{vk} \sum_{z=3}^{\infty} (N - I^b) z^{-1} {{(-x_{0 k}^{-1} (x_{g vk} +  \tilde{x}_{vk}))}^z  }}
\nonumber \\
  x_{0 k} &= &  \acute{L}_{k} \sqrt{1 - I^b N^{-1}} \,\, e^{\imath \acute{L}_k \phi_{k}} ,  \;\; \eta_{k} = 2 \cos \phi_{k} \, \exp(- \imath \acute{L}_k \phi_{k})
  \nonumber \\
 \sin \phi^f_i & =&  2^{-1} \tilde{\epsilon}^{-1} (1 - I^b N^{-1})^{-1/2} ({U_0^{\dagger} {\hat{E}}^f U_0})_{ii},\;\;  \sin \phi^b_j = 2^{-1} \tilde{\epsilon}^{-1} (1 - I^b N^{-1})^{-1/2}  (T^{-1}_0  {\hat{E}}^b T_0)_{jj}
  \nonumber \\
  x_{g vi}^f & \equiv &  \imath \eta_{i}^{-1} \tilde{\epsilon}^{-1} (U_0^{\dagger}(U_v^{\dagger} \hat{E}^f U_v - \hat{E}^f)U_0 + \imath \tilde{\epsilon} \sum_{v_2} (k/(1-k))_{v v_2} U^{\dagger}_v U_{v_2} x^f_0 U^{\dagger}_{v_2} U_v)_{ii} \ll N^{-1/2}
  \nonumber \\
  x_{g vj}^b & \equiv & \imath  \eta_{j}^{-1} \tilde{\epsilon}^{-1} (T_0^{-1} (T_v^{-1} \hat{E}^b T_v - \hat{E}^b)T_0 - \imath \tilde{\epsilon} \sum_{v_2} k_{vv_2} T_v^{-1} T_{v_2} x^b_0 T^{-1}_{v_2} T_v)_{jj} \ll N^{-1/2}
  \nonumber \\
\gamma & = & {  N^{{{N I^{b} V } } + {I^{f} I^{f} V / 2}}  {(\det L)}^{V I^b}} (\det \acute{L})^{V(N - I^b)} { {{{(\prod_{l=N-I^{b}}^{N-1}{l!})}}^{-V}}  2^{ - {I^{f} V / 2}}   {\pi}^{{-I^{b}{(I^{b}-1)}V/2} - {I^{f} I^{f} V / 2}}   }  
\nonumber \\
& \times & e^{-2^{-1}NV(I^b + I^f) + {2^{-1} {NV}{ {\tilde{\epsilon}}^{-2}} V^{V I^f} {Tr}({\hat{E}}^f {\hat{E}}^f)}   } \imath^{-VNI^b - VNI^f}  
 \nonumber \\
{\frac{d}{d{\tilde{J}}^{b}_{vj}}} &\rightarrow & -\imath N  {\tilde{\epsilon}}^{-1} ( T (x^b_0 + x^b_g + \tilde{x}^b) T^{-1})_{v jj}, \;\; \frac{d^2}{{d{\tilde{J}}^{b}_{v_1 j_1}}{d{\tilde{J}}^{b}_{v_2 j_2}}}  \rightarrow  {{\frac{d}{d{\tilde{J}}^{b}_{v_1 j_1}}} \otimes {\frac{d}{d{\tilde{J}}^{b}_{v_2 j_2}}}} 
\label{EigAngleSaddlePointLagrangian}
\end{eqnarray}

 Several terms in these equations are controlled by $N^{-1/2}$ and can therefore be neglected: these include the $x^b_g + \tilde{x}^b$ term which multiplies the kinetic term within the determinant, the logarithmic vertices, and all instances of $I^b N^{-1}$.  Additional $N^{-1/2}$ terms occur within the Van der Monde determinants, which are multiples of many differences $x_{k_1} - x_{k_2}$.  If a pair of signs $\acute{L}_{k_1}$ and $\acute{L}_{k_2} $ are the same then the corresponding factors of $x_{k_1} - x_{k_2}$ nearly cancel; otherwise these factors are of order ${\acute{L}_{k_1} 2 \cos \phi}$.  Therefore we decompose  the Van der Monde determinants into the multiple of two parts: a part $E_0$ containing cosine factors selected by the condition $\acute{L}_{k_1} = - \acute{L}_{k_2} $, and a second part $\tilde{E}$ composed of the small factors selected by the condition $\acute{L}_{k_1} = \acute{L}_{k_2} $.  We neglect the instances of $x_g + \tilde{x}$ occuring in $E_0$.    
 
$e^{\mathcal{L}_I}$ contains kinetic terms which are quadratic  in $\tilde{x}$. These contribute to the Hessian, which is $H = {d^2 \mathcal{L}/d\tilde{x}^2} = N \eta_{j_1} \delta_{v_1 v_2} \delta_{j_1 j_2} + N \xi,$ where $ \xi_{v_1 v_2 j_1 j_2} \equiv k_{v_1 v_2} (T^{-1}_{v_1} T_{v_2})_{j_1 j_2} (T^{-1}_{v_2} T_{v_1})_{j_2 j_1}$.  (The $Q^f$  sector is similar.)   The saddle point approximation generally requires that the entire Hessian should be included in the Gaussian kernel of the $e^{\mathcal{L}_I}$ integral.  In our case $\xi$ is controlled by the small parameter $k$, implying that the eigenvalue dynamics are almost local.  Therefore we will treat $\xi$ perturbatively, leaving it out of the kernel and including it in $e^{\mathcal{L}_I}$.  The validity of this perturbative expansion could be analyzed easily if $e^{\mathcal{L}_I}$ were simply a Gaussian integral without the determinants and the logarithmic vertices: it would be a question of whether $H^{-1}$ could be computed perturbatively.  If fluctuations in $T$ and $U$ are small then $\xi \approx k \ll \eta$ and perturbation theory is justified.    The determinant and the logarithmic corrections complicate things; in this respect the perturbation theory's validity must be explored by calculating  the first order corrections etc. and seeing whether they are small.  

In future expressions we will omit the integration over $\tilde{x}$, with the understanding that sooner or later this integration must be done, using Wick's theorem and $\langle \tilde{x}_{v_1 k_1} \tilde{x}_{v_2 k_2} \rangle = \delta_{v_1 v_2} \delta_{k_1 k_2} N^{-1} \eta_{k}^{-1}$.  (If we were to include the full Hessian then we would have $\langle \tilde{x}_{v_1 k_1} \tilde{x}_{v_2 k_2} \rangle = N^{-1} (\eta + \xi)^{-1}$.)  However we already multiply the overall normalization constant by $\prod_k (\frac{2 \pi}{N \eta_k})^{V/2}$. The simplified Lagrangian is
\begin{eqnarray}
e^{\tilde{\mathcal{L}}} & = &
e^{\mathcal{L}_I} \; {\det(\alpha + \kappa   )} \; \prod_k (\frac{2 \pi}{N \eta_k})^{V/2} \;\; { \prod_v \; E_0(x_{0}) \; \tilde{E}(x_{0 } + x_{gv} + \tilde{x}_v)}   
 \nonumber \\
 \alpha &\equiv& {    {((x^f_{0} + x^f_{g v_1} +   \tilde{x}^f_{v_1} )_{i_1 }-(x^b_{0} + x^b_{g v_1} + \tilde{x}^b_{v_1} )_{j_1} ) {\delta_{i_1 i_2}}{\delta_{j_1 j_2}}{\delta_{v_1 v_2}}} }, \;\; \kappa \equiv  {  {  x^b_{0}   k_{v_1 v_2} T_{v_1} T_{v_2}^{-1} U_{v_1} U^{\dagger}_{v_2}}    }
\nonumber \\
 \mathcal{L}_I & = & - { \frac{N}{2} \sum_{v_1 v_2} {(k/(1-k))_{v_1 v_2}}{Tr}{(U_{v_1}  \tilde{x}^f_{v_1} U^{\dagger}_{v_1}  U_{v_2}   \tilde{x}^f_{v_2}   U^{\dagger}_{v_2}  )} } 
+ {\frac{N}{2}{{\sum_{v_1 v_2} {{k_{v_1 v_2}} {Tr}{(T_{v_1} \tilde{x}^b_{v_1} T_{v_1}^{-1} T_{v_2}  \tilde{x}^b_{v_2}   T_{v_2}^{-1}  )}}}}}
\end{eqnarray}

\subsection{Correspondence to the SUSY Sigma Model}
We compare the saddle point action $\mathcal{L}_1$ in equation \ref{EigAngleSaddlePointLagrangian} with the action of the SUSY sigma model\cite{Mirlin99},  $\frac{\pi \nu}{4}\int {dx} \left[-D \nabla Q \cdot \nabla Q - 2 \imath  \omega \Lambda Q \right]$.  Both Lagrangians contain kinetic terms which are quadratic in $Q$, and also mass terms which are linear in $Q$.  The correspondence can be made precise by setting $k = \frac{D}{2} \nabla \cdot \nabla, \; N \sum_v = \pi \nu \int {dr} $, and neglecting the $(1-k)^{-1}$ factor multiplying $Q^f$'s kinetics.  This will allow us to reproduce SUSY calculations of the two point correlator and of anomalously localized states.  

However there is an important difference in the way that this paper's sigma model and the SUSY sigma model treat the energy band.  The Lagrangian developed here incorporates band information explicitly. $Q^f$ and $Q^b$ are proportional to $\hat{\rho} \propto \cos \phi$, and are small close to the band edge.  Therefore the kinetics are more sensitive to the band edge than than the mass terms.    

In contrast, the SUSY matrix $Q$ is conventionally normalized so that its eigenvalues are $\pm 1$; the SUSY model's only explicit band dependence is via the overall  multiplier $\nu$.  In other words, the SUSY model packages all energy band information in the diffusion constant $D$ and the density of states $\nu$.  These are understood as phenomenological constants.  The present paper's Lagrangian  can be construed as a derivation of $D$'s band dependence: $D \nu \propto \hat{\rho}^2$.  However this interpretation mixes up two different kinds of physics which are probably best left separate: one is the Lagrangian's sensitivity to $Q$'s fluctuations; i.e. $H = d^2 \mathcal{L}/dQ^2$, and the other is the fact that $Q$ gets small close to the band edge.  In any case, presentations of the SUSY sigma model in its final form typically do not provide  explicit prescriptions for calculating how the kinetics and observables depend on $\bar{E}$.

To my best knowledge, this present article's sigma model is the first to include explicitly the kinetic term's correct band dependence.  This explicit band information will be manifested in our results for the two point correlator, anomalously localized states, and perturbative corrections.

\subsection{Saddle Point Analysis of the Wegner Model\label{WegnerSaddle}}
Analysis of the Wegner model's saddle points obtains results quite similar to those of Disertori's model.  We begin by calculating the spatially uniform saddle points.  The following identities are true by construction:
\begin{equation}
 \rho(E) = -\pi^{-1} L_j \sum_v  {Im} \, \frac{d\bar{Z}}{d \tilde{J}^b_{vvj}}, \;\; \rho(E) = \pi^{-1} L_i \sum_v {Im} \, \frac{d\bar{Z}}{d \tilde{J}^f_{vvi}}
\end{equation}
Taking the $\tilde{J}^b$ derivative produces  $ {Re}\,\langle  T_v x^b_v T_v^{-1} \rangle_{jj} = L_j  \hat{\rho}(E),$ where $ \hat{\rho} \equiv \rho(E) \pi \tilde{\epsilon} N^{-1} V^{-1}$.  $\hat{\rho}$ scales with the ratio of the disorder strength to the band width; $\hat{\rho} \propto (\epsilon k)^{-1}$ when the kinetics dominate and $\hat{\rho} \propto 1$ when the disorder dominates. This contrasts with the Disertori model's $\hat{\rho} \approx \cos \phi$ which is always of order $1$.    When evaluating the above equation's $\tilde{J}^f$ derivative  we neglect $I^b N^{-1}$ and the $Q^f - Q^b$ coupling's dependence on $\tilde{J}^f$ - this term's contribution to $\rho$ is surpressed by $N^{-1/2}$.  We obtain $ {Im} \,  {\langle (\imath U x^f U^\dagger + \epsilon k)^{-1} \rangle}_{vvii} \approx L_i  \hat{\rho}(E) $.  In general the  $U$ and $T$ averages prevent precise statements about $x^f_0$ and $x^b_0$, but $U$ and $T$ can be pinned by calculating just the density of states ($I^f = I^b = 1$) or by choosing energy levels that are far apart compared to the level spacing.  Both mechanisms allow one to drop the averages over $U$ and $T$ and obtain  $ {Re}\,\langle   x^b_v \rangle_{jj} \approx L_j  \hat{\rho}(E), \; {Im} \,  {\langle (\imath  x^f_i + \epsilon k)^{-1} \rangle}_{vv} \approx L_i  \hat{\rho}(E) $.  In the weak localization regime ($\epsilon k \gg 1$) the latter equation allows us to connect $\hat{\rho}(E)$ to the density of states $\rho_{\epsilon k}$ of the kinetic operator $\epsilon k$; if ${Re} \, x^f_i \rightarrow 0$ then ${Im}\,(\imath x^f_i + \epsilon k )^{-1} = -\pi V^{-1} \rho_{\epsilon k}(-{Im} \, x^f_i) \; sign({Re}(x^f_i))$. Corrections to this equation are controlled by ${Re} \, x^f_i / E_\Delta$, where $E_\Delta$ is the typical scale of variations in $\epsilon k$'s density of states.  (In a finite lattice $x^f_i$ should be large compared to the level spacing in order to smooth the density of states.)  We conclude that in the weak localization regime $\hat{\rho}(E) \approx \pi V^{-1} \rho_{\epsilon k}(\tilde{\epsilon}^{-1} E)$ and ${Im}\, \langle x^f_i\rangle\approx-\tilde{\epsilon}^{-1}E$. When the disorder dominates ($\epsilon k \ll 1$) $x^f_i$ is never small and is probably best represented as a phase; one obtains ${Re}\, (x^f_i)^{-1} \approx - L_i \hat{\rho}(E)$. 

This analysis can be made more precise by considering the equation governing spatially uniform saddle points $x^f$.  The saddle point equations are practically the same as Disertori's saddle point equations if the disorder is dominant; here we consider the weak localization regime $\epsilon k \gg 1$.  We leave the $Q^f - Q^b$ coupling and Van der Monde determinants out of the saddle point, just as we did with Disertori's model.
\begin{equation}
0 = -N x^f_{0i} + \imath N \tilde{\epsilon}^{-1}(U_0^\dagger \hat{E}^f U_0)_{ii} + \imath N(\imath  x^f_{0i}  + \epsilon  k )^{-1}_{vv}
\end{equation}
When $\epsilon k \gg 1$ this saddle point equation is solved formally by $x^f_{0i} = \imath r_i + \hat{\rho}(-r_i) \; sign({Re}(x^f_{0i})) $, where $r_i = \tilde{\epsilon}^{-1}(U_0^\dagger \hat{E}^f U_0)_{ii} +  {Re}(\epsilon  k - r_i \pm \imath \nu )^{-1}_{vv}$.   True solutions can be obtained numerically, and may yield non-perturbative information about the position of the mobility edge.  We expect that the solutions will be qualitatively different depending on whether or not $\bar{E}$ is well within the energy band.  The sign of ${Re}(x^f_{0i})$ is not fixed by the saddle point equations, signalling the existence of several saddle points distinguished by their sign signatures $\acute{L}$.  When $\hat{E}^f = \bar{E}$, $x^f_0$'s imaginary part is proportional to the identity and its real part is proportional to $\acute{L}$.  This is the same structure seen in Disertori's model.

Next we turn to saddle point analysis of $x^b_{0}$.  This analysis is valid only when $T$ is pinned; otherwise one must do the $T$ integration prior to the saddle point analysis. 
\begin{eqnarray}
0 &=& - N x^b_{j} + \imath N   \tilde{\epsilon}^{-1} (T_0^{-1} \hat{E}^b T_0)_{jj} + N (x^b_{j})^{-1} + \frac{d \mathcal{L}_W}{dx^b_{0j}}, \; e^{\mathcal{L}_W} \equiv \int {dW}\exp(-\imath \epsilon N {Tr}_{vj}(\hat{S}Lk))
\end{eqnarray}
We have seen that $\mathcal{L}_W$ must cancel the $Q^f$ logarithm's dependence on $\epsilon$, and that when $T$ is pinned ${Re}\,\langle   x^b \rangle_{jj} \approx L_j  \hat{\rho}(E)$; clearly the last two terms in the saddle point equation must combine to produce a resolvent like the one governing $x^f_0$.  Equation \ref{dWIntegralSimplified} gives the leading order result for $\mathcal{L}_W$ in the $\epsilon k \gg 1$ regime, which contains a logarithm $- {(N - I^b/2){Tr} \, \ln( L \epsilon k )}$.  Corrections to $\mathcal{L}_W$ must reproduce the resolvent  by adding a self-energy to the logarithm's argument.  As we have already outlined, calculation of these corrections is a rather intricate task and is outside the scope of this paper.  When $I^b = 1$  the first correction is proportional to $N^2 \epsilon x^b_0 \sum_{v_1 v_2} k_{v_1 v_2} G^2_{v_1 v_2}$, but a different formulation like the two-particle irreducible (2PI) effective action \cite{Cornwall74, Luttinger60} may be necessary.   More general considerations indicate that when $\hat{E}^b = \bar{E}$, $x^b_{0j}$ can have only one of two values selected by the value of $L_j$. We expect that $x^b_0$ will change qualitatively when $\bar{E}$ approaches the band edge.  The biggest outstanding questions are whether $x^b_0$'s imaginary part is insensitive to $L_j$, and is the same as $x^f_0$'s imaginary part.  Both questions are answered affirmatively in Disertori's model, and also in Wegner's model when the disorder dominates.  

We conclude that the Wegner model and the Disertori model have the same saddle point structure: $x_0 \approx \acute{L} \hat{\rho}(E) + \imath s(\acute{L}, E)$.  The fermionic sector of $s$ is proportional to the identity and is approximately equal to $-\tilde{\epsilon}^{-1} E$.  The same may be true for $s$ taken as a whole.  The main difference between the two models is $\hat{\rho}$, which in Disertori's model is controlled by the simple semicircular energy band.  In the Wegner model $\hat{\rho}$ is controlled by the spectrum of $\epsilon k$ when $\epsilon k \gg 1$, and is the semicircular band when $\epsilon k \ll 1$.  

\subsubsection{The Hessian}
Fluctuations in the $x^f$ eigenvalues are controlled by the Hessian, the second derivative  of the Lagrangian with respect to $x^f_{v i}$:
\begin{equation}
H_{v_1 v_2 i_1 i_2} = -N \delta_{i_1 i_2} \delta_{v_1 v_2}  - N \delta_{i_1 i_2} (\imath  x^f_i + \epsilon  k )^{-1}_{v_1 v_2 } (\imath  x^f_i + \epsilon  k )^{-1}_{v_2 v_1 }
\label{HessianInPositionBasis}
\end{equation}
Eigenvalue fluctuations scale with the inverse square root of the Hessian; we see immediately that they are controlled by $N^{-1/2}$, just as in Disertori's model.  There the Hessian is $N \eta \,\delta_{v_1 v_2} + N k$; it is almost local, which allowed us to decouple the eigenvalue integrals site by site.   The Wegner Hessian  is almost local both when $\epsilon k \gg 1$ and when $\epsilon k \ll 1$.  To establish this fact we must analyze the second term, which is the product of two Green's functions.   It may evaluated numerically if accuracy is desired, but in the two limits $\epsilon k \ll 1, \epsilon k \gg 1$ analytical techniques may be used instead.    In the $\epsilon k \ll 1$ regime $x^f_i$ is of order $1$ and the two Green's functions may be expanded in powers of $\epsilon k / x^f_i$; at leading order the resulting Hessian is the same as is found in the Disertori model.

The opposite limit $\epsilon k \gg 1$ is less trivial.  Direct examination of equation  \ref{HessianInPositionBasis} is invalid because $k$ has eigenvalues that are arbitrarily small, including of course the zero mode $k|\vec{0}\rangle = 0$. The standard analysis runs as follows.  We switch to a momentum basis, in which the Hessian is
\begin{equation}
H_{v_1 v_2}(\vec{s}_1, \vec{s}_2) = -N \delta_{i_1 i_2} \delta(\vec{s}_1 - \vec{s}_2)  - N \delta_{i_1 i_2} \delta(\vec{s}_1 - \vec{s}_2) V^{-1} \sum_{\vec{s}_3} (\imath  x^f_i + \epsilon  k(\vec{s}_3) )^{-1}  (\imath  x^f_i + \epsilon  k(\vec{s}_1 - \vec{s}_3) )^{-1}
\end{equation}
The second term may be rewritten in terms of the density of states: 
\begin{eqnarray}
- N V^{-1} \delta_{i_1 i_2} \delta(\vec{s}_1 - \vec{s}_2) &\;& \int {d\varsigma} \rho_{\epsilon k}(\varsigma) (\imath  x^f_i +  \varsigma )^{-1}  \langle (\imath  x^f_i + \epsilon  k(\vec{s}_1 - \vec{s}_3) )^{-1} \rangle,\; 
\nonumber \\ 
\langle (\imath  x^f_i + \epsilon  k(\vec{s}_1 - \vec{s}_3) )^{-1} \rangle &\equiv& \rho_{\epsilon k}^{-1}(\varsigma)   \sum_{\vec{s}_3} \delta(\varsigma - \epsilon k(\vec{s}_3)) (\imath  x^f_i +  \varsigma + (\epsilon  k(\vec{s}_1 - \vec{s}_3) - \epsilon k(\vec{s}_3)) )^{-1}
\label{OneLoopContourIntegral}
\end{eqnarray}
In order to make further progress analytically one must restrict the momentum $\vec{s}_1$ to obey the condition $\epsilon  k(\vec{s}_1 - \vec{s}_3) - \epsilon k(\vec{s}_3) \ll {Re} \, x^f_i \; \forall \, \vec{s}_3 \, \exists \,\epsilon k(\vec{s}_3) = \varsigma $.  This equality is certainly true when $\vec{s}_1 = 0$ and may be true for a range of small momenta.  The restriction means that our analysis will only give the long-distance behavior of the Hessian; shorter distances require numerical evaluation.  Next we expand $(\imath  x^f_i +  \varsigma + (\epsilon  k(\vec{s}_1 - \vec{s}_3) - \epsilon k(\vec{s}_3)) )^{-1}$ in powers of $-(\epsilon  k(\vec{s}_1 - \vec{s}_3) - \varsigma)/(\imath  x^f_i +  \varsigma)$.  In the $\epsilon k \gg 1$ regime ${Re} \, x^f_i/ E_\Delta \ll 1$,  equation \ref{OneLoopContourIntegral} may be converted to a contour integral, and one obtains
\begin{equation}
H_{v_1 v_2}(\vec{s}_1, \vec{s}_2) = -N V^{-1} \delta_{i_1 i_2} \delta(\vec{s}_1 - \vec{s}_2) (V +  2 \pi \imath \frac{d \rho_{\epsilon k}(\varsigma)}{d \varsigma} - \pi \imath \frac{d^2}{d\varsigma^2}(\rho_{\epsilon k}(\varsigma) \langle \epsilon  k(\vec{s}_1 - \vec{s}_3) - \varsigma \rangle ))_{\varsigma = -\imath x^f_i}
\label{ApproximatedHessian}
\end{equation}
The first two terms are mass terms, while the last is kinetic.  ${Re} \, x^f_i \approx \hat{\rho} \approx \pi V^{-1} \rho_{\epsilon k} \propto (\epsilon k)^{-1}$, so the first term is roughly  $(\epsilon k)^{2}$ larger than the last two terms, and the Hessian is nearly local.  This same reasoning provides the standard derivation of the low energy Lagrangian governing Goldstone bosons both in  our models and in the SUSY sigma model.  In these cases the mass terms are  either absent or  exactly cancelled by another term in the Lagrangian.  Therefore the Goldstone bosons are massless and their low-momentum Lagrangian is proportional to $\epsilon k \propto D \nabla \cdot \nabla$.

The obvious difficulty with this reasoning is that as $\epsilon k$ becomes larger our $\epsilon  k(\vec{s}_1 - \vec{s}_3) - \epsilon k(\vec{s}_3) \ll {Re} \, x^f_i$ condition becomes ever more restrictive, and for large enough $\epsilon k$ becomes the same as the simple restriction $\vec{s}_1 = 0$.  One must ask at what point does the restriction make the entire analysis meaningless; how small  can we make the low-momentum Lagrangian's range of validity?  It is not clear whether moving to the continuum really answers this question.  Probably equation \ref{ApproximatedHessian} should be interpreted as a reliable analysis  of  orders of magnitude at small momenta, but not as a recipe for calculating either the Hessian or the Lagrangian.  If either is required one should evaluate the Hessian numerically.

Turning to $x^b$, its Hessian is proportional to $N$, and therefore its fluctuations  are controlled by $N^{-1/2}$.   Analysis of the Hessian's locality requires computation of the second derivative of $\mathcal{L}_W$.  We saw that $\mathcal{L}_W$'s first derivative must produce a resolvent, so it is likely that its second derivative is a product of two Green's functions, the same as in the $x^f$ sector.  

\subsubsection{Corrections to the Sigma Model Approximation}
Here we analyze what happens to the saddle point when $U$ and $T$ are not spatially uniform.   As with Disertori's model, we do not allow the saddle point to fluctuate but instead introduce a perturbation theory in  $x_g = H^{-1} \delta$, where $\delta$ is the difference between the spatially uniform saddle point equation and the true saddle point equation:
\begin{equation}
H x_g^f = \imath N \tilde{\epsilon}^{-1}(U_0^\dagger(U_v^\dagger \hat{E}^f U_v - \hat{E}^f) U_0)_{ii} + \imath N(U^\dagger(\imath U x^f U^\dagger + \epsilon  k )^{-1}U)_{vvii} - \imath N(\imath  x^f_{0i}  + \epsilon  k )^{-1}_{vv}
\end{equation}
The $N$'s on the right side cancel the $N$ in the Hessian.  If fluctuations in $U$ are small then the last two terms may be expanded in a Taylor series; one obtains a form analogous to the Disertori model's $ x_g$, which is
\begin{equation}
  x_{g vi}^f  =  \imath \eta_{i}^{-1} \tilde{\epsilon}^{-1} (U_0^{\dagger}(U_v^{\dagger} \hat{E}^f U_v - \hat{E}^f)U_0 + \imath \tilde{\epsilon} \sum_{v_2} (k/(1-k))_{v v_2} U^{\dagger}_v U_{v_2} x^f_0 U^{\dagger}_{v_2} U_v)_{ii} 
\end{equation}
Similar results can be obtained for the bosonic correction $x_g^b$.  The vertices of the $x_g$ perturbation theory correspond to third and higher derivatives of the Lagrangian's logarithms and of $\mathcal{L}_W$.  These vertices are proportional to $N$, so the $x_g$ perturbation theory is in control only if $x_g \ll N^{-1/2}$.  Any non-locality of Wegner's Hessian should not change this requirement significantly.  Because of the complete correspondence between Wegner's $x_g$ perturbation theory and Disertori's perturbation theory, all subsequent  convergence analysis applies equally to both models.  In particular, we will find that the sigma model approximation can not be controlled except in the spontaneously broken phase where fluctuations in $U$ and $T$ are small.  

\subsubsection{The Sigma Model Lagrangian}
We write Wegner's sigma model Lagrangian in terms of a part $\mathcal{L}_1$ which controls spatial fluctuations of $U, T$ and a part $\mathcal{L}_0$ which depends only on $U_0, T_0$:
\begin{eqnarray}
{\mathcal{L}_1}  & = &   
 {{\imath N}{\tilde{\epsilon}}^{-1} \sum_v {Tr}(   x_0^f U_0^\dagger (U^{\dagger}_{v}  {{\hat{E}}^{f}} U_{v} - \hat{E}^f) U_0 )} 
 + {{\imath N}{\tilde{\epsilon}}^{-1} \sum_v {Tr}( x_0^b T_0^{-1} (T_{v}^{-1} {{\hat{E}}^{b}} T_{v} - \hat{E}^b) T_0 )}
 \nonumber \\
 & + &    (N - I^b) {Tr}_{vi}(\ln (U x^f_0 U^\dagger - \imath \epsilon k -\imath \tilde{\epsilon}^{-1} \tilde{J}^f) - \ln ( x^f_0  - \imath \epsilon k ))
    + \mathcal{L}_W(\tilde{J}^b) - \mathcal{L}_W(\tilde{J}^b = 0, T=1) 
\nonumber \\
\mathcal{L}_0 & = & - {\frac{N}{2}{{\sum_{vk}   x_{vk}^2  }}}  + {{\imath N V}{\tilde{\epsilon}}^{-1}  {Tr}(  U_0 x_0^f U_0^{\dagger}  {{\hat{E}}^{f}} )} + {{\imath N V}{\tilde{\epsilon}}^{-1}  {Tr}(T_0 x_0^b T_0^{-1} {{\hat{E}}^{b}}   )}  
\nonumber \\
&+& (N - I^b) {Tr}_{vi}(\ln ( x^f_0  - \imath \epsilon k ))  + {(N - I^b) \sum_{vj} \ln  x^b_{vj}}  +  \mathcal{L}_W(\tilde{J}^b = 0, T=1)
 \nonumber \\
\hat{S}_{v_1 v_2} L & = &  C^\dagger_{v_1} s_{v_1} W^\dagger_{v_1} W_{v_2} s_{v_2} C_{v_2} L, \;\;  C^\dagger_v s^2_v C_v L = T_v x^b_0 T_v^{-1}, \; e^{\mathcal{L}_W} = \int {dW} \exp( -\imath N  {Tr}_{vj}(\hat{S} L {( \epsilon k + \tilde{\epsilon}^{-1} {\tilde{J}}^b)} )  )
\end{eqnarray}
The saddle point equation ensures that $\mathcal{L}_0$'s leading dependence on $\hat{E} - \bar{E}$ is just ${{\imath N V}{\tilde{\epsilon}}^{-1}  {Tr}(  U_0 x_0^f U_0^{\dagger}  {{\hat{E}}^{f}} )} + {{\imath N V}{\tilde{\epsilon}}^{-1}  {Tr}(T_0 x_0^b T_0^{-1} {{\hat{E}}^{b}}   )}  $.  

Unlike the SUSY sigma model, the present sigma model is not fully described by the Lagrangian.  Much of the interesting physics lies in the $Q^f - Q^b$ determinant, and also one  must take into account the Van der Monde determinants, as described in equation \ref{TransformedWegnerModel}.  We have glossed over the $Q^f - Q^b$ determinant's dependence on $W$, which causes a coupling between the determinant and the sigma model Lagrangian.

In the $\epsilon k \gg 1$ regime this sigma model is capable of producing accurate predictions incorporating full information about the energy band.   The logarithm controlling $U$'s kinetics can be expanded in a Taylor series whose coefficients may be determined either numerically or via analytic approximations.  $\mathcal{L}_W$'s Taylor series coefficients are equivalent to certain moments of the $dW$ integral, and these also may be calculated numerically or approximately.  An important outstanding problem is whether ${Im}\, x_0^b$ is the same as ${Im}\,x^f_0$.  If so, the kinetic terms  and the $\hat{E}$ terms in $\mathcal{L}_1$ are respectively quadratic and linear in $\hat{\rho}$, the same as in Disertori's model. If not, in subsequent formulas one must substitute $x^b_0$ for $\hat{\rho}$.   Later we will calculate observables in Disertori's model; the reader should bear in mind that the Wegner Lagrangian is very similar and produces analogous results.
  
\section{The $Q^f - Q^b$ Coupling in the Weak Localization Regime \label{WeakLocalization}}
We now narrow our focus to weak localization phenomena; i.e. the regime in which the kinetics dominate the disorder.  More precisely, in Disertori's model the weak localization regime is characterized by the condition\footnote{The SUSY literature generally has a looser definition of the weak localization regime.  We have chosen this strict definition in order to preserve mathematical rigor.} that  $N^{-1/2} \ll k_0, k \ll 1$. $E_{Th} = \tilde{\epsilon} k_0$ is the Thouless energy, the smallest non-zero eigenvalue of the kinetic operator $\tilde{\epsilon} k$.   We will use the small parameter $(N^{1/2} k_0)^{-1}$ to control the theory.  

We will also assume  spontaneous symmetry breaking of the continuous global symmetries of $U$ and $T$, implying that the system is in the delocalized phase.   In $D = \{1,2\}$ dimensions the Mermin-Wagner theorem \cite{Mermin66} prohibits SSB in very large volumes; however in finite volumes one can still see an effective SSB \cite{Posazhennikova06}.   We will outline a way of calculating corrections to the SSB assumption in powers of the inverse conductance, preparing the way for rigorous proofs about observables in the weak localization regime.

$Q^f$ and $Q^b$ are coupled by a determinant $\det(\alpha + \kappa)$, where
\begin{equation} \alpha  \equiv {    {((x^f_{0} + x^f_{g v_1} +   \tilde{x}^f_{v_1} )_{i_1 }-(x^b_{0} + x^b_{g v_1} + \tilde{x}^b_{v_1} )_{j_1} ) {\delta_{i_1 i_2}}{\delta_{j_1 j_2}}{\delta_{v_1 v_2}}} }, \;\; \kappa \equiv  {  { ( x^b_{0}  + x^f_{g v_1} + \tilde{x}_{v_1})  k_{v_1 v_2} T_{v_1} T_{v_2}^{-1} U_{v_1} U^{\dagger}_{v_2}}}
\end{equation}  
Our assumption of spontaneous symmetry breaking implies that fluctuations in $U$ and $T$ are small; as a result $\kappa$'s spectrum is controlled by $k$.  In particular $k$ has a zero eigenvalue $k |\vec{0}\rangle = 0$; therefore $A \equiv \alpha + \kappa$ has near-zero eigenvalues that are particularly sensitive to fluctuations in $x,\,U$, and $T$.  In order to capture this physics we will project out the zero-momentum sector of $A \equiv \alpha + \kappa$ and treat it separately from the other momenta.  We define two projection operators: $P_0$ selects out the zero-momentum modes, and $P_+ = 1 - P_0$.  The following formulas describe the decoupling:
\begin{eqnarray}
\det( A  ) &=& \det(P_{+} A P_{+}) \; \det({P_0 A P_0} - P_0 A (P_+ A P_+)^{-1} A P_0),
\nonumber \\
P_0 A P_0 & = & (x^f_0 + \bar{x}^f_g + \bar{\tilde{x}}^f) - (x^b_0  + \bar{x}^b_g  + \bar{\tilde{x}}^b) + {V^{-1} x^b_0 \sum_{v_1 v_2} (k T T^{-1} U U^{\dagger})_{v_1 v_2}}
\nonumber \\
V^{1/2} (P_0 A P_+)_{v_1 v_2}   & = &  (x^f_{g v_2} + \tilde{x}^f_{v_2} - \bar{x}^f_g - \bar{\tilde{x}}^f) - (x^b_{g v_2} + \tilde{x}^b_{v_2} - \bar{x}^b_g  - \bar{\tilde{x}}^b) 
\nonumber \\
&+& {x^b_0 \sum_{v_3}   (k T T^{-1} U U^{\dagger})_{v_3 v_2}} - {V^{-1} x^b_0 \sum_{v_3 v_4} (k T T^{-1} U U^{\dagger})_{v_3 v_4}}
\label{ZeroMomentumDecoupling1}
\end{eqnarray}
The new $\bar{x}$ notation prescribes spatial averaging $V^{-1} \sum_v$. The $kTT^{-1} U U^{\dagger}$ terms are exactly zero when $U$ and $T$ are constant; they are proportional to the fluctuations in $U$ and $T$, the same as $x_g$.  If $k$ is long-ranged then the sums over $v$ are regulated by factors of $V^{-1/2}$ caused by the absence of long range correlations in $U$ and $T$. 

\subsection{The $P_+$ Sector}
In the following paragraphs we will analyze the spectrum of $P_+ A P_+$ when  $x_g = \tilde{x} = 0$, starting with $\alpha$'s spectrum and later taking  $\kappa$ into account.   $\alpha$ is a local operator; its eigenvalues are just $\delta x_{0 ij} = x^f_{0 i} - x^b_{0 j}$.  $x^f_{0 i}$ and $x^b_{0j}$ nearly cancel each other if $i$ and $j$ satisfy $\acute{L}_i = \acute{L}_j$; in the opposite case they add to each other.  The small eigenvalues of $\alpha$ are quite small, of order $\delta x_0 \propto  (\hat{E}_i - \hat{E}_j)/(2 \eta \tilde{\epsilon})$ if $\hat{E}_k$ doesn't depend much on $k$.  The large eigenvalues are approximately $\acute{L}_i 2 \cos(\bar{E})$.   We separate the two sectors of $\alpha$ with projection operators $P_{small} + P_l = 1$; the projection operator for the small eigenvalues is $P_{small} = P_s = \delta_{\acute{L}_i \acute{L}_j} \delta_{i_1 i_2} {\delta_{j_1 j_2}}{\delta_{v_1 v_2}}$.  
  
The $Q^f - Q^b$ coupling includes an additional term $ \kappa $ which mixes the eigenstates of $\alpha$.  We estimate the effect of mixing on the large eigenvalues by applying the identity $a |\psi\rangle = ({P_{+l} A P_{+l}} - {P_{+l} A (P_{+s} A P_{+s})^{-1} A P_{+l}}) |\psi\rangle$.   The $P_{+l} A P_{+l}$ term is of order  $\acute{L}_i 2 \cos(\bar{E}) + O(k)$, while the last term in the identity is of order   $k^2 \times (P_{+s} A P_{+s})^{-1}$.  We have assumed that fluctuations in $U$ and $T$ are small; therefore $P_{+s} A P_{+s} \propto max(k, \, \delta x_0)$; in consequence $\kappa$ makes only perturbative changes in the large eigenvalues, with the changes controlled by the small parameter $k$.  Similarly, we analyze the effect of the mixing on the small eigenvalues with the identity $a |\psi\rangle = ({P_{+s} A P_{+s}} -{P_{+s}A (P_{+l} A P_{+l})^{-1} A P_{+s}}) |\psi\rangle$.   The $P_{+s} A P_{+s}$ term is of order $max(k, \, \delta x_0)$, while the last term is of order $k^2$, so its effects on the spectrum are controlled by the small parameter $k$.   We conclude that $P_+ A P_+ $ has a sector well described by $P_{+l}$ whose eigenvalues are all of order $ \acute{L}_i 2 \cos(\bar{E})$ and another sector well described by $P_{+s}$ whose eigenvalues are all of order $k$.  Corrections to this picture of the spectrum are controlled by the small parameter $k$, and are very small compared to the saddle point action (equation \ref{EigAngleSaddlePointLagrangian}) which contains a factor of $N$, so we will neglect them.
  
 Because the smallest eigenvalue in $P_+ A P_+$ is of order $k_0$, the effects on its spectrum of $x_g$, $\tilde{x}$, and fluctuations in $U$ and $T$ are perturbative, controlled by $(N^{1/2} k_0)^{-1} \ll 1$.  Expanding the $P_+ A P_+$ determinant perturbatively produces terms like $\exp({Tr}(k^{-1} (x_g + \tilde{x})))$. These  terms are very small compared to the saddle point action, so we will neglect them, approximating $\det(P_+ A P_+)$ as $\det(P_{+s} (x_0^f - x_0^b + x_0^b k) P_{+s} ) \;\; \det(P_{+l} (x_0^f - x_0^b) P_{+l})$.

\subsection{The Zero Momentum Sector}
We turn to the determinant which controls the zero momentum sector, $\det({P_0 A P_0} - P_0 A (P_+ A P_+)^{-1} A P_0)$.   We introduce projection operators $P_{0l}, \; P_{0s}$ satisfying $P_0 = P_{0l} + P_{0s}$ and then analyze the determinant in terms of the two sectors.  $g^{-1}$ signifies the $kTT^{-1} U U^{\dagger}$ terms.  For simplicity we assume that these terms are smaller than $N^{-1/2}$.  The following matrices give orders of magnitude; the upper left entries correspond to the $P_{0l}$ sector while the lower right entries correspond to the $P_{0s}$ sector.
\begin{eqnarray}
P_0 A P_0 &\propto & \begin{bmatrix} 1 & g^{-1} \\  g^{-1} & {\delta x_0} +N^{-1/2} + g^{-1}  \end{bmatrix}, 
\;\; P_0 A P_+ \propto   \begin{bmatrix} N^{-1/2} + g^{-1} & g^{-1} \\ g^{-1} & N^{-1/2} + g^{-1} \end{bmatrix},
\;\; (P_+ A P_+)^{-1} \propto \begin{bmatrix} 1 & 1 \\ 1 & k^{-1} \end{bmatrix} 
\\ \nonumber 
{P_0 A P_0} & - & P_0 A (P_+ A P_+)^{-1} A P_0 \propto 
\begin{bmatrix} 1  &  g^{-1} - g^{-1} N^{-1/2} k^{-1} \\ g^{-1}  - g^{-1} N^{-1/2} k^{-1} \;\; & {\delta x_0} + N^{-1/2} + g^{-1} - N^{-1} k^{-1}  - g^{-1} N^{-1/2} k^{-1} \end{bmatrix}
\end{eqnarray}
We have dropped terms whose relative magnitude is $N^{-1/2}$ or $k$ or smaller, and also some terms of order $g^{-2} k^{-1}$.  In the second line the matrix elements coupling $P_{0s}$ with $P_{0l}$ are less than $N^{-1/2}$, while the $P_{0s} - P_{0s}$ matrix elements are of order $N^{-1/2} + {\delta x_0}$.  Therefore the coupling between the two sectors is less than $N^{-1/2}$ and can be neglected, producing $\det(P_{0l} (x^f_0 - x^b_0) P_{0l}) \;\; \det({P_{0s} A P_{0s}}  -  P_{0s} A (P_+ A P_+)^{-1} A P_{0s})$.  The contribution to the $P_{0s}$ determinant from mixing with the $P_+$ sector is controlled by $(N^{1/2} k_0)^{-1}$, which is small in the weak localization regime.   We neglect this contribution, arriving at a final form for the $Q^f - Q^b$ coupling in the weak localization regime:
\begin{eqnarray}
\det( \alpha + \kappa  ) &=& \det(P_{+s} (x_0^f - x_0^b + x_0^b k) P_{+s}) \; \; \det(P_{l} (x^f_0 - x^b_0) P_{l}) \;\; \det({P_{0s} A P_{0s}}) 
\nonumber \\
P_{0} (\alpha + \kappa) P_0 & = & (x^f_0 + \bar{x}^f_g + \bar{\tilde{x}}^f) - (x^b_0  + \bar{x}^b_g  + \bar{\tilde{x}}^b) + {V^{-1} x^b_0 \sum_{v_1 v_2} (k T T^{-1} U U^{\dagger})_{v_1 v_2}}
\label{ZeroMomentumDecoupling2}
\end{eqnarray}

We have derived this form under the implicit assumption that we will not be calculating correlations between $Q^f$ and $Q^b$.  The $Q^f - Q^b$ determinant is the only source of such correlations in this theory, and should be treated more gently when they are being computed.

\subsection{Decoupling of $U$ and $T$}

The mathematically controlled reasoning leading to equation  \ref{ZeroMomentumDecoupling2} is extremely interesting because it sketches a proof that $U$ and $T$ decouple almost completely in the SSB phase: the only exception is the zero-mode coupling $\det({P_{0s} A P_{0s}})$.  Consider the case of $I^b = I^f = 2$, which is appropriate for computing two point correlators.   If $U$ and $T$ were decoupled completely, the $U$ sector would be just the classical Heisenberg model, while the $T$ sector would be a hyperbolic sigma model introduced by Spencer and Zirnbauer.   The hyperbolic model exhibits SSB  unconditionally in $D > 2$; there is no phase transition \cite{Spencer04}.  The Heisenberg model in $D > 2$ has been proven rigorously to exhibit spontaneous symmetry breaking \cite{Frohlich76}.  It is also believed that these models display an effective SSB even in $D = \{1, 2\}$ when the system is sufficiently small. 

The $Q^f - Q^b$ zero-mode coupling contains only a few factors multiplying the path integral, while the saddle point action (equation \ref{EigAngleSaddlePointLagrangian}) is proportional to the volume $NV$.  Roughly speaking, the $Q^f- Q^b$ coupling is a factor of $(NV)^{-1}$ smaller than the action, and is best treated as a prefactor.  This implies that the zero-mode coupling does not change the SSB behavior of the underlying Heisenberg and hyperbolic models, and that the SSB assumption is self-consistent in $D > 2$ dimensions.  

In order to prove SSB in $D > 2$ dimensions one would have to prove bounds on the spectra  of $P_{+s} A P_{+s}$ and $P_{+l} A P_{+l}$ which are stringent enough to show that $P_{+} A P_{+}$ is bounded below $O(k_0)$. Such bounds may be available from probabilistic arguments based the fact that the theory penalizes configurations in which the determinant has one or more small eigenvalues.
 
 \subsection{The Wegner Model\label{WegnerDeterminant}}
 Wegner's $Q^f - Q^b$ coupling is a bit different than Disertori's:
 \begin{eqnarray}
 & \, & \det(- \imath \epsilon k + Q^f - A^1(\hat{S})), \;\; A^1 \equiv \sum_{i_1, v_1} {(Q^f - \imath \epsilon k)_{i_0 i_1 v_0 v_1}} {(Q^f - \imath \epsilon k)^{-1}_{i_1 i_2 v_1 v_2}}  {\hat{S}_{v_1 v_2 j_1 j_2}} {L_{j_{1}}} , \;\; 
 \nonumber \\
\hat{S}_{v_1 v_2} L & = &  C^\dagger_{v_1} s_{v_1} \langle W^\dagger_{v_1} W_{v_2} \rangle s_{v_2} C_{v_2} L, \;\;  C^\dagger_v s^2_v C_v L = T_v x^b_v T_v^{-1}
\end{eqnarray}
Ignoring fluctuations in $W$, at the saddle point $\hat{S}$ is spatially constant and the determinant simplifies to $\det(-\imath \epsilon k + x^f_0 - x^b_0)$.  This should be compared to the Disertori model's $\det(x^b_0 k + x^f_0 - x^b_0)$.  We see again the same possibility of cancellation vs addition between $x^f_0$ and $x^b_0$, which distinguishes the $P_s$ sector from the $P_l$ sector.  When the disorder is small ($\epsilon k \gg 1$) the saddle point solutions are roughly ${Re}\, x_0 \approx \hat{\rho} \propto (\epsilon k)^{-1}$, so the $P_l$ sector's eigenvalues are not much larger than the $P_s$ eigenvalues. The only real distinction between the two sectors is that the $P_{+l}$ eigenvalues have a real component of order $2 \hat{\rho}$ while the real component of the $P_{+s}$ eigenvalues is much smaller and perhaps even zero.  The real component corresponds to a mass, and it is remarkable the mass in Wegner's $Q^f - Q^b$ determinant are rather small compared to the kinetic energy ($2 \hat{\rho} + \imath \epsilon k$), unlike the Disertori model where the $P_l$ sector's masses completely dominate the kinetic energy.

The rough equality of the $P_{+l}$ and $P_{+s}$ eigenvalues simplifies our argument that none of the $P_+$ sector's eigenvalues is much smaller than $k_0$, which is the basis our factorization of the $P_0$ sector of the determinant from the $P_+$ sector.  Within the $P_0$ sector $P_{0s}$ can still be factorized from $P_{0l}$, so that a controlled analysis of the $Q^f - Q^b$ coupling in the $\epsilon k \gg 1$ regime should obtain a form similar to Disertori's coupling given in \ref{ZeroMomentumDecoupling2}, and eventually arrive at identical predictions for observables at leading order.
  
\section{\label{Dimensionality}Fluctuations in $Q^f$ and $Q^b$}
When the global symmetries in $Q^f$ and $Q^b$ are spontaneously broken, fluctuations in these fields are small and weakly interacting.  The next step is to integrate these fluctuations.  To simplify the arithmetic, we specialize this paper's mathematical formulas to calculate two point correlators.  We set $I^b = I^f = 2$, so that $Q^f$ and $Q^b$ are both $2\times 2$ matrices.  The dominant saddle points, i.e. the ones which minimize the number of near-zero factors in the determinants, are $\acute{L}^f_i = (\sigma_3)_{ii}, \; \acute{L}^b_j = (\sigma_3)_{jj}$ for the Retarded-Advanced correlator, and $\acute{L}_k = \{1, 1, -1, -1\}$ for the Advanced-Advanced correlator.   ($\sigma_3 $ is the Pauli matrix.)  At the $\acute{L}_k = \{1, 1, -1, -1\}$ saddle point all of the $U$ and $T$ terms disappear from the saddle point Lagrangian (equation \ref{EigAngleSaddlePointLagrangian}) leaving only $\frac{N}{2}\sum_{vk} \eta_k x^2_{gvk}$; this saddle point describes almost local dynamics.  Therefore in $D > 0$ dimensions the Advanced-Advanced correlator is at first order purely local, the same as the zero-dimensional result.  We will focus on the Advanced-Retarded correlator.

We adopt parameterizations of $T$ and $U$  which are tailored for perturbation theory:
\begin{equation}
T = \begin{bmatrix} {\sqrt{1 + (y^b)^2 + (z^b)^2}} & {y^b - \imath z^b} \\ {y^b + \imath z^b} & {\sqrt{1 + (y^b)^2 + (z^b)^2}}  \end{bmatrix}, 
\;\; U =   \begin{bmatrix} {\sqrt{1 - (y^f)^2 - (z^f)^2}} & {\imath y^f + z^f}  \\ {\imath y^f - z^f}  & {\sqrt{1 - (y^f)^2 - (z^f)^2}} \end{bmatrix}
\end{equation}
The inverses of $T$ and $U$ can be obtained by inverting the sign of both $y$ and $z$.
$y^b$ and $z^b$ vary from $-\infty$ to $\infty$, while $y^f$ and $z^f$ vary within the unit circle defined by $(y^f)^2 + (z^f)^2 = 1$.  The  integration measures are $ {dy^f} {dz^f} $ and  $  {dy^b} {dz^b}$.  

 We parameterize $U_0$ with $\lambda^f_0, \theta^f_0$ such that $\lambda^f_0 = 1 - 2(y^f_0)^2 - 2 (z^f_0)^2,\; \tan \theta^f_0 = z^f_0/y^f_0$.  Similarly we parameterize $T_0$ with $\lambda^f_0 = 1 + 2(y^b_0)^2 + 2 (z^b_0)^2,\; \tan \theta^b_0 = z^b_0/y^b_0$.  The Jacobian for $U_0, T_0$ in these coordinates is  $2^{-4} {d\lambda_0^f} {d\lambda_0^b} {d\theta_0^f} {d\theta_0^b}$.  The limits of the $\lambda^b_0$ integration are from $1$ to $\infty$, while the $\lambda^f_0$ integration is from $-1$ to $1$. In these coordinates the following relations are very useful:
\begin{eqnarray}
(U^{\dagger}_2 U_1 {\sigma}_3 U^{\dagger}_1 U_2)_{11} & = & -(U^{\dagger}_2 U_1 {\sigma}_3 U^{\dagger}_1 U_2)_{22} 
\nonumber \\
&= & {\lambda_2 (1 -2 y_1^2 - 2 z_1^2)} + {2 \sqrt{1 - \lambda_2^2} \sqrt{1 - y_1^2 - z_1^2} (y_1 \cos \theta_2 + z_1 \sin \theta_2)}
\nonumber \\
&= & {(2 y_1^2 + 2 z_1^2 -1)(2 y_2^2 + 2 x_2^2 - 1)} + {4 \sqrt{1 - y_1^2 - z_1^2} \sqrt{1 - y_2^2 - z_2^2} (y_1 y_2 + z_1 z_2)}
\nonumber \\
& \approx & 1 - {2 (y_1 - y_2)^2} - {2 (z_1 - z_2)^2} + {4(y_1^2 + z_1^2)(y_2^2 + z_2^2)} - {2(y_1^2 + z_1^2 + y_2^2 + z_2^2)(y_1 y_2 + z_1 z_2)}
\nonumber \\
(T^{-1}_2 T_1 {\sigma}_3 T^{-1}_1 T_2)_{11} & = & -(T^{-1}_2 T_1 {\sigma}_3 T^{-1}_1 T_2)_{22}
\nonumber \\
 & = & {\lambda_2 (2 y_1^2 + 2 z_1^2 +1)} - {2 \sqrt{\lambda_2^2 - 1} \sqrt{1 + y_1^2 + z_1^2}  (y_1 \cos \theta_2 + z_1 \sin \theta_2)}
\nonumber \\
 & = & {(2 y_1^2 + 2 z_1^2 +1)(2 y_2^2 + 2 x_2^2 + 1)} - {4 \sqrt{1 + y_1^2 + z_1^2} \sqrt{1 + y_2^2 + z_2^2} (y_1 y_2 + z_1 z_2)}
 \\ \nonumber
& \approx & 1 + {2 (y_1 - y_2)^2} + {2 (z_1 - z_2)^2} + {4(y_1^2 + z_1^2)(y_2^2 + z_2^2)} - {2(y_1^2 + z_1^2 + y_2^2 + z_2^2)(y_1 y_2 + z_1 z_2)}
\end{eqnarray}

We define $\hat{E} = \bar{E} +  (\omega/2) \sigma_3, \;\omega \equiv E_1 - E_2$.  The saddle point equations are
 $2 \tilde{\epsilon} \sin \phi^f_i  =    ({U_0^{\dagger} {\hat{E}}^f U_0})_{ii} = \bar{E} + \frac{1}{2} \omega^b \lambda^b (\sigma_3)_{ii},\;\; 2 \tilde{\epsilon} \sin \phi^b_j =    (T^{-1}_0  {\hat{E}}^b T_0)_{jj} = \bar{E} + \frac{1}{2} \omega^b \lambda^b (\sigma_3)_{jj}$.  Their solutions are
 \begin{eqnarray}  
 x_0 &=& \sigma_3 ( e^{\imath \sigma_3 \bar{\phi}} + \imath  ({2 \tilde{\epsilon} \eta_k})^{-1}{ \lambda \omega}) + O(\omega^2) = \sigma_3 \hat{\rho} + s, \; \; \; \sin \bar{\phi} \equiv \frac{\bar{E}}{2 \tilde{\epsilon}}
 \nonumber \\
 \hat{\rho} & = & \cos {\bar{\phi}} + \frac{\imath \lambda \omega}{4 \tilde{\epsilon}} + O(\omega^2), \; \; s = \imath \sin {\bar{\phi}} -  \frac{\lambda \omega}{4 \tilde{\epsilon}}  \tan {\bar{\phi}} + O(\omega^2) ;
 \nonumber \\
  \mathcal{L}_0 & = &\imath NV \sum_k \acute{L}_k  (  \phi_k + \cos\phi_k \sin \phi_k) - NV \sum_k \sin^2 \phi_k
 \nonumber \\
& = &  - NV 2^{-1}\tilde{\epsilon}^{-2}((\bar{E}^f)^2 + (\bar{E}^b)^2) + \imath  NV \tilde{\epsilon}^{-1} (\omega^b \lambda^b \cos \bar{\phi}^b + \omega^f \lambda^f \cos \bar{\phi}^f) - NV2^{-3} \tilde{\epsilon}^{-2} ( (\omega^b \lambda^b)^2 + (\omega^f \lambda^f)^2) + O(\omega^3)
\nonumber \\
& = &  - NV 2^{-1}\tilde{\epsilon}^{-2}((\bar{E}^f)^2 + (\bar{E}^b)^2) + \imath  NV \tilde{\epsilon}^{-1} (\omega^b \lambda^b \hat{\rho}^b + \omega^f \lambda^f \hat{\rho}^f) + NV2^{-3} \tilde{\epsilon}^{-2} ( (\omega^b \lambda^b)^2 + (\omega^f \lambda^f)^2) + O(\omega^3)
  \end{eqnarray}
Under the assumption that $ \omega \tilde{\epsilon}^{-1} \ll 1$, we drop the quadratic term in the last equation, set ${Tr}(\hat{E}^f \hat{E}^f) - (\bar{E}^f)^2 - (\bar{E}^b)^2 \approx 0$, set $\hat{\rho} = \cos \bar{\phi}$, and set $(2 \hat{\rho})^2(\prod_k \eta_k)^{-1/2} \approx 1$.   We also use the Stirling approximation $(N-1)! (N-2)! \approx 2 \pi N^{2N-2} e^{-2N} $ to simplify;
\begin{eqnarray}
\gamma e^{\mathcal{L}_0 + \tilde{\mathcal{L}}} & = &  {  N^{2NV + 2V}  }   {{{(\prod_{l=N-2}^{N-1}{l!})}}^{-V}}    2^{ -V -4}   {\pi}^{-3V}  V^{2V} \; e^{\mathcal{L}_I} \; {\det(\alpha + \kappa   )} \; \prod_k (\frac{2 \pi}{N \eta_k})^{V/2} \;\; (4 \hat{\rho}^2)^{2V}
\nonumber \\
& \times & \exp{(-2NV + {2^{-1} {NV}{ {\tilde{\epsilon}}^{-2}} {Tr}({\hat{E}}^f {\hat{E}}^f)}  -2^{-1} NV \tilde{\epsilon}^{-2}((\bar{E}^f)^2 + (\bar{E}^b)^2) + NV \imath \tilde{\epsilon}^{-1} (\omega^b \lambda^b \hat{\rho} + \omega^f \lambda^f \hat{\rho})  )}  
\nonumber \\
& \approx &  2^{ -4}  N^{2V} {\pi}^{-2V}  V^{2V}  (2 \hat{\rho})^{2V}
\; \exp{( \imath NV \tilde{\epsilon}^{-1} (\omega^b \lambda^b \hat{\rho} + \omega^f \lambda^f \hat{\rho})  )}  \; e^{\mathcal{L}_I} \; {\det(\alpha + \kappa   )} ,
\nonumber \\
\det( \alpha + \kappa  ) &=& (2 \hat{\rho})^{2V} \det(P_{+s}  ({\imath  (2 \tilde{\epsilon} \eta_k)^{-1} (\lambda^f \omega^f - \lambda^b \omega^b)} + x_0^b \sigma_3 k) P_{+s})  \;\; \det({P_{0s} \sigma_3 (\alpha + \kappa) P_{0s}}) 
\nonumber \\
P_{0s} \sigma_3 (\alpha + \kappa) P_{0s} & = & {\imath  (2 \tilde{\epsilon} \eta_k)^{-1} (\lambda^f \omega^f - \lambda^b \omega^b)} + \sigma_3 (\bar{x}^f_g + \bar{\tilde{x}}^f) - \sigma_3 (\bar{x}^b_g  + \bar{\tilde{x}}^b) + {V^{-1} x^b_0 \sigma_3 \sum_{v_1 v_2} (k T T^{-1} U U^{\dagger})_{v_1 v_2}}
\end{eqnarray}
In calculations of the two point correlator $P_s$ merges the $i$ and $j$ indices: $P_s = {\delta_{i_1 j_1}} {\delta_{i_1 i_2}}{\delta_{j_1 j_2}}{\delta_{v_1 v_2}}$.  In some places we have neglected the difference between $\bar{E}^f$ and $\bar{E}^b$, but this can be restored easily if needed.    

\subsection{Integration of the Fluctuations in $U$ and $T$}
We have already assumed spontaneous symmetry breaking, implying that fluctuations in $U$ and $T$ are small.  We now integrate the small fluctuations, reducing the model to zero-dimensional integrals.  For the moment we ignore $e^{\mathcal{L}_I}$'s dependence on these variables, and will return to this issue later.  Earlier we moved the zero-momentum component of $U, T$ into $U_0, T_0$; therefore $\sum_v y_v, z_v = 0$ and all of the terms in the Lagrangian which are linear in $y, z$ come to exactly zero.  The equations are further simplified by noting that the part of $x_0$ which is proportional to the identity does not play any role in $U$ and $T$'s dynamics.  To second order in $y, z$ the integrand and action are
 \begin{eqnarray} 
  \mathcal{L}_1 &=&  {{-2 \imath N \omega^f \lambda^f \hat{\rho}} {\tilde{\epsilon}}^{-1}  \sum_{v }  ((y^f_{v})^2 + (z^f_{v})^2)} 
  + {{2 \imath N \omega^b \lambda^b \hat{\rho}} {\tilde{\epsilon}}^{-1}   \sum_{v }   ((y^b_{v})^2 + (z^b_{v})^2)  } +{ O(\imath N \omega {\tilde{\epsilon}}^{-1} \sqrt{1 - \lambda^2}  (y^3, z^3))}
 \\
 & - & { \frac{8N}{2} \hat{\rho}^2 \sum_{v_1 v_2} {(k/(1-k))_{v_1 v_2} (y_{v_1}^f y_{v_2}^f + z_{v_1}^f z_{v_2}^f)} }
 - {\frac{8N}{2} \hat{\rho}^2 {{\sum_{v_1 v_2} {{k_{v_1 v_2} (y_{v_1}^b y_{v_2}^b + z_{v_1}^b z_{v_2}^b)} }}}} + O(Nky^4,\, Nkz^4)
\nonumber \\
&\,& {\frac{d}{d{\tilde{J}}^{b}_{vj}}} \rightarrow - \imath N  {\tilde{\epsilon}}^{-1} (\grave{s}_v^b +  \grave{c}_v^b \sigma_3 ( \lambda^b (1 + 2 y^b_v y^b_v + 2 z^b_v z^b_v) -  2 \sqrt{(\lambda^b)^2 - 1} (y_v \cos \theta_0^b + z_v \sin \theta_0^b)(1 + O(y^2, z^2))))_{jj},
\nonumber \\
\nonumber
&\,& \frac{d^2}{{d{\tilde{J}}^{b}_{v_1 j_1}}{d{\tilde{J}}^{b}_{v_2 j_2}}}  \rightarrow  {{\frac{d}{d{\tilde{J}}^{b}_{v_1 j_1}}} \otimes {\frac{d}{d{\tilde{J}}^{b}_{v_2 j_2}}}}, 
 \; \grave{c}_v = \hat{\rho} + (x^b_{gv1} - x^b_{gv2} + \tilde{x}_{v1}^b - \tilde{x}_{v2}^b )/2, \; \grave{s}_v = s +  (x^b_{gv1} + x^b_{gv2} + \tilde{x}_{v1}^b + \tilde{x}_{v2}^b)/2
\end{eqnarray}

We neglect the $(1-k)^{-1}$ controlling $Q^f$'s kinetics and perform the Gaussian integration, which generates a multiplicative constant equal to $(4 N  \hat{\rho}^2 / \pi)^{2 - 2V}$, plus two determinants.   The $c$ subscript on the expectation value in the following equation denotes the fully connected part of the expectation value. 
\begin{eqnarray}
{\bar{Z}} & = &  {  N^2  }   {    {\pi}^{ - 2}   }   \; \hat{\rho}^4 \; \int_{-1}^{+1} {d\lambda^f} \int_{1}^{\infty} {d\lambda^b}  \int {d\theta^f} {d\theta^b}  \;	 e^{ \imath \pi \rho (\omega^b \lambda^b  + \omega^f \lambda^f )} \;  \;  
\nonumber \\
& \times & e^{ \tilde{\mathcal{L}}} \; {\det}^{-1}(k P_+ + {\imath P_+ \omega^f \lambda^f }  (2 \hat{\rho} \tilde{\epsilon} )^{-1} ) \; {\det}^{-1}(k  P_+ - \imath P_+ \omega^b \lambda^b  (2 \hat{\rho} \tilde{\epsilon} )^{-1} ) 
\nonumber \\
e^{\tilde{\mathcal{L}}} & = & \langle \det(\alpha + \kappa) \;  e^{\mathcal{L}_I + \frac{N}{2}\sum_{vk} \eta_k x^2_{gvk}} \rangle_{U, T} \;,
\nonumber \\
\det( \alpha + \kappa  ) &=&  \det(P_{+s}  ({\imath  (2 \tilde{\epsilon} \eta_k)^{-1} (\lambda^f \omega^f - \lambda^b \omega^b)} + x_0^b \sigma_3 k) P_{+s})  \;\; \det({P_{0s} \sigma_3 (\alpha + \kappa) P_{0s}}) 
\nonumber \\
P_{0s} \sigma_3 (\alpha + \kappa) P_{0s} & = & {\imath  (2 \tilde{\epsilon} \eta_k)^{-1} (\lambda^f \omega^f - \lambda^b \omega^b)} + \sigma_3 (\bar{x}^f_g + \bar{\tilde{x}}^f - \bar{x}^b_g  - \bar{\tilde{x}}^b) + O(g^{-1})
\nonumber \\
\mathcal{L}_I &=& - { \frac{N}{2} \sum_{v_1 v_2} {(k/(1-k))_{v_1 v_2}}{Tr}{(U_{v_1}  \tilde{x}^f_{v_1} U^{\dagger}_{v_1}  U_{v_2}  \tilde{x}^f_{v_2}  U^{\dagger}_{v_2}  )} } 
+  {\frac{N}{2}{{\sum_{v_1 v_2} {{k_{v_1 v_2}} {Tr}{(T_{v_1}  \tilde{x}^b_{v_1} T_{v_1}^{-1} T_{v_2}  \tilde{x}^b_{v_2}   T_{v_2}^{-1}  )}}}}}
\nonumber \\
{\frac{d}{d{\tilde{J}}^{b}_{vj}}} &\rightarrow & - \imath N  {\tilde{\epsilon}}^{-1} (\grave{s}_v^b +  \sigma_3 (  \grave{c}_v^b \lambda^b (1 + 2 \langle (y_v^b)^2 +  (z_v^b)^2 \rangle)  ))_{jj} -2\sqrt{(\lambda^b)^2 - 1}\,\times O(g^{-3})
\nonumber \\
 \frac{d^2}{{d{\tilde{J}}^{b}_{v_1 j_1 = 1}}{d{\tilde{J}}^{b}_{v_2 j_2 = 2}}}  & \rightarrow &
-N^2  {\tilde{\epsilon}}^{-2} (\grave{s}_{v_1}^b +    \grave{c}_{v_1}^b \lambda_0^b (1 + 2 \langle (y_{v_1}^b)^2 + (z_{v_1}^b)^2 \rangle)    )  (   \grave{s}_{v_2}^b -  \grave{c}_{v_2}^b \lambda_0^b (1 + 2 \langle (y_{v_2}^b)^2 + (z_{v_2}^b)^2 \rangle)  )
 \nonumber \\ 
 &+& 4 N^2  {\tilde{\epsilon}}^{-2} \grave{c}_{v_1}^b \grave{c}_{v_2}^b  ((\lambda_0^b)^2 - 1) ( \langle y_{v_1}^b y_{v_2}^b \rangle_c \cos^2 \theta^b_0 + \langle z_{v_1}^b z_{v_2}^b \rangle_c \sin^2 \theta^b_0)(1 + O(g^{-2}))
 \nonumber \\
 & + & 4 N^2 \tilde{\epsilon}^{-2} \grave{c}_{v_1}^b \grave{c}_{v_2}^b(\lambda_0^b)^2 \langle (y_{v_1}^b)^2 (y_{v_2}^b)^2 + (z_{v_1}^b)^2 (z_{v_2}^b)^2 \rangle_c
\nonumber \\
\langle y_{v_1}^b y_{v_2}^b \rangle &=& (8 N  \hat{\rho}^2)^{-1} \langle v_1 | \Pi(-(2 \hat{\rho}\tilde{\epsilon})^{-1}\omega^b \lambda^b )  | v_2 \rangle (1 + O(g^{-2})), 
\nonumber \\
 \langle y_{v_1}^f y_{v_2}^f \rangle & = &  (8 N  \hat{\rho}^2)^{-1} \langle v_1 |  \Pi((2 \hat{\rho} \tilde{\epsilon})^{-1} \omega^f \lambda^f  ) | v_2 \rangle (1 +  O(g^{-2})),
\;\; \Pi(\gamma)  \equiv  (P_+ k + P_+ \imath \gamma)^{-1}
\label{SigmaModelReducedToZeroDims1}
\end{eqnarray}

$\Pi$ (conventionally called the diffuson propagator\cite{Andreev96, Mirlin99}) describes propagation of only non-zero-momentum states.  We will verify that $\rho = NV \hat{\rho} \pi^{-1} \tilde{\epsilon}^{-1}$ is the density of states.  Equation \ref{SigmaModelReducedToZeroDims1} is valid only if SSB occurs; in particular if SSB does not occur then we expect $\langle y_{v_1} y_{v_2} \rangle \propto 1$.  Moreover this equation is only the leading order result, and is subject to corrections:
\begin{itemize}
\item The kinetic term $\frac{N }{2} \hat{\rho}^2 \sum_{v_1 v_2} k_{v_1 v_2} {Tr}(T_{v_1} \sigma_3 T_{v_1}^{-1} T_{v_2} \sigma_3 T_{v_2}^{-1})$  contains every even power of $y^b, z^b$.   $U$'s kinetics are similar.  These terms generate perturbation theory vertices with even numbers of legs; every vertex is proportional to $N \hat{\rho}^2 k$.    The vertices, which we will call $1/g\hat{\rho}$ corrections, modify the free energy density and change the value of correlators like $\langle y_{v_1} y_{v_2} \rangle$.   (The conductance $g$ is defined\cite{GangOfFour} as $g \equiv E_{Th} / \Delta$, where $\Delta$ is the level spacing.  In Disertori's model $g = \hat{\rho} k_0 $.)
\item The observable $\frac{d}{d\tilde{J}^b}$ contains every odd power of $y^b, z^b$, causing more $1/g \hat{\rho}$ corrections.
\item The mass terms generate vertices with every odd power of $y, z$, and all are proportional to $N \hat{\rho} \omega \tilde{\epsilon}^{-1}$.  Like the $1/g\hat{\rho}$ vertices, these mass vertices modify both the free energy and observables.
\item $x_g$ occurs in many places, both in the Lagrangian and in observables.  It contains a kinetic term which is  proportional to $\eta^{-1} \hat{\rho} k$ and contains every even power of $y, z$, and also a mass term which  is proportional to $\eta^{-1} \tilde{\epsilon}^{-1} \omega$ and contains every odd power of $y, z$.  
\item $\mathcal{L}_I$ contributes additional terms with every even power of $y, z$, multiplied by $N k (x_g + \tilde{x})^2$. 
\end{itemize}

First we consider the $1/g\hat{\rho}$ corrections.  All of the vertices are proportional to $N \hat{\rho}^2 k$, while the bare propagator $\langle y_{v_1} y_{v_2} \rangle \propto \langle v_1 | (N \hat{\rho}^2 k)^{-1} | v_2 \rangle$.  Assuming that in perturbation theory diagrams each vertex with its associated $N \hat{\rho}^2 k$ roughly cancels a single propagator $\langle y_{v_1} y_{v_2} \rangle$, diagrams with $l$ loops are controlled by ${\langle y_{v_1} y_{v_2} \rangle}^{l-1} \propto (N \hat{\rho}^2 k)^{1 -l}$.  This inverse assumption breaks down in $D = \{1, 2 \}$ where SSB does not occur \cite{Mermin66} -   the diagonal elements of the propagator diverge in the $V \rightarrow \infty$ limit.  However, in finite volumes one still has $(N \hat{\rho}^2)^{1-l}$ scaling, so for large enough $N \hat{\rho}^2$ one obtains a controlled perturbative expansion.  

As long as $\omega (2 \hat{\rho} \tilde{\epsilon})^{-1} \ll k$, the mass vertices are much smaller than the $1/g\hat{\rho}$ vertices.  However once $\omega \tilde{\epsilon}^{-1}$ is of the same order as the conductance $g$ the mass vertices become more important.  At the same time the low-momentum behavior of the propagator changes to $\propto (4 N \hat{\rho} \, \omega / \tilde{\epsilon})^{-1}$.  In this regime the Feynman diagrams are controlled by powers of $(4 N \hat{\rho} \, \omega / \tilde{\epsilon})^{-1}$ instead of $1/g \hat{\rho}$.

We can now check the validity of the SSB assumption and the expansion in powers of $y, z$.   The leading $1/g\hat{\rho}$ correction to the free energy density is proportional to $ N \hat{\rho}^2 V^{-1} \sum_{v_1 v_2} k_{v_1 v_2} \langle x_{v_1} x_{v_2} \rangle^2 \propto (N \hat{\rho}^2 k)^{-1}$.  As long as this diagram is small compared to $1$, we may conclude that the perturbation theory is justified, and that the SSB assumption is correct.  In fact we have already seen that in $D > 2$ dimensions SSB does occur in the weak localization regime.  Even when $D = \{1, 2\}$ one still sees an effective symmetry breaking in small enough volumes \cite{Posazhennikova06}; the free energy density correction is a good gauge of how small is small enough. The $\hat{\rho}^{-2}$ in the correction is a sign that close to the band edge the $1/g\hat{\rho}$ perturbation theory breaks down and fluctuations become large. 
 
 Turning to $x_g$, we have already seen that $x_g / x_0 \ll N^{-1/2}$ is a necessary condition for control of the logarithmic vertices; if this condition is violated then the spatially uniform saddle point for the eigenvalues is invalid, and one must use a spatially fluctuating saddle point.  We can now estimate the magnitude of $x_g / x_0$ - its kinetic part is of order $ k \langle y_{v_1} y_{v_2} \rangle$.  If SSB is observed then this is $ \propto (N \hat{\rho}^2)^{-1} \ll N^{-1/2}$; however if SSB does not occur then the kinetic part is of order $k$ which may be much larger than $N^{-1/2}$.  Estimating the magnitude of $x_g / x_0$'s mass term requires a little more thought, since it contains odd powers of $y, z$ and therefore must always be paired with another $x_g$ or with a mass term.  The simplest estimate is $x_g^2 / x_0^2 \propto (2 \hat{\rho} \tilde{\epsilon})^{-2} \omega^2 \langle y_{v_1} y_{v_2} \rangle$; in the SSB regime one has the condition $\omega \tilde{\epsilon}^{-1} \ll 2^4 \hat{\rho}^3$.  
 
Lastly,   $\mathcal{L}_I$ is proportional to $k$ times all even powers of $y, z$.  As long as one is not calculating correlations of $\tilde{x}$, its effects are $1/N$ smaller than the other kinetic terms, and may be neglected.

\subsection{Correspondence to the SUSY Sigma Model}
The supersymmetric sigma model has been used to make detailed calculations of corrections to leading order results; see for example Blantner and Mirlin's work on correlations of eigenfunctions  [\onlinecite{Blantner97}].  These corrections were in powers of $1/g$, and correspond to our $1/g\hat{\rho}$ expansion.  The extra $\hat{\rho}$ in this paper's perturbation theory signals the fact that we have explicitly included band information while the SUSY model does not.

In the present model both $\mathcal{L}_I$ and the $x_g$ terms  arise from fluctuations in the eigenvalues; probably SUSY papers have not calculated them until now.  It seems likely that they have confined themselves to the present model's $1/g\hat{\rho}$ vertices and mass vertices.

\subsection{Fluctuations of the Saddle Point Signature\label{SaddlePointFluctuations}}
 Our saddle point analysis indicated the possibility that each site would separately choose its own saddle point signature $\acute{L}$, leading to a sort of Potts model.  Further analysis of this scenario seems hopeless, since one would have to integrate the $U$ fluctuations in the presence of a disordered background corresponding to the various site-wise configurations of $\acute{L}$.  Here we will establish that there is a single optimal value of $\acute{L}$ and that deviations from that value are penalized on a per-site basis by a free energy cost proportional to $\ln  k$.  
 
 Consider the two determinants in the denominator of equation \ref{SigmaModelReducedToZeroDims1}, which were obtained by integrating out the $U,T$ fluctuations.  They were obtained for the Retarded-Advanced saddle point  $\acute{L} = \{1, -1, 1, -1 \}$, which is equivalent to the $\{-1, 1, 1, -1 \}$ saddle point.    Now consider the other Retarded-Advanced saddle point $\acute{L} = \{1, 1, 1, -1\}$; at this saddle point the $U$ fluctuations are local and do not produce a determinant in the denominator.  The difference means that the former saddle point is exponentially favored  over the latter saddle point by $ ({ k + \imath(2 \hat{\rho} )^{-1} \omega \lambda \tilde{\epsilon}^{-1} })^{-V + 1}$.  
 
 Now consider site-wise fluctuations away from the favored $\acute{L} = \{1, -1, 1, -1 \}$ saddle point.  If a single site switches to the disfavored saddle point then the entire path integral is multiplied by ${ k + \imath(2 \hat{\rho} )^{-1} \omega \lambda \tilde{\epsilon}^{-1} }\ll 1$.  This per-site penalty controls the fluctuations and avoids the Potts model scenario. 

This argument can be generalized easily  to other saddle points.  When calculating the Advanced-Advanced correlator the favored saddle point is $\acute{L} = \{1, 1, -1 , -1\}$.  I have checked all saddle points required for three and four point correlators ($I^f = I^b = 4$).  The general rule seems to be that the favored saddle point minimizes $|{Tr}\acute{L}|$, and that the per-site free-energy cost  is roughly ${\frac{1}{4}({Tr}\acute{L})^2 \ln k}$.  These results were obtained for the Disertori model; the Wegner model requires more careful analysis before determining whether and how its saddle point signatures are regulated.  This is likely the only major remaining conceptual challenge concerning the Wegner model's weak localization regime.

\section{Observables\label{Observables}}
The supersymmetric sigma model has been used to calculate many different observables.  Here we calculate just a few, to check on agreement between the two models, and to finish our analysis of how to maintain mathematical control. We will drop $x_g$ and $1/g\hat{\rho}$ corrections.  Blantner and Mirlin\cite{Blantner97} found that while $1/g$ corrections are important for eigenfunction correlations, they are do not enter at lowest order into  the level correlator $R_2$.

\subsection{Large $\omega$ approximation}
Equation \ref{SigmaModelReducedToZeroDims1} has the following form:
\begin{equation}
{\bar{Z}}  =   {  N^2  }   {    {\pi}^{ - 2}   }   \; \hat{\rho}^4 \; \int_{-1}^{+1} {d\lambda^f} \int_{1}^{\infty} {d\lambda^b}  \int {d\theta^f} {d\theta^b}  \;	 e^{ \imath \pi \rho (\omega^b \lambda^b  + \omega^f \lambda^f )} \; Z_\lambda(\lambda^f, \lambda^b, \theta^f, \theta^b)
\end{equation} 
$Z_\lambda$ was calculated using a saddle point approximation which fails when $\omega^b \lambda^b$ is large enough to move the saddle point to the band edge. Our solution to this problem is to "pin" $\lambda^b$ by making $\omega^b$ large.  This does not actually restrict the $d\lambda^b$ integration; the real logic is as follows.  We assume that the $Z_\lambda$ is much larger inside the band than elsewhere, and that the proper way to control the $\lambda^b$ divergence is by losing information about the band edge.  Therefore we perform a Taylor series expansion of $Z_\lambda$ in powers of $\lambda^b - 1$ and then integrate.   When $ \pi \rho \omega^b \gg 1$ the exponential oscillates  very quickly, so that the $d\lambda^b$ integral is dominated by the $\lambda = 1$ limit of integration.  In this scenario the band edge really is unimportant, and we obtain a  systematic expansion of $\bar{Z}$ controlled by powers of $(\pi \rho \omega^b )^{-1}$. 

One may easily verify that this procedure is equivalent to changing coordinates to $\lambda^b = 1 + 2 (y^b_0)^2 + 2 (z^b_0)^2$, expanding $Z_\lambda$ in powers of $y$ and $z$, and then performing the resulting Gaussian integral. These are exactly the same coordinates which we used to integrate $T$'s spatial fluctuations; from a certain viewpoint we have simply done a Gaussian integration of all of the $T$ degrees of freedom together, including $T_0$.  Based on this viewpoint, Andreev and Altshuler\cite{Andreev95, Andreev96} claimed that they had applied the saddle point approximation to all of the $T$ degrees of freedom including $T_0$.  This explanation is a bit misleading, since properly speaking the saddle point approximation is a Taylor series expansion of the exponent, while in the $T_0$ integration the exponent is treated exactly at leading order in $1/g\hat{\rho}$.  The real $T_0$ physics is in the Taylor series expansion of $Z_\lambda$.

Andreev and Altshuler \cite{Andreev95, Andreev96} used this Taylor series expansion to perform both the $U_0$ and $T_0$ integrals.  When calculating two point correlators ($I^b = I^f = 2$) there are two saddle points corresponding to the two permutations of $\hat{E}^f_1, \hat{E}^f_2$; in the angular coordinates which we have chosen these permutations are represented by $\lambda_f = \pm 1$.  Altshuler and Shklovskii \cite{Altshuler86} computed the $\lambda^f = -1$ contribution nine years earlier than the Andreev-Altshuler publications.  They avoided  generating functions or sigma models and instead resummed perturbative expansions of the Green's functions, finding a smooth $\omega^{-2}$ behavior.  Later Andreev and Altshuler used the supersymmetric sigma model  to calculate $R_2$ completely, including both the smooth $\omega^{-2}$ part and an oscillatory part coming from the $\lambda_f = 1$ saddle point. Here we will reproduce and extend their results by applying the same saddle point approximation to the present sigma model.

The new coordinates are $\lambda^b = 1 + 2(y_0^b)^2 + 2(z_0^b)^2, \; \lambda^f = \pm ( 1 - 2(y_0^f)^2 - 2(z_0^f)^2)$, where the $\pm$ specifies the saddle point. The $y_0, z_0$ integrations should be done at the same time as the $y_v, z_v$ integrations. We revise equation \ref{SigmaModelReducedToZeroDims1} to match the altered procedure;
\begin{eqnarray}
{\bar{Z}} & = &   {\det}^{-1}(k  \pm {\imath   \omega^f  }  (2 \hat{\rho} \tilde{\epsilon} )^{-1} ) \; {\det}^{-1}(k   - \imath \omega^b  (2 \hat{\rho} \tilde{\epsilon} )^{-1} ) e^{\imath N V \hat{\rho} \tilde{\epsilon}^{-1} (\omega^b  \pm \omega^f )}
\nonumber \\
 &\times&  \det(P_{+s}  ({\imath  (2 \tilde{\epsilon} \eta_k)^{-1} (\lambda^f \omega^f - \lambda^b \omega^b)} + x_0^b \sigma_3 k) P_{+s})  \;\; \det({\imath  (2 \tilde{\epsilon} \eta_k)^{-1} (\lambda^f \omega^f - \lambda^b \omega^b)} + \sigma_3 ( \bar{\tilde{x}}^f  - \bar{\tilde{x}}^b)) 
\nonumber \\
& \,& {\frac{d}{d{\tilde{J}}^{b}_{v,j=2}}} \rightarrow  - \imath N  {\tilde{\epsilon}}^{-1} (\grave{s}_v^b -  (  \grave{c}_v^b  (1 + 2 \langle (y_v^b)^2 +  (z_v^b)^2 \rangle)  ))
\nonumber \\
& \, & \frac{d^2}{{d{\tilde{J}}^{b}_{v_1 j_1 = 1}}{d{\tilde{J}}^{b}_{v_2 j_2 = 2}}}   \rightarrow 
-N^2  {\tilde{\epsilon}}^{-2} (\grave{s}_{v_1}^b +    \grave{c}_{v_1}^b  (1 + 2 \langle (y_{v_1}^b)^2 + (z_{v_1}^b)^2 \rangle)     )  ( \grave{s}_{v_2}^b - \grave{c}_{v_2}^b  (1 + 2 \langle (y_{v_2}^b)^2 + (z_{v_2}^b)^2 \rangle)   )
\nonumber \\ 
 &+& 4 N^2 \tilde{\epsilon}^{-2} \grave{c}_{v_1}^b \grave{c}_{v_2}^b \langle (y_{v_1}^b)^2 (y_{v_2}^b)^2 + (z_{v_1}^b)^2 (z_{v_2}^b)^2 \rangle_c 
 \\ \nonumber
\langle y_{v_1}^b y_{v_2}^b \rangle &=& (8 N  \hat{\rho}^2)^{-1} \langle v_1 | (k -\imath(2 \hat{\rho} \tilde{\epsilon})^{-1}\omega^b  )^{-1}  | v_2 \rangle, 
\; \langle y_{v_1}^f y_{v_2}^f \rangle  =  (8 N  \hat{\rho}^2)^{-1} \langle v_1 |  (k \pm \imath ( 2 \hat{\rho} \tilde{\epsilon})^{-1} \omega^f  )^{-1} | v_2 \rangle,
\label{SigmaModelReducedToZeroDims2}
\end{eqnarray}

This can be simplified  a bit, dropping  $\tilde{x} \langle y^2 \rangle$ terms because they are $1/g\hat{\rho}$ smaller than the $\tilde{x}$ terms, and factoring out the zero modes.
\begin{eqnarray}
{\bar{Z}} & = &  4 \tilde{\epsilon}^2 \hat{\rho}^2 (\pm \omega^f \omega^b)^{-1} {\det}^{-1}(k P_+ \pm {\imath P_+ \omega^f  }  (2 \hat{\rho} \tilde{\epsilon} )^{-1} ) \; {\det}^{-1}(k  P_+ - \imath  P_+ \omega^b  (2 \hat{\rho} \tilde{\epsilon} )^{-1} )  e^{\imath N V \hat{\rho} \tilde{\epsilon}^{-1} (\omega^b  \pm \omega^f)}
\nonumber \\ 
 & \times &
 {\det({    {\imath P_+ (2 \tilde{\epsilon} \eta_1)^{-1} (\pm \omega^f - \omega^b)   } } +   { P_+  {   (\hat{\rho} + s)  k_{v_1 v_2} }    }   )} \; {\det({    {\imath P_+ (2 \tilde{\epsilon} \eta_2)^{-1} (\pm \omega^f - \omega^b)   } } +   {  { P_+  (\hat{\rho} - s)  k_{v_1 v_2} }    }   )}
 \nonumber \\
 & \times & 
 (\imath (2 \tilde{\epsilon} \eta_1)^{-1} (\pm \omega^f - \omega^b) + \bar{x}^f_1 - \bar{x}^b_1) \;
 (\imath (2 \tilde{\epsilon} \eta_2)^{-1} (\pm \omega^f - \omega^b) - \bar{x}^f_2 + \bar{x}^b_2)
\nonumber \\
& \,& {\frac{d}{d{\tilde{J}}^{b}_{v,j=2}}} \rightarrow  - \imath N  {\tilde{\epsilon}}^{-1} (x_{02} + \tilde{x}^b_{2v}   - 2 \hat{\rho} \langle (y_v^b)^2 +  (z_v^b)^2 \rangle)  
\nonumber \\
& \, & \frac{d^2}{{d{\tilde{J}}^{b}_{v_1 j_1 = 1}}{d{\tilde{J}}^{b}_{v_2 j_2 = 2}}}   \rightarrow 
 -N^2  {\tilde{\epsilon}}^{-2} (x_{01} + \tilde{x}_{1 v_1}^b + 2\hat{\rho}     \langle y_v^b y_v^b +  z_v^b z_v^b \rangle )
\; (x_{02} + \tilde{x}_{2 v_2 }^b - 2\hat{\rho}     \langle y_v^b y_v^b +  z_v^b z_v^b \rangle )
\nonumber \\
& + & 4 N^2 \tilde{\epsilon}^{-2} \hat{\rho}^2  \langle y_{v_1}^2 y_{v_2}^2 + z_{v_1}^2 z_{v_2}^2 \rangle_c
 \end{eqnarray}

Next we do the $\tilde{x}$ integrations, set $\omega^f = \omega^b$,  and make the $\omega \tilde{\epsilon}^{-1} \ll 1$ approximations $\eta_1 \eta_2 = 4 \hat{\rho}^2,\; x^b_{01} x^b_{02} =(\hat{\rho}+s)(-\hat{\rho}+s) \approx -1$.    The density of states can be determined using  $\rho = \pi^{-1} \sum_v {Im}( {dZ}/{dJ_{v, j=2}})$, and at leading order is given by $\rho = N V \hat{\rho} \pi^{-1} \tilde{\epsilon}^{-1} $.  We have specialized  the formalism to calculate the two point correlator ($I^f=I^b=2$), and in this case $\rho$ seems to be dressed by  corrections in powers of  $1/g\hat{\rho}$.  These $1/g\hat{\rho}$  corrections must  completely cancel, because the density may be calculated in the $I^f = I^b = 1$ case where there are no angular variables.  However the density may be dressed by $x_g$ corrections to the saddle point and by terms proportional to $k$ from the determinant. The second derivative is
\begin{eqnarray}
 & \, & V^{-2} \omega^{-2} \exp(\imath 2  \pi \rho \, \omega) \frac{{{\det}^2(   {2 \hat{\rho} \, \tilde{\epsilon} k P_+          }   )} }{{\det}(2 \hat{\rho} \, \tilde{\epsilon} k P_+ + {\imath \omega P_+   }  ) \; {\det}(2 \hat{\rho}  \, \tilde{\epsilon} k  P_+ - \imath  \omega  P_+    )}
 \nonumber \\
 & + &          ( (2 \hat{\rho} \tilde{\epsilon} k - \imath  \omega )^{-1}_{v_1 v_2})^2   +  N^2  \tilde{\epsilon}^{-2}   
-\imath 2  N   \tilde{\epsilon}^{-1}  \omega^{-1}    \hat{\rho}  ( 1 - \imath (2 \hat{\rho} \tilde{\epsilon})^{-1} \omega)^{-1} \langle v | (k - \imath (2 \hat{\rho} \tilde{\epsilon})^{-1} \omega)^{-1} | v \rangle    
\nonumber \\
&+ &    V^{-2}  ( {Tr}  (2 \hat{\rho} \tilde{\epsilon} k P_+ - \imath \omega P_+)^{-1}  )^2   
\end{eqnarray}
 Next we subtract the second derivative  at the Advanced-Advanced saddle point.  As we mentioned earlier, this saddle point does not involve any kinetics at highest order; it produces the zero-dimensional random matrix result, $-N^2 \tilde{\epsilon}^{-2} e^{-2 \imath \bar{\phi}}$.  To obtain the two point correlator $R_{2}(\omega)$, we take the real part,  sum over $v_1$ and $v_2$, and divide  by $2 \pi^2 \rho^2$:
\begin{eqnarray}
R_{2}(\omega)  - 1& = & (2 \pi^2 \rho^2)^{-1}  {Re} \,\,   {Tr}     ( (  2 \hat{\rho}\, \tilde{\epsilon} k - \imath  \omega  )^{-2}) 
\nonumber \\
&+& (2 \pi^2 \rho^2)^{-1}  \omega^{-2}    \cos(2  \pi \rho \, \omega) \frac{{{\det}^2(   {2 \hat{\rho} \, \tilde{\epsilon} k P_+          }   )} }{{\det}(2 \hat{\rho} \, \tilde{\epsilon} k P_+ + {\imath \omega P_+   }  ) \; {\det}(2 \hat{\rho}  \, \tilde{\epsilon} k  P_+ - \imath  \omega  P_+    )}
\nonumber \\
&+&   (2 \pi^2 \rho^2)^{-1}    {Re}\,  ( {Tr}  (2 \hat{\rho} \tilde{\epsilon} k P_+ - \imath \omega P_+)^{-1}  )^2 , \;\;\; \hat{\rho}  =  \cos \bar{\phi} + O(\omega)
\label{AndreevAltshulerCorrelator}
\end{eqnarray}
 To make contact with the SUSY result, we choose the kinetic operator to be $ k = -\frac{D}{2} \nabla^2$.  Except for the last term, our result for $R_{2}$ is the same as that obtained already by Andreev, Altshuler, and Shklovskii \cite{Altshuler86, Andreev95, Andreev96, Mirlin99}, but with added information about the energy band: we have a new $\hat{\rho}$ multiplying the kinetics.   The kinetics  are more sensitive to the band edge than the mass terms, and as a consequence $R_2$'s oscillatory component is amplified near the band edge.  To my best knowledge, this present article is the first derivation of the two point correlator which explicitly includes this energy band information. 
 
 The oscillatory term characteristic of Wigner-Dyson level repulsion would not exist if the eigenvalues obeyed the Gaussian statistics one would naively expect of massive modes in a sigma model; in that case the $\tilde{x}^b_{1v_1}$ and $\tilde{x}^b_{2 v_2}$ in the observables would integrate to zero.  Instead the $Q^f - Q^b$ determinant's factors of $\bar{x}^f - \bar{x}^b$ distort the probability distribution of the eigenvalues, causing a repulsive force between $\bar{x}^f$  and $\bar{x}^b$ which pushes $\bar{x}^f$ toward positive values and $\bar{x}^b$ toward negative values.  The origin of Wigner-Dyson level repulsion is in eigenvalue repulsion\footnote{This is likely the reason why replica calculations of the two point correlator are unable to obtain the oscillatory component unless they take into account replica symmetry breaking.}.  This non-trivial eigenvalue dynamics is hidden in the SUSY approach where eigenvalues are integrated quite early, but it is still there, working through the Grassman variables.

 The oscillatory term comes from the $\lambda = +1$ saddle point, and has a $\lambda = -1$ sister which contributes to the last term in $R_2$.  This sister term is of the same order as the other terms, and is controlled by the trace of $y^b, z^b$'s propagator.  It was not reported by Andreev and Altshuler, who may have believed that a factorizable term can not contribute to $R_2(\omega) - 1$.  Despite its appearances it is produced in a non-factorizable way: part of it comes from the $\langle \tilde{x}^b_{1v_1} \tilde{x}^b_{2v_2} \bar{x}^b_1 \bar{x}^b_2 \rangle$ expectation value, and can not be decomposed into $\langle \tilde{x}^b_{1v_1}  \bar{x}^b_1 \bar{x}^b_2 \rangle \times \langle  \tilde{x}^b_{2v_2} \bar{x}^b_1 \bar{x}^b_2 \rangle = 0 \times 0$.   Verification of whether the extra term is a bona fide improvement of Andreev and Altshuler's results would require evaluation of $1/g$ corrections which might be able to cancel it.   The extra term's structure can not be reproduced by $1/g$ corrections to the prefactors or the determinant, which leaves perturbative corrections from the Lagrangian as the only available cancellation mechanism.  Evaluation of such corrections is outside the scope of this paper.
 
\subsubsection{The Wegner Model}
The ratio of determinants controlling $R_2$'s oscillatory term is particularly interesting because it is sensitive to all of the theory's degrees of freedom.  The determinants in the numerator originate in the $P_{+s}$ sector of the $Q^f - Q^b$ determinant, while the determinants in the denominator come from integration of the the $U, T$ degrees of freedom.  The eigenvalue degrees of freedom also contribute to the denominator, but in Disertori's model these are exactly cancelled by the $P_{+l}$ sector of the $Q^f - Q^b$ coupling, so the net contribution of the massive modes is a factor of $1$.   This agrees with the SUSY treatment of the sigma model approximation, in which the massive modes integrate to exactly $1$.  Our partial analysis of the Wegner model leaves open the possibility that massive modes may contribute non-trivially to that model, especially if the $\omega$-induced mass of the $P_{+l}$ sector of the $Q^f - Q^b$ coupling could become larger than its small natural mass $2\hat{\rho} \ll 1$.   In this case one would see a knee in the oscillations' magnitude around the point where $\omega$ becomes a significant contribution to the eigenvalue mass.  Understanding this issue would require a careful solution of the Wegner saddle point equations and evaluation of Wegner's Hessian and determinant.

\subsection{Small $\omega$ approximation \label{SmallOmega}}
When $\omega $ is not much larger than the level spacing the band edge is important, or in other words the complete non-compact $T_0$ manifold must taken into account.   Therefore one must perform the $T_0$ integration prior to the saddle point approximation.  The only term in the Lagrangian (equation \ref{DisertoriModel}) which depends on $T_0$ is \begin{eqnarray}
{\imath N}{\tilde{\epsilon}}^{-1} & \sum_v &{Tr}(T_{v} T_0 x_v^b T_0^{-1} T_{v}^{-1} ({\hat{E}}^{b} - \tilde{J}^b)  )
\nonumber \\
&=& {\imath N}{\tilde{\epsilon}}^{-1}   \sum_v (\bar{E}^b + \bar{J}^b_v) {(x^b_{v1} + x^b_{v2})} 
+ {{\imath N}{\tilde{\epsilon}}^{-1} 2^{-1} \lambda^b_0 \sum_v (\omega^b - \delta \hat{J}^b_v) (x^b_{v1} - x^b_{v2}) (1 + 2 y^b_v y^b_v + 2 z^b_v z^b_v)} 
\nonumber \\
&-& {{\imath N}{\tilde{\epsilon}}^{-1}  \sqrt{\lambda^b_0 \lambda^b_0 - 1} \sum_v (\omega^b - \delta \hat{J}^b_v) (x^b_{v1} - x^b_{v2})  \sqrt{ 1 + y^b_v y^b_v + z^b_v z^b_v}\, (\cos \theta^b_0 y^b_v + \sin \theta^b_0 z^b_v)}
\end{eqnarray} 
Integration of $T_0$ at an early stage would result in an infinite-range interaction with all powers of $y,z$, causing problems for further analysis.  Therefore Kravtsov and Mirlin \cite{Kravtsov94} began by integrating the spatial fluctuations, producing a zero-dimensional model with special terms in the Lagrangian reflecting the fluctuations.  Afterwards they integrated the spatially uniform component of the SUSY matrix $Q$ over the complete domain of integration.  Kravtsov and Mirlin's results are complementary to Andreev and Altshuler's results; corrections to the former are well controlled when  $ \omega \tilde{\epsilon}^{-1}  \ll 2 \hat{\rho}  k_0$, while corrections to the latter are well controlled when $(N V \hat{\rho})^{-1} \ll \omega \tilde{\epsilon}^{-1}$.  Together they provide a complete description of the two point correlator for all values of $\omega$ which are well within the band edge.  Kravtsov and Mirlin's technique has been reused for calculating many other observables, including eigenfunction correlations, multi-fractal statistics of eigenfunctions, and $1/g$ corrections \cite{Mirlin00}.   Here we will reproduce their two point correlator by integrating $U_0$ and $T_0$ over the whole domain of integration. 

Kravtsov and Mirlin's approach requires doing the saddle point approximation last, and therefore is difficult to reconcile with the controlled approach we have adopted in this paper.   The saddle point approximation supplied information which allowed us to control the $Q^f - Q^b$ determinant and other issues, and was a cornerstone of our integration of the spatial fluctuations.  This is not a difficulty in the conventional SUSY derivation, which is not as attentive to control issues.  For us it is a difficulty.  It might be possible to simply assume the saddle point results, perform the integrations, and then justify the saddle point assumptions a posteriori.  More likely the correct approach is to start with the $T_0$ integration and then find some rigorous way of controlling the global interaction, perhaps by treating it separately from the rest of the theory.  Here we will not pursue such possibilities but instead imitate the SUSY approach, foregoing mathematical control.  

The main purpose of this calculation is to demonstrate that the $Q^f - Q^b$ models can reproduce the standard SUSY two point correlator at small $\omega$.  In particular we are neither looking for additional terms like the one we found in the previous section, nor searching for any other improvements.  Such efforts belong to a more considered and rigorous treatment.  Kravtsov and Mirlin's result depends only on the four point correlator $\langle y_{v_1}^2 y_{v_2}^2 \rangle$.  Since we only want to reproduce  Kravtsov and Mirlin's result, we will drop all higher order correlators and all occurences of the on-site two point correlator $\langle y_v^2 \rangle$.

We start with Disertori's model, as given in equation \ref{DisertoriModel}, and develop it as follows.  Using the Stirling approximation, the normalization constant becomes  $\gamma = N^{4V} 2^{-2V} \pi^{-4V}  e^{{\frac{NV}{2 {\tilde{\epsilon}}^2} {Tr}({\hat{E}} {\hat{E}})}   } e^{2NV}$.  It is necessary to delay the integration of spatially uniform values of $x^f$ and $x^b$ until the last moment,  but the spatial fluctuations in these variables can be integrated immediately, multiplying the path integral by $\prod_k (\frac{2 \pi}{N \eta_k})^{V/2} \approx (\pi/N\hat{\rho})^{2V-2}$. The angular variables are accompanied by a $2^{-4}$ Jacobian.   We assume that $x^f, x^b$ are close to the Retarded-Advanced saddle point $\acute{L}_k = \{1, 1, -1, -1\}$, and therefore set $x^f_1 - x^f_2 = x^b_1 - x^b_2 = 2\hat{\rho}$ everywhere except in the spatially uniform part of the mass term.  We use the weak localization form of the $Q^f - Q^b$ determinant given in  equation \ref{ZeroMomentumDecoupling2}.  We neglect corrections to the saddle point approximation, i.e. $x_g$ and $\mathcal{L}_I$, set  $\hat{E}^f = \hat{E}^b, \omega^f = \omega^b$, neglect $I^b N^{-1}$, and make several $\omega \tilde{\epsilon} \ll 1$ approximations.  The resulting path integral, prior to the next steps, is probably equivalent to the SUSY sigma model. 

Turning to the $Q^f - Q^b$ determinant, we assume that $k_0 \gg \{|x^f_1 - x^b_1|, |x^f_2 - x^b_2|\}$  and we expand the the $P_{+s}$ determinant in powers of $x^f - x^b$, keeping the second order contribution and dropping all other terms.  (These steps require that $ \omega \tilde{\epsilon}^{-1}  \ll 2 \hat{\rho}  k_0$.) The determinant becomes  $-(2 \hat{\rho})^{6V} V^2 \det^2(P_+ k) (x_1^f - x_1^b) (x_2^f - x_2^b) \exp(- 2^{-1} {Tr}(P_+ k^{-2}) \sum_i (x^b_i)^{-2} (x^f_i - x^b_i)^2 )$. Lastly we expand $U$ and $T$ in powers of $y, z$, and neglect all cubic and higher powers; these correspond to $1/ g\hat{\rho}$ corrections and may be easily restored if needed.  The result of these steps is   
  \begin{eqnarray}
{\bar{Z}} & = &- N^{2V+2} e^{2NV}     2^{ -6}   {\pi}^{ - 2V -2}     V^{2 } {\det}^2(P_+ k)   (2 \hat{\rho})^{4V+2}  e^{{\frac{NV}{2 {\tilde{\epsilon}}^2} {Tr}({\hat{E}} {\hat{E}})}   }
\nonumber \\
& \times & \int_{-1}^{1} { d\lambda_0^f } \int_{1}^{\infty} {d\lambda_0^b} \int_0^{2\pi} {d\theta_0^f} {d\theta_0^b} \;  {\int_{x^b L \geq 0} { dy^f } {dz^f} {dy^b} {dz^b} \;  {dx^b_1 dx^b_2 dx^f_1 dx^f_2}   \;     (x^f_1 - x^b_1) (x^f_2 - x^b_2) \;   e^{\mathcal{L}}  }
\nonumber \\
{\mathcal{L}}  & = &  - {\frac{NV}{2}{{\sum_{k}   x_{k}^2  }}} + {N V \sum_{k} \ln  x_{k}} - 2^{-1} {Tr}(P_+ k^{-2}) \sum_i (x^b_i)^{-2} (x^f_i - x^b_i)^2 
\nonumber \\
& + & \imath N V \tilde{\epsilon}^{-1} 2^{-1} (x^b_1 + x^b_2) (\hat{E}_1 - \bar{\tilde{J}}^b_1 + \hat{E}_2  - \bar{\tilde{J}}^b_2)  + \imath N V \tilde{\epsilon}^{-1} 2^{-1} (x^b_1 - x^b_2) (\omega - \delta \bar{\tilde{J}}^b)\lambda_0^b
\nonumber \\
& + & \imath N V \tilde{\epsilon}^{-1} 2^{-1} (x^f_1 + x^f_2) (\hat{E}_1 + \hat{E}_2 )  + \imath N V \tilde{\epsilon}^{-1} 2^{-1} (x^f_1 - x^f_2) (\hat{E}_1  - \hat{E}_2  )\lambda_0^f
 \nonumber \\ 
 & - & {4 N \hat{\rho}^2 \sum_{v_1 v_2} {k_{v_1 v_2} (y_{v_1}^f y_{v_2}^f + z_{v_1}^f z_{v_2}^f)} }
 - {4 N \hat{\rho}^2 {{\sum_{v_1 v_2} {{k_{v_1 v_2} (y_{v_1}^b y_{v_2}^b + z_{v_1}^b z_{v_2}^b)} }}}}
\nonumber \\
&+& {{ \imath 2 N \hat{\rho}}{\tilde{\epsilon}}^{-1}  \lambda^b_0 \sum_v (\omega - \delta \tilde{J}^b_v)  ( y^b_v y^b_v + z^b_v z^b_v)} 
- {{ \imath 2 N \hat{\rho}}{\tilde{\epsilon}}^{-1}   \sqrt{(\lambda^b_0)^2  - 1} \sum_v (\omega - \delta \tilde{J}^b_v)   (\cos \theta^b_0 y^b_v + \sin \theta^b_0 z^b_v)}
\nonumber \\
&-&  {{\imath 2 N \hat{\rho}}{\tilde{\epsilon}}^{-1}  \lambda^f_0 \omega  \sum_v   (  y^f_v y^f_v +  z^f_v z^f_v)} 
+ {{ \imath 2 N \hat{\rho}}{\tilde{\epsilon}}^{-1}  \omega \sqrt{1 - (\lambda^f_0)^2} \, \sum_v    (\cos \theta^f_0 y^f_v + \sin \theta^f_0 z^f_v)}
\end{eqnarray}
Next we integrate the spatial fluctuations $y, z$.  Because we have omitted all cubic and higher terms this Gaussian integral may be done exactly, but we follow Kravtsov and Mirlin and treat the $\omega$ terms as perturbations, assuming again that $ \omega \tilde{\epsilon}^{-1}  \ll 2 \hat{\rho}  k_0$.  In any case these terms must be treated perturbatively when we do the $\lambda$ integrals.  The integration multiplies the path integral by $\pi^{2(V-1)}  (2\hat{\rho})^{-4(V-1)} N^{-2(V-1)} {\det}^{-2}(P_+ k)$, and correlators are determined by $\langle y_{v_1} y_{v_2} \rangle = (8 N  \hat{\rho}^2)^{-1} \langle v_1 | (P_+ k )^{-1}  | v_2 \rangle$ and Wick's theorem.  To  determine the effective Lagrangian induced by the integration one calculates fully connected diagrams with no dangling legs.  As announced earlier we keep  only diagrams proportional to the fourth moment $\langle y_{v_1}^2 y_{v_2}^2 \rangle$.  The diagram with two factors of $\sqrt{1 - (\lambda^f_0)^2}$ is exactly zero because $\sum_{v_1} \langle y_{v_1} y_{v_2} \rangle = 0$.   
\begin{eqnarray}
{\bar{Z}} & = &- N^{4} e^{2NV}        {\pi}^{ -4}     V^{2 }    \hat{\rho}^{6}  e^{{\frac{NV}{2 {\tilde{\epsilon}}^2} {Tr}({\hat{E}} {\hat{E}})}   } \int_{-1}^{1} { d\lambda_0^f } \int_{1}^{\infty} {d\lambda_0^b} \int_0^{2\pi} {d\theta_0^f} {d\theta_0^b} \; {\int_{x^b L \geq 0} {dx^b_1 dx^b_2 dx^f_1 dx^f_2}   \;     (x^f_1 - x^b_1) (x^f_2 - x^b_2) \;   e^{\mathcal{L}}  }
\nonumber \\
{\mathcal{L}}  & = &  - {\frac{NV}{2}{{\sum_{k}   x_{k}^2  }}} + {N V \sum_{k} \ln  x_{k}} - 2^{-1} {Tr}(P_+ k^{-2}) \sum_i (x^b_i)^{-2} (x^f_i - x^b_i)^2 
\nonumber \\
& + & \imath N V \tilde{\epsilon}^{-1} 2^{-1} (x^b_1 + x^b_2) (\hat{E}_1 - \bar{\tilde{J}}^b_1 + \hat{E}_2  - \bar{\tilde{J}}^b_2)  + \imath N V \tilde{\epsilon}^{-1} 2^{-1} (x^b_1 - x^b_2) (\omega - \delta \bar{\tilde{J}}^b_v) \lambda_0^b 
\nonumber \\
& + & \imath N V \tilde{\epsilon}^{-1} 2^{-1} (x^f_1 + x^f_2) (\hat{E}_1 + \hat{E}_2 )  + \imath N V \tilde{\epsilon}^{-1} 2^{-1} (x^f_1 - x^f_2) \omega \lambda_0^f 
 \nonumber \\ 
&-& 2^{-3} \hat{\rho}^{-2} {\tilde{\epsilon}}^{-2}  \sum_{v_1 v_2} ( \omega^2 (\lambda^f_0)^2+ (\omega - \delta \tilde{J}^b_{v_1}) (\omega - \delta \tilde{J}^b_{v_2}) (\lambda^b_0)^2)  \langle v_1 | (P_+ k )^{-1}  | v_2 \rangle^2 
\nonumber \\
&-& 2^{-2} N {\tilde{\epsilon}}^{-2}  \sum_{v_1 v_2} (\omega - \delta \tilde{J}^b_{v_1}) (\omega - \delta \tilde{J}^b_{v_2}) ((\lambda^b_0)^2 - 1)  \langle v_1 | (P_+ k )^{-1}  | v_2 \rangle
\label{KravtsovIntermediate} 
\end{eqnarray}
We first outline the correct procedure for doing the angular integrals, and then take instead a shortcut.
One should first turn the terms which are quadratic in $\lambda$ into derivatives with respect to sources, then do the remaining trivial integral, and then take the derivatives, as follows: 
\begin{equation}
\int_1^\infty d\lambda^b_0 e^{\imath a \lambda^b_0 +  b (\lambda^b_0)^2} \approx \left[ \exp(b \, d^2/dl^2) \int_1^\infty d\lambda^b_0 e^{\imath a \lambda^b_0 + l \lambda^b_0}\right]_{l = 0}
 = \left[ -e^{b \, d^2/dl^2} (\imath a + l)^{-1} e^{\imath a  + l} \right]_{l = 0}
\label{CorrectSources}
\end{equation}
Applying this technique to equation \ref{KravtsovIntermediate}, one produces two factors in the denominator: $(l^b + \imath \pi \rho (\omega - \delta \bar{\tilde{J}}^b_v)  )^{-1} (l^f + \imath  \pi \rho \omega )^{-1}, \; \rho = N V \hat{\rho} \tilde{\epsilon}^{-1} \pi^{-1}$.  The $l$'s and $J$'s in these factors dress the observables, multiplying the number of terms which should be evaluated.  We simply neglect these terms, which are not necessary to reproduce Kravtsov and Mirlin's result.  Applied to equation \ref{CorrectSources}, neglecting the sources in the denominator produces the final result $\imath a^{-1} \exp(\imath a  + b)$. This procedure is justified only when $a = \pi \rho \omega \gg 1$, but one may verify that in the zero-dimensional case the real parts of the neglected terms completely cancel, so there is no effect on the $D = 0$ final result.   
\begin{eqnarray}
{\bar{Z}} & = &  N^2 V^2       2^2   {\pi}^{ -2}       \hat{\rho}^{4} \tilde{\epsilon}^{2} \omega^{-2} e^{2NV} e^{{\frac{NV}{2 {\tilde{\epsilon}}^2} {Tr}({\hat{E}} {\hat{E}})}   }
 \sum_{\pm} \mp   {\int_{x^b L \geq 0} {dx^b_1 dx^b_2 dx^f_1 dx^f_2}  \;       (x^f_1 - x^b_1) (x^f_2 - x^b_2) \;  e^{\mathcal{L} + \mathcal{L}_{eff}}  } \;
 \nonumber \\ 
\mathcal{L}_{eff} &=& - {2^{-3} \hat{\rho}^{-2}  {\tilde{\epsilon}}^{-2}    \sum_{v_1 v_2}(\omega^2 + (\omega - \delta \tilde{J}^b_{v_1}) (\omega - \delta \tilde{J}^b_{v_2})) \langle v_1 | (P_+ k )^{-1}  | v_2 \rangle^2 } 
- 2^{-1} {Tr}(P_+ k^{-2}) \sum_i (x^b_i)^{-2} (x^f_i - x^b_i)^2 
\nonumber \\
{\mathcal{L}}  & = &  - {\frac{NV}{2}{{\sum_{k}   x_{k}^2  }}} + {N V \sum_{k} \ln  x_{k}} 
- \imath N V \tilde{\epsilon}^{-1}   (\bar{\tilde{J}}^b_1 x^b_1 +   \bar{\tilde{J}}^b_2 x^b_2) 
\nonumber \\ 
&+ &\imath N V \tilde{\epsilon}^{-1}   ((\bar{E} + \omega/2) x^b_1 + (\bar{E} - \omega /2) x^b_2)  
 +  \imath N V \tilde{\epsilon}^{-1} ((\bar{E} \pm \omega/2) x^f_1 + (\bar{E} \mp \omega /2) x^f_2)  
\end{eqnarray}
Next we take derivatives with respect to the sources.  Corrections to the first derivative may be dropped because in the $I^f = I^b = 1$ calculation of the same quantity there are no angular variables and therefore no corrections; $ \frac{d}{dJ^b_{jv}}  \rightarrow  - \imath N \tilde{\epsilon}^{-1}   x^b_j $.   Taking the second derivative produces a term which is proportional to $\omega$ and can be neglected because it is purely imaginary, and another term proportional to $\omega^2$ which we neglect because it contains the fourth power of $\langle v_1 | (P_+ k )^{-1}  | v_2 \rangle$.  The remaining terms are
\begin{eqnarray}
\frac{d^2}{dJ^b_{1v_1}dJ^b_{2v_2}} & \rightarrow &  2^{-2}{\tilde{\epsilon}}^{-2} \hat{\rho}^{-2}    \langle v_1 | (P_+ k )^{-1}  | v_2 \rangle^2 
-N^2  \tilde{\epsilon}^{-2}   x^b_1 x^b_2  
\end{eqnarray}
Lastly we perform the saddle point approximation.  We neglect the effective Lagrangian's contribution to the saddle point equations, which is supressed by $N^{-1}$.  The saddle point solutions are $x^b_1 = \exp(\imath \phi_+), x^b_2 = -\exp(-\imath \phi_-), x^f_1 = \exp(\imath \phi_\pm), x^f_2 = -\exp(-\imath \phi_\mp)$, where $\sin \phi_\pm = (\bar{E} \pm \omega /2)/(2 \tilde{\epsilon})$.  The saddle point action is $-2NV - NV\bar{E}^2 \tilde{\epsilon}^{-2} + \imath (1 \pm 1) \rho \omega $. We shift $x^f$ and $x^b$ to match the saddle points;  $x^f_i - x^b_i \rightarrow x^f_i - x^b_i +  \imath (\sigma_3)_{ii} (1 \mp 1) \omega (2 \tilde{\epsilon} \eta_i)^{-1}$.   After performing this shift we neglect the remaining $x^f - x^b$ occuring in $\mathcal{L}_{eff}$ and simplify; in the $+$ term $\mathcal{L}_{eff} = -  {  \omega^2 (2 \hat{\rho} {\tilde{\epsilon}})^{-2}    {Tr}(P_+ k^{-2}) }$, while in the $-$ term it is zero.        
\begin{eqnarray}
{\bar{Z}} & = & N^{2}  V^2   2^{2}   {\pi}^{ -2}        \hat{\rho}^{4} \tilde{\epsilon}^{2} \omega^{-2} \sum_{\pm} \mp e^{\imath(1 \pm 1) \rho \omega} \int_{x^b L \geq 0} {dx^b_1 dx^b_2 dx^f_1 dx^f_2}  \;       (x^f_1 - x^b_1 + (1 \mp 1)\imath \omega (2 \tilde{\epsilon} \eta_1)^{-1})
\nonumber \\
& \times &     (x^f_2 - x^b_2 -  (1 \mp 1)\imath \omega (2 \tilde{\epsilon}\eta_2)^{-1} ) \;  \exp(\mathcal{L}_{eff} - {2^{-1}{NV}{{\sum_{k} \eta_k  x_{k}^2  }}} )
 \nonumber \\ 
\nonumber \\
 \frac{d^2}{dJ^b_{1v_1}dJ^b_{2v_2}}  &\rightarrow & 2^{-2}{\tilde{\epsilon}}^{-2} \hat{\rho}^{-2}   \langle v_1 | (P_+ k )^{-1}  | v_2 \rangle^2 
-N^2  \tilde{\epsilon}^{-2}  (\exp(\imath \phi) + x^b_1) (-\exp(-\imath \phi) +  x^b_2)  
\end{eqnarray}
The remaining Gaussian integrals multiply the result by $ \pi^2 N^{-2}  {\hat{\rho}}^{-2}$.  The terms which are proportional to $\omega$ are purely imaginary and may neglected.   We expand $\exp(\mathcal{L}_{eff}) \approx 1 + \mathcal{L}_{eff}$, drop a term proportional to the fourth power of $\langle v_1 | (P_+ k )^{-1}  | v_2 \rangle$, and simplify.
 \begin{eqnarray}
\rho  & = &  \pi^{-1} \sum_v {Im} \frac{d\bar{Z}}{dJ^b_{2v}}=  \pi^{-1} \sum_v {Im}( \imath  N  \tilde{\epsilon}^{-1}   \exp(-\imath  \phi) ) =  N V \hat{\rho} \pi^{-1} \tilde{\epsilon}^{-1} 
\\ \nonumber 
  \frac{d^2 \bar{Z}}{dJ^b_{1v_1}dJ^b_{2v_2}} &= &   (2   \tilde{\epsilon} \hat{\rho})^{-2}    {\langle v_1 | (P_+ k)^{-1} | v_2 \rangle}^2  
- V^{-2}        \omega^{-2}   
+    N^2     \tilde{\epsilon}^{-2}      
 -   V^{-2}  e^{2\imath \rho \omega}  (2   \tilde{\epsilon} \hat{\rho})^{-2}  {Tr}(P_+ k^{-2})
+ V^{-2}        \omega^{-2} e^{2\imath \rho \omega}  
\end{eqnarray}
To obtain the two point correlator we subtract the Advanced-Advanced second derivative $-N^2 \tilde{\epsilon}^{-2} e^{-2 \imath \bar{\phi}} $, take the real part, sum over $v_1, v_2$, and divide by $2 \pi^2 \rho^2$:
 \begin{eqnarray}
R_2(\omega) - 1& = &     (2 \pi^2 \rho^2)^{-1}  (\cos(2 \rho \omega) - 1)  (\omega^{-2}  - (2   \tilde{\epsilon} \hat{\rho})^{-2}    {Tr}(P_+ k^{-2}) )        
\end{eqnarray}
We choose $k =\frac{D}{2}\nabla^2$ to make contact with Kravtsov and Mirlin's result\cite{Kravtsov94}, which is identical to ours except for the extra band information contained in $\hat{\rho}$.

\subsection{Anomalously Localized States}
Until now we have considered only spatially uniform saddle points.  Other saddle points which are not spatially uniform may also be important - for instance instantons, skyrmions, etc.  Muzykantskii and Khmelnitskii studied non-uniform saddle points in the supersymmetric sigma model and found that they control the conductance in the long time limit \cite{Muzykantskii95}.  Their non-uniform saddle points correspond to statistically rare eigenfunctions whose weight is concentrated in a small volume - these are called anomalously localized states \cite{Mirlin97, Nikolic01a, Uski00}.  In the weak localization regime  ALS  control the asymptotic tails of the probability distributions of many observables; see Ref. \onlinecite{Mirlin00} for a review.  Here we will show that the $Q^f-Q^b$ Lagrangian produces the same ALS already seen in the SUSY sigma model.  We will derive the same equations controlling the ALS which Muzykantskii and Khmelnitskii used as the starting point for their calculations.

We begin by fixing the saddle point and parameterizing $U, \, T, \, Q^f = U x_0^f U^\dagger,$ and $Q^b L = T x_0^b T^{-1}$ as
\begin{eqnarray}
U &=&  \begin{bmatrix} \cos \psi^f & \imath \sin \psi^f e^{-\imath \theta^f}  \\ \imath \sin \psi^f e^{\imath \theta^f}  & \cos \psi^f \end{bmatrix}, \;\;
T = \begin{bmatrix} \cosh \psi^b & \sinh \psi^b e^{-\imath \theta^b} \\ \sinh \psi^b e^{\imath \theta^b} & \cosh \psi^b  \end{bmatrix}
\nonumber \\
Q^f &=& s + \hat{\rho} \begin{bmatrix} \cos (2 \psi^f) & -\imath \sin (2\psi^f)  e^{-\imath \theta^f} \\ \imath \sin ( 2 \psi^f) e^{\imath \theta^f} & - \cos (2 \psi^f) \end{bmatrix}, 
\;\; Q^b L =   s + \hat{\rho} \begin{bmatrix} \cosh (2 \psi^b) & -\sinh (2\psi^b)  e^{-\imath \theta^b} \\ \sinh ( 2 \psi^b) e^{\imath \theta^b} & - \cosh (2 \psi^b) \end{bmatrix}
\end{eqnarray}
The inverses of $U$ and $T$ can be obtained by adding $\pi$ to $\theta^f$, $\theta^b$.  $\psi^b$ varies from $1$ to $\infty$,  $\psi^f$ varies from $0$ to $\pi /2$, and $\theta^f,\, \theta^b$ vary from $0$ to $2 \pi$.    The  Jacobian is $\frac{1}{4}\sin(2 \psi^f) \sinh (2 \psi^b)$, and this must be multiplied by another $1/2$ because the $Q^f$ parameterization covers its domain twice.

We neglect contributions to the Lagrangian (equation \ref{EigAngleSaddlePointLagrangian}) from the Jacobian, from fluctuations in the eigenvalues, and from the $(1-k)^{-1}$ factor controlling $Q^f$'s kinetics.  We have seen that in the weak localization regime the $Q^f - Q^b$ coupling reduces to a zero-momentum coupling, so we neglect its effects on the ALS. Lastly, we set $\hat{E}^f = \hat{E}^b$. The remaining Lagrangian is
\begin{eqnarray}
\mathcal{L} & = & \imath N \tilde{\epsilon}^{-1} \sum_v {Tr}(Q^f_v \hat{E}) + \imath N \tilde{\epsilon}^{-1} \sum_v {Tr}(Q^b_v L \hat{E})
- {\frac{N}{2}\sum_{v_1 v_2} k_{v_1 v_2} {Tr}(Q^f_{v_1} Q^f_{v_2})} + {\frac{N}{2}\sum_{v_1 v_2} k_{v_1 v_2} {Tr}(Q^b_{v_1} L Q^b_{v_2} L)}
\end{eqnarray}

Following Efetov, we choose $k = \frac{D}{2} \nabla \cdot \nabla, \; N \sum_v = \pi \nu \int {dr} $.  The Lagrangian becomes
\begin{eqnarray}
-\mathcal{L} & = & \int {dr} \left[{\frac{\pi \nu D}{2} \hat{\rho}^2 (2 \nabla \psi^b )\cdot (2 \nabla \psi^b)}  - {\imath {\pi \nu } \hat{\rho} \, \omega  \tilde{\epsilon}^{-1}  \cosh(2 \psi^b)} + {\frac{ \pi \nu D}{2} \hat{\rho}^2 (2 \nabla \psi^f) \cdot (2 \nabla \psi^f)} - {\imath {\pi \nu } \hat{\rho} \, \omega   \tilde{\epsilon}^{-1}  \cos(2 \psi^f) } \right ]
\nonumber \\
& + &  \frac{\pi \nu D}{2} \hat{\rho}^2 {\int {dr} \sin^2 (2 \psi^f) \nabla \theta^f \cdot \nabla \theta^f + \sinh^2 (2 \psi^b) \nabla \theta^b\cdot \nabla \theta^b} 
\label{ALSAction}
\end{eqnarray}
Clearly  $-\mathcal{L}$ is minimized if $\theta^f$ and $\theta^b$ are constant, in which case the last line of equation \ref{ALSAction} is exactly zero.  $-\mathcal{L}$ should be compared to Muzykantskii and Khmelnitskii's equation 7:
\begin{equation}
A = \frac{\pi \nu}{2}\int {dr}  \{ \left[D (\nabla \theta)^2 - 2 \imath \omega \cosh \theta \right] + \left[D (\nabla \theta_1)^2 + 2 \imath \omega \cos \theta_1 \right] \}
\end{equation}
Their $\theta$, $\theta_1$, and $\omega$ should be identified with this paper's $2 \psi^b$, $2 \psi^f - \pi$, and $\omega \tilde{\epsilon}^{-1}$ respectively.  Except for the band information contained in $\hat{\rho}$, the two equations are identical.  Furthermore Muzykantskii and Khmelnitskii's ${Str}(\Lambda Q) = 4(\cos \theta_1 - \cosh \theta)$ is equal to this paper's $\gamma ({Tr} (Q^f_v \sigma_3) +  {Tr}(Q^b_v \sigma_3)) = 2 \gamma \hat{\rho}(\cos(2 \psi^f) -\cosh(2 \psi^b))$, where $\gamma$ is a positive multiplicative constant.  They impose the boundary conditions $\theta = \theta_1 = 0$ at the interface with an ideal conductor, and it seems reasonable to conclude that the corresponding boundary condition $\theta^b = 0, \, \theta^f  = \pi / 2$ should be used in this paper's model.  With these correspondences in place, every result in  Muzykantskii and Khmelnitskii's article on ALS \cite{Muzykantskii95} follows from this paper's model.  

\section{Assumptions and Mathematical Control\label{MathematicalControl}}

Having completed the process of calculating observables, we now review our assumptions and the small parameters which we used to maintain mathematical control.  The really weighty control issues were centered on two tasks: deriving a sigma model via controlled integration of the eigenvalues, and controlling the $Q^f - Q^b$ determinant. 

Everything rests on the assumption that a good saddle point can be obtained while naively neglecting the $Q^f - Q^b$ determinant and Van der Monde determinants.  (One can perform a saddle point analysis including the determinants by moving their logarithm into the Lagrangian and then performing a Taylor series expansion.  While the results are qualitatively correct, $N^{-1/2}$ no longer regulates the eigenvalue integration and one loses mathematical control.  Mathematical control requires that the determinants be treated separately from the saddle point approximation.) Inclusion of the determinants in the integral must not cause changes larger than $N^{-1/2}$ in the mean value of $x^b, x^f$.  Our later analysis justified this assumption in the weak localization regime.  Outside of this regime two mechanisms protect it: (1) $N$ multiplies the saddle point action but does not occur in the determinants. (2) The determinants always increase the free energy; they create repulsive not attractive forces.  The true ground state is likely to be one that minimizes the free energy associated with the determinants, and therefore should have some clear connection with the naive ground state.

Our controlled derivation of the sigma model was based on six small parameters:
\begin{itemize}
\item $(N \eta)^{-1} \ll 1$. Consequently the sigma model breaks down when $\hat{E}$ is close to the band edge. As a further consequence we are required to control the non-compact $T$ integrals.  This is only possible in the SSB regime.
\item $ (\pi \omega \rho)^{-1} \ll 1$.  This constraint controls the non-compact integration of $T$'s spatially uniform component.   If it is violated the band edge becomes important and one must integrate $T_0$ prior to the saddle point approximation.  This second approach works only when $ \omega  (2 \hat{\rho} E_{Th} )^{-1} \ll 1$.   
\item Spontaneous symmetry breaking; $  \langle y_{v_1} y_{v_2} \rangle \propto 1/g\hat{\rho} \ll 1$; validity of perturbation theory in $y, z$.  This mechanism controls the non-compact integration of $T$'s spatial fluctuations.
\item The diffusive regime, $k \ll 1$.  The condition $1 - k > 0$ was  essential to the Hubbard-Stratonovich transformation of the fermionic sector.  Additionally the small parameter $k$ allowed us to approximate the eigenvalue fluctuations as being local, and to control deviations from locality with perturbation theory.  These issues were packaged in $\mathcal{L}_I$.  We eventually argued that $\mathcal{L}_I$ does not contribute at leading order in $N$ unless one is calculating eigenvalue correlations.  While eigenvalue locality may not be absolutely essential to mathematical control, it circumvents a host of complications concerning the integration of eigenvalue fluctuations within the determinants and observables. 
\item $\omega (2 \hat{\rho}  \tilde{\epsilon})^{-1} \ll 1, \; k \ll 1 $.  Our naive saddle point analysis predicted several saddle points at each site corresponding to different values of the saddle point signature ${Tr}(\acute{L})$. In order to avoid a Potts model scenario we had to argue that all sites share the same global saddle point.  Our argument rested heavily on the assumption that $\omega$ and $k$ are small compared to the band width, in which case the determinant resulting from integrating $U$'s fluctuations selects a single optimal saddle point and imposes a per-site energy penalty for deviations from that saddle point.  If $\omega$ or $k$ were of the same order as the band width then this argument would collapse and we would have a Potts model.  
\item $x_g / x_0 \ll N^{-1/2}$.  The previous small parameters completely justify a saddle point integration, but the saddle point is not spatially uniform; it depends on the configuration of the angular variables $U$ and $ T$.    Wanting to maintain contact with the SUSY sigma model and to avoid rather daunting mathematical complications, we made a further approximation and chose a uniform saddle point.  The difference between this saddle point and the true saddle point can be managed with perturbation theory only if the two are reasonably close to each other.  $x_g$, which quantifies the fluctuating forces caused by $U$ and $T$, must be small. This requirement requires spontaneous symmetry breaking and constrains $\omega$.   If these requirements are not fulfilled then one loses mathematical control and must revert to the non-uniform saddle point, obtaining a different and much more complicated sigma model.
\end{itemize}

Having derived the sigma model, we turned to the task of controlling the Disertori model's $Q^f - Q^b$ determinant.  The small $k$ and $\omega$ parameters  are the cornerstones of this analysis,  allowing us to identify two sectors in the determinant and establish that the two don't mix significantly. There was also an auxiliary assumption of spontaneous symmetry breaking.  The small parameter $N^{-1}$, in conjunction with our choice not to calculate $Q^f - Q^b$ correlations, allowed us to neglect the determinant's contributions to the action.   With this foundation in place, we introduced a new small parameter:
\begin{itemize}
\item Weak Localization Regime: $(N^{1/2} E_{Th} / \tilde{\epsilon})^{-1}  = (N^{1/2} k_0)^{-1} \ll 1$.  In this regime the determinant is dominated by the kinetic operator $k$; it simplifies to a zero-momentum coupling responsible for Wigner-Dyson level spacing statistics, plus a normalization constant. 
\end{itemize}
It was also convenient to take $(g \hat{\rho} N^{1/2})^{-1} \ll 1$, but this assumption is not necessary for control of the $Q^f - Q^b$ determinant.

In the weak localization regime the $Q^f-Q^b$ model is dominated by the physics of  $U$ and $T$. Our SSB assumption gave birth to a machinery for calculating observables.  We saw that in $D > 2$ dimensions the SSB assumption is self-consistent in the weak localization regime, since the $Q^f - Q^b$ model is (excepting prefactors from the zero-momentum sector of the determinant) a product of two theories which have each been rigorously proved to exhibit SSB.  In $D = \{1,2 \}$  dimensions SSB is expected to still occur in small enough volumes, and we have provided a formula for estimating the numerical accuracy of the SSB assumption in any particular system.

During calculation of the two level correlator we made additional assumptions:
\begin{itemize}
\item We dropped all instances of $x_g$ and $1/g\hat{\rho}$.  These corrections can be computed easily enough, as demonstrated by analogous SUSY calculations of $1/g$ corrections.
\item We again exploited $\omega \tilde{\epsilon}^{-1} \ll 1, \, k \ll 1$. These are conveniences introduced for the purpose of analytical treatment, and could be easily avoided at the cost of complicated formulas or numerical integration. 
\end{itemize}

\section{Outlook}
The SUSY sigma model is widely understood as an effective field theory describing completely the long-wavelength physics of disordered systems.  While this model has had its greatest successes in the weak localization regime and in the exactly solvable cases of $D = \{0,1\}$ dimensions, it also gives a picture of the  more strongly disordered conducting regime and of the Anderson transition.  Its picture describes the weak localization regime in terms of spontaneous symmetry breaking and small fluctuations of the order parameter.  In one or two dimensions the Mermin-Wagner theorem \cite{Mermin66} guarantees that fluctuations will become large in the $V \rightarrow \infty$ limit, corresponding to Anderson localization.  In $D > 2$ dimensions the Anderson transition is expected to be analogous to the SSB phase transition that occurs in other $D > 2$ systems; at low temperature/disorder SSB occurs, while at high temperature/disorder SSB is frustrated. Within the framework of the SUSY sigma model, this intuition suggests that under renormalization group transformations (i.e. increasing the volume) the model will evolve from the weak localization regime into one of two stable fixed points: a localized fixed point and a delocalized fixed point.  The disorder strength determines the flow's starting point, while the SUSY group structure and the system geometry control the flow.  The SSB/renormalization picture is complicated by the expectation that the order parameter is a function \cite{Evers08}, and by debate about whether scaling is controlled by a single parameter \cite{Queiroz02, Kantelhardt02, Prior05, Evers08}.

This present work gives a different picture of disordered systems.  First of all, the sigma model approximation breaks down when SSB is frustrated, including at the band edge.  More important are our results about the weak localization regime in $D > 2$ dimensions.  We have seen that in this regime renormalization group transformations do not under any conditions evolve into a localized fixed point, which is especially striking because the same theory reproduces and extends the SUSY weak localization results, including anomalously localized states.  This conflicts with the conventional wisdom  that ALS in a given dimension $D$ are precursors of localization; i.e. signals of a competing localized fixed point existing in the same dimension $D$.  We conclude that while the SSB/renormalization picture is able to give an account of localization in $D = \{1,2\}$ dimensions and near the band edge in higher dimensions, it does not give the full story about localization in the bulk of the spectrum.

The most novel feature of the $Q^f - Q^b$ models is  the spectral determinant coupling $Q^f$ with $Q^b$.  We have seen that the weak localization regime is characterized by the condition that this determinant be dominated by the kinetic operator $k$.  $k$'s smallest eigenvalues scale as $|\vec{s}|^2 \propto V^{-2/D}$;  as the system size is increased their control of the determinant will cease.  In the $V \rightarrow \infty$ limit the low end of the determinant's spectrum will be controlled by $Q$, not $k$.   This suggests a new mechanism of localization in the bulk of the spectrum: the determinant's small eigenvalues may amplify $Q$'s long-wavelength fluctuations.

\begin{acknowledgments}
I would like to express thanks for stimulating conversations with Margherita Disertori, Yan Fyodorov, Martin Zirnbauer, Jacobus Verbaarschot, M.A. Skvortsov, Alexander Mirlin, John Keating, and Konstantin Efetov.  I am grateful for support from Giorgio Parisi, Tom Spencer, the Isaac Newton Institute, the Universit\`a degli Studi di Roma "La Sapienza", the ICTP, and the APCTP. 
\end{acknowledgments}

\bibliography{vincent}

\end{document}